\documentclass[11pt]{article}
\usepackage{axodraw}
\usepackage{epsfig}
\usepackage{amsfonts}
\usepackage{amsmath}
\usepackage{bbm}
\input pix.sty

\renewcommand{\eq}{eq.~}
\renewcommand{\eqs}{eqs.~}
\renewcommand{\se}{sec.~}

\renewcommand{\fig}{fig.~}
\renewcommand{\figs}{figs.~}
\newcommand{\dd}{\mathrm{d}}
\newcommand{\tinymsbar}{{\overline{\mbox{\tiny\rm{MS}}}}}
\newcommand{\Lambdamsbar}{{\Lambda_\tinymsbar}}
\newcommand{\mD}{m_\rmi{D}}

\newcommand{\Nf}{N_{\rm f}}
\newcommand{\Nc}{N_{\rm c}}

\newcommand{\rmO}{{\mathcal{O}}}
\newcommand{\bmu}{\bar\mu}

\def\lsi{\raise0.3ex\hbox{$<$\kern-0.75em\raise-1.1ex\hbox{$\sim$}}}
\def\gsi{\raise0.3ex\hbox{$>$\kern-0.75em\raise-1.1ex\hbox{$\sim$}}}
\newcommand{\lsim}{\mathop{\lsi}}

\newcommand{\disc}{\mathop{\mbox{Disc\,}}}

\newcommand{\nF}{n_\rmii{F}}
\newcommand{\nB}{n_\rmii{B}}
 \renewcommand{\nF}[1]{n_\rmii{F{#1}}}
 \renewcommand{\nB}[1]{n_\rmii{B{#1}}}
\newcommand{\rmii}[1]{{\mbox{\tiny\rm{#1}}}}
\newcommand{\re}{\mathop{\mbox{Re}}}

\newcommand{\Tint}[1]{{\hbox{$\sum$}\!\!\!\!\!\!\!\int\,}_{\!\!\!\!\raise-0.9ex\hbox{$\scriptstyle{#1}$}}}
\newcommand{\Tinti}[1]{{{\Sigma}\!\!\!\!\raise0.3ex\hbox{$\int$}_\rmii{${#1}$}}}

\newcommand{\ZZ}{{\mathbb{Z}}}

\newcommand{\bi}{\begin{itemize}}
\newcommand{\ei}{\end{itemize}}

\newcommand{\hide}[1]{ }
\newcommand{\bsl}[1]{\,\slash\!\!\!\!{#1}\,}

\def\TAsc(#1,#2)(#3,#4,#5)%
{\SetWidth{2.0}\CArc(#1,#2)(#3,#4,#5)\SetWidth{1.0}}
\def\Lwidth{3}

\def\TAgl(#1,#2)(#3,#4,#5){\SetWidth{2.0}\PhotonArc(#1,#2)(#3,#4,#5){\Lwidth}%
{6.283 #3 mul 360 div #4 #5 sub #4 #5 sub mul sqrt mul Tdensity mul}%
\SetWidth{1.0}}
\def\TLgl(#1,#2)(#3,#4){\SetWidth{2.0}\Photon(#1,#2)(#3,#4){\Lwidth}
{#1 #3 sub #1 #3 sub mul #2 #4 sub #2 #4 sub mul add sqrt Tdensity mul}%
\SetWidth{1.0}}
\newcommand{\piC}[1]{\;\parbox[c]{40pt}{\begin{picture}(120,60)(0,-20)
\SetWidth{1.0}\SetScale{0.35} #1 \end{picture}}\;}
\def\ConnectedA(#1,#2,#3){\piC{#1(60,-15)(75,34,146) #2(60,75)(75,214,326)%
 #3(60,60)(20,190,350)%
 \GBoxc(0,30)(10,10){1} \GBoxc(120,30)(10,10){1}%
  }}
\def\ConnectedB(#1,#2,#3){\piC{#1(60,-15)(75,34,146) #2(60,75)(75,214,326)%
 #3(60,60)(60,0)%
 \GBoxc(0,30)(10,10){1} \GBoxc(120,30)(10,10){1}%
  }}
\def\ConnectedC(#1,#2){\piC{#1(60,-15)(75,34,146) #2(60,75)(75,214,326)%
 \GBoxc(0,30)(10,10){1} \GBoxc(120,30)(10,10){1}%
  }}
\def\ConnectedD(#1,#2){\piC{#1(60,-15)(75,34,146) #2(60,75)(75,214,326)%
 \GBoxc(0,30)(10,10){1} \GBoxc(120,30)(10,10){1}%
 \SetWidth{2.0} 
 \Line(55,55)(65,65)%
 \Line(55,65)(65,55)
  }}

\makeatletter \@addtoreset{equation}{section} \makeatother
\renewcommand{\theequation}{\arabic{section}.\arabic{equation}}
\makeatletter
\renewcommand\section{\@startsection {section}{1}{\z@}%
                                   {-5.5ex \@plus -1ex \@minus -.2ex}
                                   {2.3ex \@plus.2ex}%
                                   {\normalfont\large\bfseries}}
\renewcommand\subsection{\@startsection{subsection}{2}{\z@}%
                                     {-3.25ex\@plus -1ex \@minus -.2ex}%
                                     {1.5ex \@plus .2ex}%
                                     {\normalfont\normalsize\bfseries}}
\renewcommand\thesection {\@arabic\c@section}
\renewcommand\thesubsection   {\thesection.\@arabic\c@subsection}
\renewcommand{\@seccntformat}[1]{%
\csname the#1\endcsname.\hspace{1.0em}}
\makeatother

\begin{document}

\begin{titlepage}
\begin{flushright}
BI-TP 2008/40\\
arXiv:0812.2105\\ 
\vspace*{1cm}
\end{flushright}
\begin{centering}
\vfill

{\Large{\bf
Heavy quark medium polarization at next-to-leading order
}} 

\vspace{0.8cm}

Y.~Burnier, 
M.~Laine, 
M.~Veps\"al\"ainen 

\vspace{0.8cm}

{\em
Faculty of Physics, University of Bielefeld, 
D-33501 Bielefeld, Germany\\}

\vspace*{0.8cm}

\mbox{\bf Abstract}
 
\end{centering}

\vspace*{0.3cm}
 
\noindent
We compute the imaginary part of 
the heavy quark contribution to the photon polarization 
tensor, i.e.\ the quarkonium spectral function in the vector channel, 
at next-to-leading order in thermal QCD. Matching our result, 
which is valid sufficiently far away from the two-quark threshold,  
with a previously determined resummed expression, which
is valid close to the threshold, we obtain a phenomenological
estimate for the spectral function valid for all non-zero energies. 
In particular, the new expression allows to fix the overall 
normalization of the previous resummed one. Our result may be helpful 
for lattice reconstructions of the spectral function (near the continuum 
limit), which necessitate its high energy behaviour as input, 
and can in principle also be compared with the dilepton production 
rate measured in heavy ion collision experiments. In an appendix
analogous results are given for the scalar channel.

\vfill

 
\vspace*{1cm}
  
\noindent
January 2009

\vfill

\end{titlepage}

%
\section{Introduction}

Heavy fermion vacuum polarization, i.e.\ the contribution 
of a massive fermion species to the (imaginary part of the)
photon polarization tensor, 
or to the spectral function of the electromagnetic 
current, is one of the classic observables 
of relativistic quantum field theory: the result has been known
up to 2-loop, or next-to-leading, or $\rmO(\alpha_s)$ 
level already since the 1950s~\cite{old}.\footnote{%
  The older computations were formulated within QED, 
  but at this order the results carry over directly to QCD, 
  whose notation we adopt.  
  } 
Nevertheless, significant new 
insights were still obtained in the 1970s~\cite{br} and even in 
the 1990s~\cite{db}. 
By now a lot of information is also 
available concerning corrections of $\rmO(\alpha_s^2)$ and 
$\rmO(\alpha_s^3)$ (for recent work and references, see ref.~\cite{ah}).
The physics motivation for the continued interest is related, 
for example, to determining the heavy quark production cross section, 
$\sigma(e^-e^+\longrightarrow c\,\bar c)$, often expressed 
through the $R$-ratio, as well as to computing the heavy
quarkonium decay width. 

In the present paper, we consider essentially
the same observable as in the
classic works,  but in a situation where the heavy
quarks live at a finite temperature, $T$, rather than 
in the vacuum. We refer
to this observable as the ``heavy quark medium polarization''. 
Again the result has direct physical significance, in that it 
determines the heavy quark contribution to the 
production rate of lepton--antilepton pairs from 
the thermal plasma~\cite{dilepton} (cf.\ \eq\nr{dilepton} below). 
There has been considerable phenomenological interest
particularly in what a finite temperature does to the  
resonance peaks in the spectral function, given that this might yield 
a gauge for the formation of a deconfined partonic medium~\cite{ms}. 
Some recent work on the resonance region within the 
weak-coupling expansion, taking steps towards a systematic 
use of effective field theory 
techniques to resum appropriate classes of higher loop orders, 
can be found in refs.~\cite{static}--\cite{nb3} 
(see ref.~\cite{sewm} for a review and ref.~\cite{col}
for an alternative approach with similar results), 
and recent reviews on some of the phenomenological approaches 
on the market can be found in refs.~\cite{phi_revs,phi_revs2}
(possible pitfalls of {\em ad hoc} potential models
at finite temperatures have been reviewed in ref.~\cite{op}, and
underlined from a different perspective in ref.~\cite{col}). 
Analogous spectral functions can also be determined for 
theories with gravity duals~\cite{ads}.

Unfortunately, it appears that ultimately weak-coupling (and related)
techniques will be insufficient for determining quantitatively the shape
of the spectral function around the resonance region. The reason is that 
field theory at finite temperatures suffers from infrared problems, 
implying that the weak-coupling series goes in powers of 
$(\alpha_s/\pi)^{1/2}$ rather than $\alpha_s/\pi$, 
often with large (sometimes non-perturbative~\cite{ir}) coefficients; 
see, e.g., ref.~\cite{nspt_mass} and references therein. 
Therefore, particularly for the case of charmonium where even
at zero temperature weak-coupling computations can hardly be trusted, 
it appears that non-perturbative techniques are a must. Even though 
the situation should be somewhat better under control for bottomonium, 
a crosscheck by lattice methods would still be more than welcome. 

On the point of lattice techniques, however, 
we are faced with a rather fundamental problem. Lattice
techniques are applicable in Euclidean spacetime, while the spectral function
is an inherently Minkowskian object. In principle the spectral
function {\em can} be obtained through a certain analytic continuation
of the Euclidean correlator; however, as a mathematical operation,
at finite temperatures 
such an analytic continuation is unique only if the asymptotic behaviour 
for large Minkowskian arguments is known (see, e.g., ref.~\cite{fund}).
In a practical setting, many further problems arise  because
lattice data is not analytic in nature; yet the need to input 
outside information (``priors'') to the analysis
certainly remains a central issue
(see refs.~\cite{latt} for recent lattice results, 
and ref.~\cite{latt_rev} for an overview). 

It is at this point that weak-coupling techniques may again become
helpful. The goal would now be not so much to determine the spectral 
function around the resonance region, but to determine it at 
very high energies, which information should be relatively reliable, 
thanks to asymptotic freedom. Indeed, the {\em free} thermal 
spectral function has been studied in great detail previously, 
even at a finite lattice 
spacing~\cite{free_spectral}. Nevertheless, loop corrections are expected
to remain substantial even up to energy scales of 
several tens of GeV, so it is important
to account for them, and this is the basic goal of the present
work. As a more refined goal, we wish to demonstrate
that thermal corrections are small far away from the threshold; 
thereby knowledge of the asymptotic behaviour could be taken 
to higher orders, by employing well-known numerically-implemented
results from zero temperature~\cite{hs} (this assumes, 
of course, that a continuum extrapolation can be carried out
on the lattice).
    
Apart from this lattice-related goal, we also wish to pursue the complementary
goal of treating the bottomonium spectral function without any exposure to the
often hard-to-control systematic uncertainties of lattice simulations. 
This can be achieved by constructing an 
interpolation between the asymptotic result determined in the 
present paper, and the near-threshold behaviour estimated within 
a resummed framework in ref.~\cite{peskin}. 

The plan of the paper is the following. In \se\ref{se:setup} we define
the observable to be computed. The general strategy of the computation
is discussed in \se\ref{se:details}, and the 
main results are summarized in \se\ref{se:final}. A phenomenological
reconstruction of the spectral function in the whole energy range is 
carried out in \se\ref{se:num}, while \se\ref{se:concl} lists our 
conclusions. In appendix A, we display in some detail the intermediate
steps entering the determination of one of the ``master'' sum-integrals
appearing in the computation; in appendix B, we list the final results for
all the master sum-integrals; and in appendix C we provide results
for the spectral function in the scalar channel, discussing briefly also 
the ambiguities that hamper this case. 

%
\section{Basic definitions}
\la{se:setup}

The heavy quark contribution to the spectral function of the electromagnetic
current can be defined as 
\be
 \rho_V (\omega) 
 \equiv 
 \int_{-\infty}^\infty 
 \!\! {\rm d}t \,  e^{i \omega t} \!
 \int \! {\rm d}^{3-2\epsilon} \vec{x}\,
 \left\langle
  \fr12 {[ 
  \hat {\cal{J}}^\mu(t,\vec{x}), 
  \hat {\cal{J}}_\mu(0,\vec{0})
   ]}
 \right\rangle
 \;, \la{rhoV} 
\ee
where 
$
 \hat {\mathcal{J}}^\mu
 \equiv
 \hat{\bar\psi}\, \gamma^\mu\, \hat\psi  
$; 
$\hat\psi$ is the heavy quark field operator in the Heisenberg picture; 
$
 \langle \ldots \rangle \equiv {\mathcal{Z}^{-1}} 
 \tr[ (...) e^{-\beta\hat H}]
$
is the thermal expectation value; 
$
 \beta \equiv {1}/{T} 
$
is the inverse temperature; 
and we assume the metric convention
($+$$-$$-$$-$). 
This spectral function 
determines the production rate of muon--antimuon pairs
from the system~\cite{dilepton}, 
\be
 \frac{{\rm d} N_{\mu^-\mu^+}}{{\rm d}^4 x\,{\rm d}^4 q} =  
 \frac{ -2 e^4 Z^2}{3 (2\pi)^5 q^2} 
 \biggl( 1 + \frac{2 m_\mu^2}{q^2}
 \biggr)
 \biggl(
 1 - \frac{4 m_\mu^2}{q^2} 
 \biggr)^\fr12 n_\rmi{\,B}(\omega) \rho_V(\omega)
 \;, 
 \la{dilepton} 
\ee
where $Z$ is the heavy quark electric charge in units of $e$,
and $n_\rmi{\,B}$ is the Bose-Einstein distribution function.
In defining \eq\nr{rhoV} and the argument of $\rho_V$ in 
\eq\nr{dilepton}, we have assumed
that the muon--antimuon pair is at rest with respect to the thermal 
medium, i.e.\ $q \equiv (\omega, \vec{0})$.\footnote{%
 For a non-zero total spatial momentum $\vec{q}$, with $0<|\vec{q}| \ll M$, 
 the main modification of our results would be a shift of the two-particle
 threshold from $\omega \approx 2 M$ to 
 $\omega \approx 2 M + \vec{q}^2 / 4 M$.
 } 
The pole mass of the heavy quark  (charm, bottom) is denoted by $M$.

The parametric temperature range we concentrate on 
in this paper is the one 
where the ``quarkonium'' resonance peak 
disappears from the spectral function $\rho_V$~\cite{peskin}:
\be
 g^2 M < T < gM
 \;. \la{range} 
\ee
This implies that in any case $T \ll M$, so that exponentially 
small corrections, 
$\sim \exp(-\beta M)$, can well be omitted. The thermal effects
come thereby exclusively from the gluonic sector, where no exponential
suppression takes place. 

In order to compute the spectral function $\rho_V$ of \eq\nr{rhoV}, 
we start by determining the corresponding Euclidean correlator,  
\be
 C_E (\omega_n) 
 \equiv 
 \int_{0}^{\beta}
 \!\! {\rm d}\tau \,  e^{i \omega_n \tau} \!
 \int \! {\rm d}^{3-2\epsilon} \vec{x}\,
 \left\langle
  \hat {\cal{J}}^\mu(\tau,\vec{x}) 
  \hat {\cal{J}}_\mu(0,\vec{0})
 \right\rangle
 \;, \la{CE}
\ee
for which a regular path-integral expression can be 
given (i.e.,\ hats can be removed from the definition). 
Here $\omega_n \equiv 2\pi n T$, $n\in \ZZ$, denotes bosonic
Matsubara frequencies. The spectral function is then given by 
the discontinuity (see, e.g., refs.~\cite{leb,kg})
\be
 \rho_V(\omega) 
 = \disc \Bigl[ C_E(-i \omega) \Bigr]
 \equiv 
 \frac{1}{2i} 
 \Bigl[ 
   C_E(-i [\omega + i 0^+])
 -   C_E(-i [\omega - i 0^+])
 \Bigr]  
 \;. \la{disc_def}
\ee
In the following we denote Euclidean four-momenta 
with capital letters, in particular $Q\equiv(\omega_n,\vec{0})$.
Moreover, $\Tinti{K} \equiv T \sum_{k_n} \mu^{2\epsilon} \int 
{\rm d}^d\vec{k}/(2\pi)^d$ 
stands for a sum-integral over bosonic 
Matsubara four-momenta, while $\Tinti{\{\!P\!\}}$ signifies 
a sum-integral over fermionic ones. The space-time dimensionality
is denoted by $D\equiv 4 - 2\epsilon$, and the space dimensionality
by $d\equiv 3 - 2\epsilon$.

%
\section{Details of the computation}
\la{se:details}

%
\subsection{Propagators}

At a finite temperature $T$ it is not clear, {\em a priori}, 
whether the result of the computation will be infrared finite, given that
(after analytic continuation) the gluon propagator contains the 
Bose-enhanced factor $\nB{}(k^0) \approx T/k^0$, for $|k^0| \ll T$. 
For this reason, we carry out the analysis by using
the Hard Thermal Loop resummed~\cite{htlold,htl} form of 
the gluon propagator, which takes into account Debye screening, 
and thereby shields (part of) the infrared divergences. 
Introducing
(see, e.g., refs.~\cite{leb,kg})
\ba
 P^T_{00}(K) \!\! & = & \!\! 
 P^T_{0i}(K) = P^T_{i0}(K) \equiv 0
 \;, \quad P^T_{ij}(K) \equiv \delta_{ij} - 
     \frac{k_i k_j}{\vec{k}^2}
 \;, \la{PT} \\ 
 P^E_{\mu\nu}(K) \!\! & \equiv & \!\!
 \delta_{\mu\nu} - \frac{K_\mu K_\nu}{K^2}
 - P^T_{\mu\nu}(K)
 \;, \la{PE}
\ea
where $K=(k_n,\vec{k})$, $k_n = 2\pi n T$, 
the Euclidean gluon propagator can be written as 
\be
 \langle A^a_{\mu} (x) A^b_\nu (y) \rangle = 
 \delta^{ab} \Tint{K} e^{iK\cdot (x - y)}
 \biggl[
   \frac{P^T_{\mu\nu}(K)}{K^2 + \Pi_T(K)} + 
   \frac{P^E_{\mu\nu}(K)}{K^2 + \Pi_E(K)} + 
   \frac{\xi\, K_\mu K_\nu}{(K^2)^2} 
 \biggr] 
 \;,  \la{prop}
\ee
where $\xi$ is a gauge parameter. The projector $P^T$ is transverse
both with respect to $K$ and to the four-velocity of the heat bath 
and, in the static limit, describes colour-magnetic modes; 
the projector $P^E$ is transverse only with respect to $K$ 
and, in the static limit, describes colour-electric modes. 
The self-energies $\Pi_T, \Pi_E$ are 
well-known~\cite{htlold,htl} functions of the form $\mD^2\, f(k_n/|\vec{k}|)$,
where $\mD = (\Nc/3 + \Nf/6) gT$ is the Debye mass parameter; we will
not need their explicit expressions in the following, apart from
knowing that $f$ is an even function of its argument and regular
on the real axis. The fermion propagator has the free form, 
\be
 \langle \psi(x) \bar\psi(y) \rangle = 
  \Tint{\{P\}} \!\! e^{iP\cdot (x - y)} \, 
  \frac{- i \bsl{P} + M_B }{P^2 + M_B^2}
 \;, 
\ee
where $M_B$ is the bare heavy quark mass.


%
\subsection{Contractions}

The first step of the computation is to carry out the Wick contractions
and the Dirac traces. At 1-loop level, omitting $Q$-independent terms which 
are killed by the discontinuity in \eq\nr{disc_def}, we get
\ba
 \nn[-10mm]
 \ConnectedC(\TAsc,\TAsc) 
 & = & 
 [Q-\mbox{indep.}] + 
 2 C_A \Tint{\{P\}} \frac{(D-2) Q^2 - 4 M^2}
 {\Delta(P)\Delta(P-Q)}
 \;. \la{lo} \\[-10mm] \nonumber
\ea
Here $C_A \equiv \Nc$, and
\be
 \Delta(P) \equiv P^2 + M^2
 \;. 
\ee

At next-to-leading order (NLO), we have to evaluate the counterterm
graph as well as genuine 2-loop graphs. The counterterm graph can be
deduced from the 1-loop expression in \eq\nr{lo}, by re-interpreting
the mass parameter as the bare one, $M_B^2$, and then expanding
it in terms of the pole mass:  
\be
 M_B^2 = M^2 - \frac{6 g^2 C_F M^2}{(4\pi)^2} 
 \biggl( \frac{1}{\epsilon} + \ln\frac{\bmu^2}{M^2} + \fr43 \biggr) 
 + \rmO(g^4)
 \;,  \la{MB}
\ee
where $C_F \equiv (\Nc^2-1)/2\Nc$, and $\bmu$ is 
the scale parameter of the $\msbar$ scheme. This yields
\ba
 \nn[-10mm] 
 \ConnectedD(\TAsc,\TAsc) 
 & = & 
 [Q-\mbox{indep.}] + 
 \frac{24 g^2 C_A C_F M^2}{(4\pi)^2} 
 \biggl( \frac{1}{\epsilon} + \ln\frac{\bmu^2}{M^2} + \fr43 \biggr) 
 \nn[-5mm] & \times & 
 \Tint{\{P\}} \biggl[ 
 \frac{(D-2) Q^2 - 4 M^2}
 {\Delta^2(P)\Delta(P-Q)}
 + 
 \frac{2} {\Delta(P)\Delta(P-Q)}
 \biggr]
 \;. \la{ct}
\ea
For the genuine 2-loop graphs, we make use of the identities
\be
 K_\mu P^T_{\mu\nu}(K) 
 = Q_\mu P^T_{\mu\nu}(K) = 0  
 \;, \quad
 P^T_{\mu\mu}(K) = D - 2
 \;, \quad
 P_\mu P_\nu P^T_{\mu\nu}(K) = 
 \vec{p}^2 - (\vec{p}\cdot\hat{\vec{k}})^2
 \;,  
\ee
where $\hat{\vec{k}} \equiv \vec{k} / |\vec{k}|$, and 
the second equality follows from the fact that $Q$ is aligned with the 
heat bath. We then complete squares in the numerator, and note that 
\be
 \Tint{K\{P\}} \frac{Q\cdot K}{[K^2 + \Pi(K)] 
 \Delta(P)\Delta(P-Q)\Delta(P-K)\Delta(P-Q-K)} = 0 
 \;, 
\ee
as can be shown with the shifts $P\to -P+Q, K\to -K$.
Thereby we arrive at
\ba
 \nn[-10mm]
 && \hspace*{-2.5cm}
 \ConnectedA(\TAsc,\TAsc,\TAgl) \; + 
 \ConnectedB(\TAsc,\TAsc,\TLgl) 
 =   [Q-\mbox{indep.}] + 4 g^2 C_A C_F \Tint{K\{P\}} 
 \biggl\{  \la{nlo} \\[-5mm] 
 &&  \hspace*{-2cm}
 \biggl( \frac{1}{K^2+\Pi_T}-\frac{1}{K^2+\Pi_E} \biggr)
	[\mathbf{p}^2-(\mathbf{p}\cdot \hat{\mathbf{k}})^2 ] \times  
 \nn
 && \hspace*{-0.8cm}
 \times \left[ 
     -\frac{4[(2-D)Q^2+4M^2]}{\Delta^2(P)\Delta(P-Q)\Delta(P-K)}
	-\frac{2[(2-D)Q^2+4M^2] +4K^2}
      {\Delta(P)\Delta(P-Q)\Delta(P-K)\Delta(P-Q-K)} \right] \nn 
 &\displaystyle +\frac{D-2}{K^2+\Pi_T}& \hspace*{-3mm} 
 \left[ -\frac{2}{\Delta(P)\Delta(P-Q)} 
	+\frac{(2-D) Q^2 + 4 M^2}{\Delta^2(P)\Delta(P-Q)}
 \right. \nn 
 && \hspace*{-1cm} \left. {}
        +\frac{2}{\Delta(P)\Delta(P-Q-K)}
	+\frac{-2(D-2)Q\cdot K + 4 K^2 }{\Delta(P)\Delta(P-Q)\Delta(P-K)}
	 \right. \nn 
 &&  \hspace*{-1cm} \left. {}
 - \frac{[ (2-D) Q^2 + 4 M^2 ] K^2}{\Delta^2(P)\Delta(P-Q)\Delta(P-K)}
 - \frac{[(6-D) Q^2/2\,  +2 M^2 ]K^2 +K^4}
  {\Delta(P)\Delta(P-Q)\Delta(P-K)\Delta(P-Q-K)}
	 \right] \nn
 &\displaystyle +\frac{1}{K^2+\Pi_E}& \hspace*{-3mm}
 \left[ 
 -\frac{4[(2-D)Q^2 +4M^2 ]}{\Delta(P)\Delta(P-Q)\Delta(P-K)} 
 + \frac{4[(2-D)Q^2 +4M^2 ] M^2 }{\Delta^2(P)\Delta(P-Q)\Delta(P-K)}
 \right. \nn
 && {}+\frac{(2-D)Q^4 
 +(8-2D)Q^2 M^2 +8M^4 +[ (2-D)Q^2  +4  M^2]K^2}
 {\Delta(P)\Delta(P-Q)\Delta(P-K)\Delta(P-Q-K)}
 \biggr] \; \biggr\}
 \;. \nonumber
\ea
Note that any dependence on the gauge parameter 
$\xi$ has disappeared; thus $P^E_{\mu\nu}$ could 
have been replaced with $\delta_{\mu\nu} - P^T_{\mu\nu}$. 

%
\subsection{Outline of the subsequent steps}
\la{se:steps}

Given \eq\nr{nlo}, we need to carry out the Matsubara sums and 
the spatial momentum integrals. More concretely, the steps (specified
in explicit detail for one example in appendix~A) are as follows: 
\begin{itemize}
\item
Writing the gluon propagator in a spectral representation, 
the Matsubara sums 
$  
 T \sum_{k_n}
$
and 
$
 T \sum_{ \{p_n\} }
$
can be carried out exactly in all cases. 

\item
The result after these steps contains 
many appearances of the Fermi distributions, 
$\nF{}(E) \equiv 1 / [\exp(\beta E) +1]$, 
where the energy $E$ is that of a heavy quark, 
$E \ge M$. All such terms are suppressed
by at least $e^{-M/T} \ll 1$, and can be omitted.

\item
The remaining temperature dependence appears as
Bose distributions with the gluon energy, $\nB{}(k^0)$.
Here the issue is the opposite: in the small energy range, 
$|k^0| \ll T$, there is an enhancement factor $T/k^0$, 
which could lead to infrared divergences. This is an 
important point, so we devote a separate subsection to it
(\se\ref{se:IR}). The upshot is that there are 
{\em no} infrared divergences at the present order.

\item
Having verified the absence of infrared divergences, we can 
forget about the HTL resummation in the gluon propagators,
i.e.\ set $\Pi_T = \Pi_E = 0$ in \eq\nr{nlo}, 
and insert the free spectral function for the gluons. Thereby
the integral over the gluon energy $k^0$ is trivially carried out. 
(In practice, we first insert the free gluon spectral function, 
integrate over $k^0$, and verify the absence of infrared divergences 
{\em a posteriori} for each 
independent (``master'') sum-integral separately.)

\item
The remaining spatial integrals, over $\vec{k}$ and $\vec{p}$, 
are effectively three-dimensional (over the absolute values of 
$\vec{k,p}$ and over the angle between $\vec{k}$ and $\vec{p}$).
Some of them are ultraviolet divergent, and require regularization. 
The integrals come in two forms, which we call ``phase space'' and 
``factorized''. We are able to carry out two of the integrations
in all cases; for the zero-temperature parts entering the final
result, all three integrations are doable~\cite{old}--\cite{db}, 
while for the finite-temperature parts an exponentially convergent
integral over $k = |\vec{k}|$ remains to be carried out numerically. 

\end{itemize}
The results obtained after these steps are listed for all the
master sum-integrals appearing in \eq\nr{nlo} in appendix B. 

%
\subsection{Absence of infrared divergences}
\la{se:IR}

Inserting the free gluon spectral function, which sets $k^0=k$, 
into any of the master sum-integrals, 
there remains an integral over the gluon momentum 
$k$ to be carried out. In principle this integral could be
infrared divergent. This turns out indeed to be the case
for the ``phase space'' and ``factorized'' parts
(for definitions, see appendix~\ref{ss:spatial}) of the integrals
separately; in fact, the integrals denoted by 
$S_5^0$ and $S_6^0$ (cf.\ \eqs\nr{S50}, \nr{S60}), have 
logarithmically divergent infrared parts even at zero temperature,  
which were an issue in the 1970s~\cite{br}. However, the infrared divergences
were found to cancel in the sum of the phase space and factorized parts. 
In our case, the logarithmic 
divergences turn into linear ones, due to the additional factor
$\nB{}(k) \approx T/k$; nevertheless, when the phase
space and factorized parts are added together, we find that both powerlike
and logarithmic divergences cancel, and the integrals become finite, 
for each master sum-integral separately. This can clearly be seen
in \eqs\nr{S50_T} and \nr{S60_T} for $S_5^0$ and $S_6^0$, respectively. 
The same is true for the integrals denoted by
${\hat S_5^0}, {\hat S_6^0}, {\hat S_6^2}$
(\eqs\nr{hS50}, \nr{hS60}, \nr{hS62}), appearing in the first term
of \eq\nr{nlo} and disappearing if the HTL self-energies are set
to zero from the outset. 
Therefore, we conclude that there are no infrared problems
in our observable at the next-to-leading order. 
(It is to be expected, though, that there are some at higher orders.)

%
\section{Final result}
\la{se:final}

Given the considerations in \se\ref{se:IR},
showing the absence of infrared divergences, 
we are free to set $\Pi_T = \Pi_E = 0$ in \eq\nr{nlo}.
Noting furthermore that 
the factorized gluon tadpole reads 
\be
 \Tint{K} \frac{1}{K^2} = \frac{T^2}{12} + \rmO(\epsilon)
 \;, 
\ee
and employing the notation of appendix~B for the sum-integrals 
$S_i^j(\omega)$, the full result can be written as 
\ba
 \left. \rho_V(\omega) \right|_\rmii{raw} \!\! & = & \!\!
 - 4 C_A (\omega^2 + 2 M^2) S_1(\omega) + 8 g^2 C_A C_F
 \biggl\{ \nn & & 
 \biggl[ \frac{T^2}{6} 
   - \frac{6 M^2}{(4\pi)^2}
   \biggl(
     \frac{1}{\epsilon} + \ln\frac{\bmu^2}{M^2} + \fr43 
   \biggr) 
 \biggr]
 \Bigl[ 
   - S_1(\omega)
   + (\omega^2 + 2 M^2 - \epsilon\,\omega^2) S_2(\omega)
 \Bigr] \nn[1mm] 
 & & +  2 S_3(\omega) 
 - 4 (\omega^2 + 2 M^2 - \epsilon\,\omega^2) S_4^0(\omega) 
 - 4 (1-2\epsilon) S_4^1(\omega)
 + 4 (1-\epsilon) S_4^2(\omega)
 \nn[2mm] 
 & & + 2 (\omega^2 + 2 M^2 - \epsilon\,\omega^2)
 \Bigl[ 2 M^2 S_5^0(\omega) - (1-\epsilon) S_5^2(\omega) \Bigr]
 - (\omega^4 - 4 M^4) S_6^0(\omega) 
 \nn  & &
 + \Bigl[ (2 - \epsilon) \omega^2 + 2 \epsilon M^2 \Bigr] S_6^2(\omega)
 - (1-\epsilon) S_6^4(\omega)
 \biggr\} + \rmO(\epsilon) \;.
   \la{full_raw}
\ea
We have set here $\epsilon\to 0$ whenever the master sum-integral
that it multiplies is finite. 

Now, the explicit thermal correction on the second line of \eq\nr{full_raw}
has a simple physical meaning: it corresponds to an expansion
of the leading-order result through a thermal 
mass shift~\cite{dhr}
\be
 M^2 \to M^2 + \frac{g^2 T^2 C_F}{6}
 \;, \la{M_thermal}
\ee
i.e.\ 
$
 \delta M = g^2 T^2 C_F / 12 M
$.
Note that this term multiplies the function 
$
 S_2(\omega) = \theta(\omega - 2M)/[ 16\pi\omega(\omega^2 - 4 M^2)^{1/2} ]  
 ( 1 + \rmO(\epsilon,e^{-\beta M}))
$ 
(cf.\ \eq\nr{S2_final}), 
which diverges at the thres\-hold, while the sum
of all the other terms turns out to remain finite. Thereby the 
thermal correction would completely dominate the result close
enough to the threshold, were it not to be resummed into a mass 
correction {\em \`a la} \eq\nr{M_thermal}. On the other hand, once
it has been resummed, this term is in general small: in the range 
that we are interested in, $g^2 M < T < gM$, it corresponds 
parametrically to a higher order contribution.
Therefore, for simplicity, 
we drop this term in the following (of course, if desired, 
it is trivial to include it as an overall mass shift), 
and reinterpret the result as
\ba
 \rho_V(\omega)  \!\! & = & \!\!
 - 4 C_A (\omega^2 + 2 M^2) S_1(\omega) + 8 g^2 C_A C_F
 \biggl\{ \nn & & 
 \biggl[ 
   - \frac{6 M^2}{(4\pi)^2}
   \biggl(
     \frac{1}{\epsilon} + \ln\frac{\bmu^2}{M^2} + \fr43 
   \biggr) 
 \biggr]
 \Bigl[ 
   - S_1(\omega)
   + (\omega^2 + 2 M^2 - \epsilon\,\omega^2) S_2(\omega)
 \Bigr] \nn[1mm] 
 & & +  2 S_3(\omega) 
 - 4 (\omega^2 + 2 M^2 - \epsilon\,\omega^2) S_4^0(\omega) 
 - 4 (1-2\epsilon) S_4^1(\omega)
 + 4 (1-\epsilon) S_4^2(\omega)
 \nn[2mm] 
 & & + 2 (\omega^2 + 2 M^2 - \epsilon\,\omega^2)
 \Bigl[ 2 M^2 S_5^0(\omega) - (1-\epsilon) S_5^2(\omega) \Bigr]
 - (\omega^4 - 4 M^4) S_6^0(\omega) 
 \nn  & &
 + \Bigl[ (2 - \epsilon) \omega^2 + 2 \epsilon M^2 \Bigr] S_6^2(\omega)
 - (1-\epsilon) S_6^4(\omega)
 \biggr\} + \rmO(\epsilon) \;.
   \la{full_raw2}
\ea
Nevertheless, 
it is perhaps appropriate to stress that only the part of the 
thermal correction multiplying the function $S_2(\omega)$ can 
be unambiguously resummed on the grounds that the result would
otherwise diverge at the threshold, while the term $\sim T^2 S_1(\omega)$
could in principle be kept explicit, and would then have an $\rmO(1)$
effect on the thermal part of the result. 

Inserting the explicit expressions for the functions 
$S_i^j(\omega)$ from
appendix B into \eq\nr{full_raw2}, 
the final result for the vacuum part becomes
\ba
 \left. \rho_V(\omega) \right|^\rmii{vac} \!\!\! & = & \!\!\!
 - \theta(\omega - 2 M)   
  \frac{C_A (\omega^2 - 4 M^2)^{\fr12}(\omega^2 + 2 M^2)}{4\pi\omega}
 + \theta(\omega - 2 M) 
 \frac{8 g^2 C_A C_F}{(4\pi)^3 \omega^2} 
 \biggl\{ \nn & & 
 (4 M^4 - \omega^4)  L_2 \biggl( \frac{\omega - \sqrt{\omega^2 - 4 M^2}}
 {\omega + \sqrt{\omega^2 - 4 M^2}} \biggr)
 + (7 M^4 + 2 M^2\omega^2 - 3 \omega^4)
  \,\mathrm{acosh} \biggl( \frac{\omega}{2 M} \biggr)
 \nn &  & 
 + \omega (\omega^2 - 4 M^2)^{\fr12}
 \biggl[
   (\omega^2 + 2 M^2) \ln \frac{\omega (\omega^2 - 4 M^2)}{M^3}
  -\fr38 (\omega^2 + 6 M^2) 
 \biggr]
 \biggr\} + \rmO(\epsilon,g^4) \;, 
  \nn \la{full_vac}
\ea
where the function $L_2$ is defined as
\be
 L_2(x) \equiv 
 4 \, \mathrm{Li}_2 (x) + 2 \, \mathrm{Li}_2(-x)
 + [2 \ln(1-x) + \ln(1+x)] \ln x
 \;. \la{L2} 
\ee
The result in \eq\nr{full_vac} agrees 
with the classic result from the literature~\cite{old}--\cite{db}.
The thermal correction, in turn, reads, 
\ba
 \left. \rho_V(\omega) \right|^\rmii{$T$} \!\!\! & = & \!\!\!
 \frac{8 g^2 C_A C_F}{(4\pi)^3 \omega^2}
 \int_0^\infty \! \dd k \, \frac{\nB{}(k)}{k} 
 \biggl\{ \nn 
 & & \hspace*{0.0cm}
   \theta(\omega) \, 
   \theta\Bigl(k- 
    \frac{4M^2-\omega^2}{2\omega}
    \Bigr)
   \biggl[ 
   2 \omega^2 k^2 \sqrt{1-\frac{4 M^2}{\omega(\omega+2 k)}}
  \nn & &  \hspace*{1cm}
  +     (\omega^2 + 2 M^2) \sqrt{\omega(\omega+2 k)}
    \sqrt{\omega(\omega+2 k)- 4M^2}
  \nn & & \hspace*{1cm}
  - 2 \Bigl(\omega^4 - 4 M^4 + 2 \omega k (\omega^2 + 2 M^2) 
  + 2 \omega^2 k^2\Bigr)
  \, \mathrm{acosh} \sqrt{\frac{\omega(\omega+2k)}{4M^2}} 
   \biggr]
   \nn  & + & \hspace*{0.0cm}
   \theta(\omega - 2M) \, 
   \theta\Bigl(\frac{\omega^2-4M^2}{2\omega}-k\Bigr)
   \biggl[ 
   2 \omega^2 k^2 \sqrt{1-\frac{4 M^2}{\omega(\omega-2 k)}}
  \nn & &  \hspace*{1cm}
  +     (\omega^2 + 2 M^2) \sqrt{\omega(\omega-2 k)}
    \sqrt{\omega(\omega-2 k)- 4M^2}
  \nn & & \hspace*{1cm}
  - 2 \Bigl(\omega^4 - 4 M^4 - 2 \omega k (\omega^2 + 2 M^2) 
  + 2 \omega^2 k^2\Bigr)
  \, \mathrm{acosh} \sqrt{\frac{\omega(\omega-2k)}{4M^2}} 
   \biggr]
   \nn & + & \hspace*{0.0cm}
  \theta(\omega - 2M) \biggl[ 
   - 2 (\omega^2 + 2 M^2)\, \omega \sqrt{\omega^2 - 4 M^2}
  \nn & & \hspace*{1cm}
   + 4 \Bigl(\omega^4 - 4 M^4 + 2 \omega^2 k^2\Bigr)
   \, \mathrm{acosh} \biggl( \frac{\omega}{2M} \biggr) 
  \biggr] \biggr\}
 + \rmO(e^{-\beta M},g^4)
 \;, \la{full_T}
\ea
where we have restricted to $\omega > 0$
($\omega < 0$ follows from antisymmetry, 
$\rho_V(-\omega) = - \rho_V(\omega)$).
Eq.~\nr{full_T} is our main result. 

A numerical evaluation of \eq\nr{full_T}, compared 
with the vacuum part in \eq\nr{full_vac}, is shown 
in \fig\ref{fig:nlo}. We note that even though 
the thermal part is not exponentially 
suppressed for $\omega > 2 M$, 
it still only amounts to a small correction
at phenomenologically interesting temperatures. On the 
other hand, the thermal part
does possess the new qualitative feature
that the result is non-zero below the threshold as well, where it 
is then the dominant effect; 
this can be traced back to reactions where a heavy quark and 
anti-quark annihilate into a gluon remaining inside the 
thermal medium, and a photon escaping from it.

\begin{figure}[t]

\centerline{%
\epsfysize=8.0cm\epsfbox{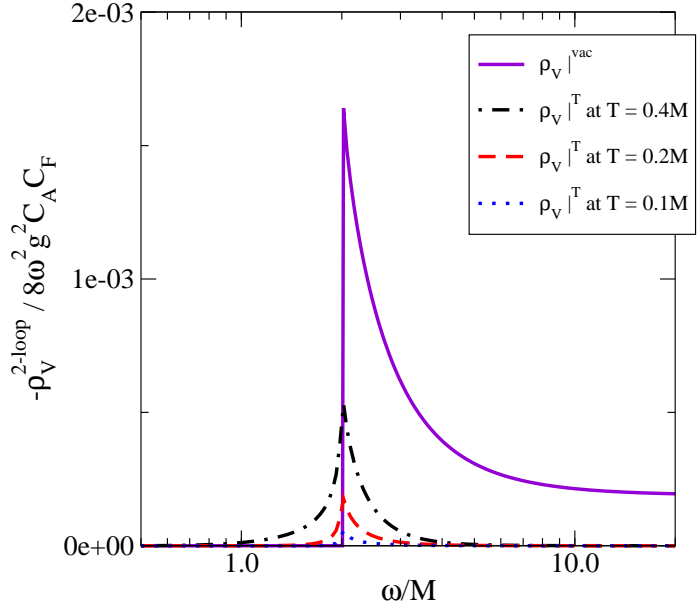}%
}

\vspace*{0.5cm}


\caption[a]{\small
The vacuum and thermal parts of the 
next-to-leading order correction in the vector channel, 
normalized by dividing with $-8 \omega^2 g^2 C_A C_F$.
The vacuum part remains finite for $\omega\to\infty$
(in units of the figure, its asymptotic value is $3/512\pi^3$), 
while the thermal part disappears fast for $\omega/M\gg 1$.}

\la{fig:nlo}
\end{figure}

As an amusing remark, we note that while the next-to-leading order 
vacuum part is discontinuous at the threshold, the next-to-leading
order thermal part appears to be continuous. A similar pattern holds
also for the scalar channel (\fig\ref{fig:nloS}): then the next-to-leading
order vacuum part is continuous, while the next-to-leading order thermal
part appears to have a continuous first derivative. These features are
perhaps a manifestation of the fact that a non-zero temperature in general
``smoothens'' the spectral function; in a resummed framework, it may then 
not be surprising if any resonance peak of the vacuum result
should disappear from the spectral function at high enough temperatures.

To summarize, the characteristic feature of 
\fig\ref{fig:nlo} is a significant ``threshold enhancement'', 
due mostly to the vacuum part at $T \ll M$. Within perturbation
theory, this is to be interpreted as a first term of a series which, 
when summed to all orders, builds up possible quarkonium resonance peaks 
at $\omega < 2 M$. At the same time, the result of a resummed computation 
(to be discussed in more detail at the beginning of the next section) 
should extrapolate towards the perturbative one at some $\omega > 2 M$.

%
\section{Phenomenological implications}
\la{se:num}

\begin{figure}[t]

\centerline{%
\epsfysize=8.0cm\epsfbox{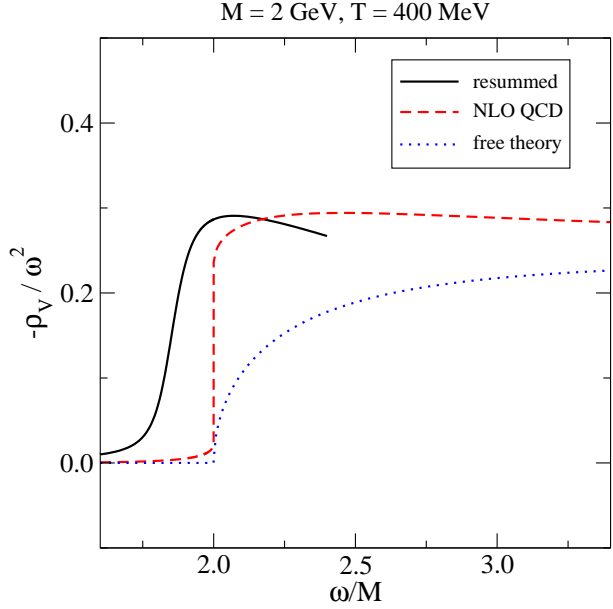}%
}

\vspace*{0.5cm}


\caption[a]{\small
A comparison of the near-threshold ``resummed'' 
result of ref.~\cite{peskin}, matched to the ``NLO QCD''
expression of the present paper through an overall normalization
factor, as discussed in the text. The difference of the NLO QCD 
result and the free theory result contains both the vacuum part
and the thermal part; the magnitude of the latter is reflected
by how much the curve deviates from zero below the threshold.}

\la{fig:interpolation}
\end{figure}

We would now like to combine our result 
with that obtained within an NRQCD~\cite{cl,kt} 
and PNRQCD~\cite{ps,nb2}
inspired resummed framework in ref.~\cite{peskin}. 
In order to do this, we need to pay attention to the 
correct normalization of the resummed result. In fact, the well-known 
(vacuum) normalization factor can be read off from \eq\nr{full_vac}: denoting
\be
 v \equiv \frac{\sqrt{\omega^2 - 4 M^2}}{\omega}
 \;, \la{v_def}
\ee
the leading order vacuum expression can be expanded near the threshold as 
\be
 \left. - \frac{\rho_V(\omega)}{\omega^2} \right|_\rmii{LO} 
 = \theta(\omega - 2 M)\biggl[ 
 \frac{3 C_A v}{8\pi} + \rmO(v^3)
 \biggr]
 \;, \la{LO_exp_V} 
\ee
while the next-to-leading order vacuum result becomes
\be
 \left. - \frac{\rho_V(\omega)}{\omega^2} \right|_\rmii{NLO} 
 = 8 g^2 C_A C_F \theta(\omega - 2 M)
 \biggl[ \frac{3}{512\pi} -  \frac{3 v}{64\pi^3} + \rmO(v^2)
 \biggr]
 \;. \la{NLO_exp_V} 
\ee
Since radiative corrections within a non-relativistic framework always 
contain a power of $v$, it is possible to account for the second term
in \eq\nr{NLO_exp_V}, equalling $-g^2 C_F/\pi^2$ times the leading
term in \eq\nr{LO_exp_V}, only by a multiplicative correction of 
the current,\footnote{%
 The same relation is valid both for NRQCD and PNRQCD~\cite{bkp}.
 } 
\be
 {\cal J}_\rmii{QCD}^\mu = {\cal J}_\rmii{NRQCD}^\mu
 \left(1 - \frac{g^2 C_F}{2\pi^2} + ... \right)
 \;. \la{V_norm}
\ee
In principle the coupling here should be evaluated at the scale 
$\sim M$~\cite{current}, but in practice our resolution is low enough
that we follow a simpler recipe (cf.\ next paragraph). 
In any case, the normalization factor is numerically significant, 
and its precise treatment plays a role; we actually do not 
impose it exactly, but rather search for a value
minimizing the squared difference of the 
two results in the range $(\omega-2M)/M = 0.0 - 0.4$, 
thereby also accounting for thermal corrections.
This results in a normalization factor in the range 
$0.7 - 0.9$, which indeed is the same ballpark as 
suggested by (the square of) \eq\nr{V_norm}, 
given our choice of $g^2$ (cf.\ next paragraph). 
The ``interpolated'', or rather ``assembled'' result, 
is subsequently defined as
$  
 \rho_\rmii{$V$}^\rmii{(assembled)} \equiv 
 \mbox{max}(\rho_\rmii{$V$}^\rmii{(QCD)},
 \rho_\rmii{$V$}^\rmii{(resummed)})
$.
An example for how the interpolation works in practice
is shown in \fig\ref{fig:interpolation}.

As far as the value of $g^2$ goes, no systematic choice is possible
in the absence of NNLO computations at finite temperature. We follow 
here a purely phenomenological recipe, whereby $g^2$ is taken from 
another context where a sufficient level has been reached~\cite{gE2}, 
and take~\cite{adjoint}
\be
 g^2 \simeq \frac{8\pi^2}{9 \ln( 9.082\, T/ \Lambdamsbar)} 
 \;,
 \qquad \mbox{for $\Nc = \Nf = 3$}
 \;. \la{numg2}
\ee 
We also fix $\Lambdamsbar \simeq 300$~MeV to be compatible
with ref.~\cite{peskin}. It should be obvious that the subsequent
results contain unknown uncertainties; still, the situation
could in principle be systematically improved upon through 
higher order computations.

\begin{figure}[t]

\centerline{%
 \epsfysize=5.0cm\epsfbox{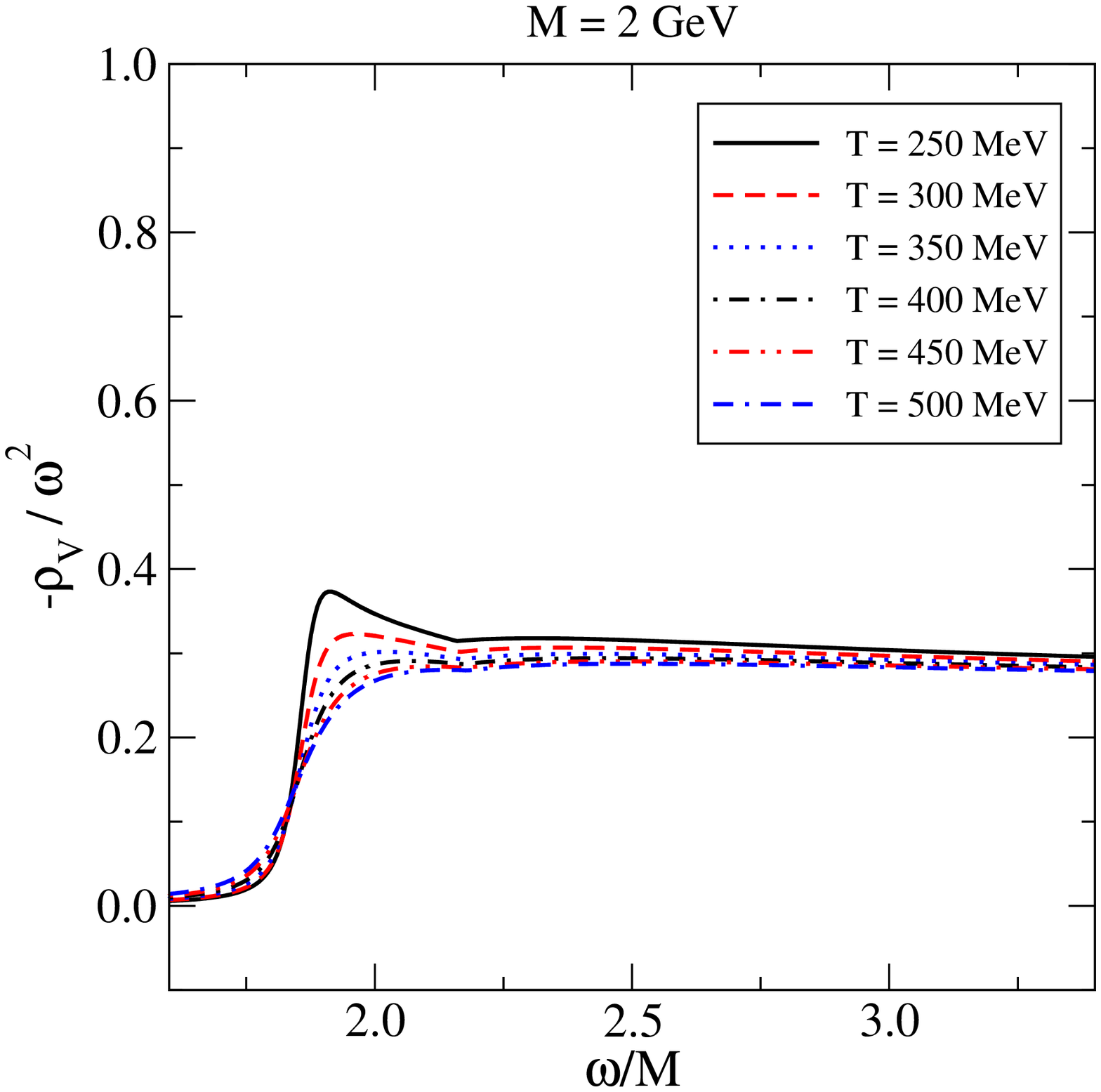}%
 ~~\epsfysize=5.0cm\epsfbox{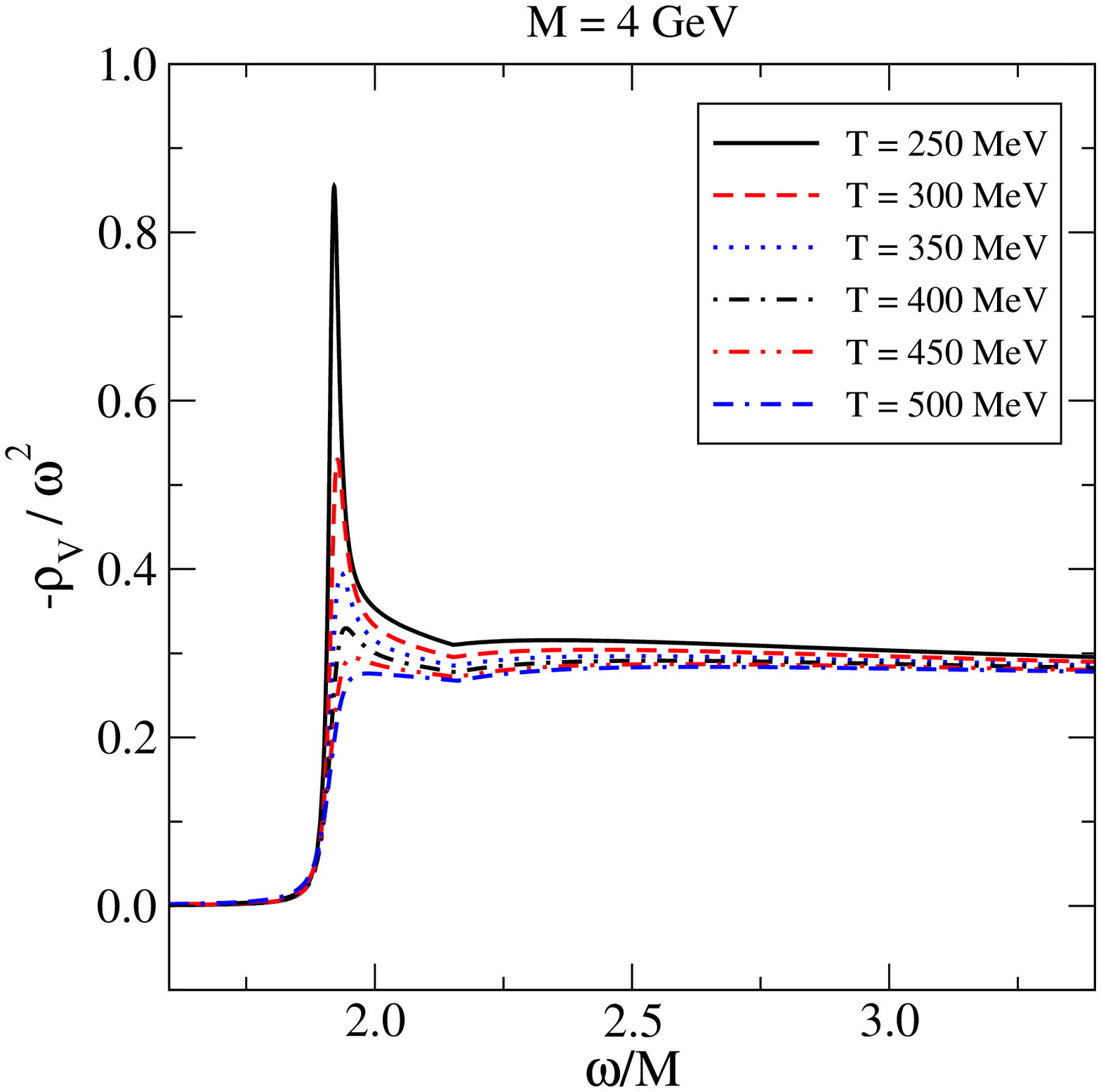}%
 ~~\epsfysize=5.0cm\epsfbox{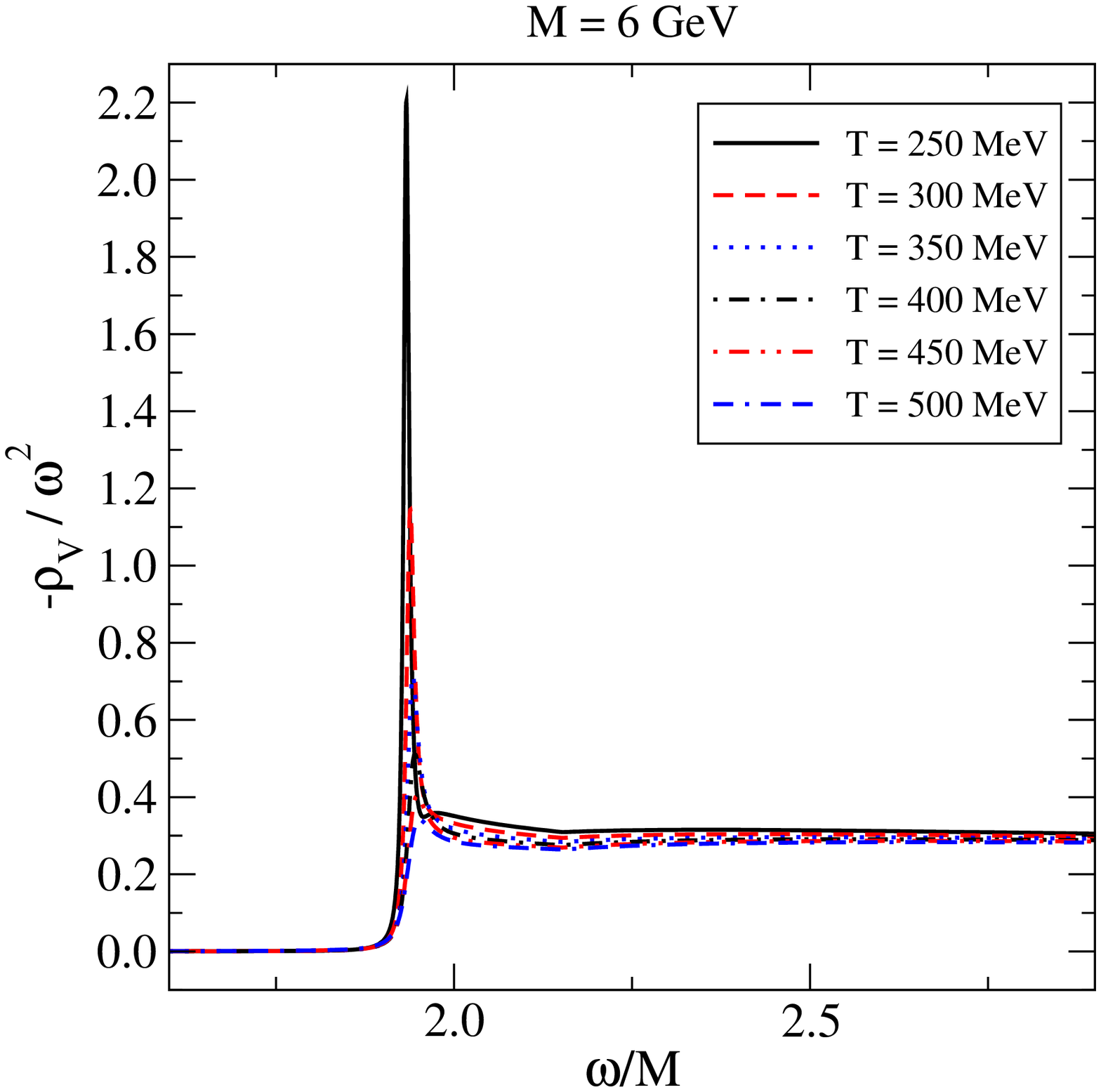}%
}

\vspace*{0.5cm}


\caption[a]{\small
The phenomenologically assembled vector channel
spectral function $\rho_V(\omega)$, in units of $-\omega^2$, 
for $M = 2, 4, 6$~GeV (from left to right).
To the order considered, $M$ is the heavy quark pole mass. Note that
for better visibility, the axis ranges are different in the rightmost 
figure. 
}

\la{fig:rhoV_M}
\end{figure}

\begin{figure}[t]

\centerline{%
 \epsfysize=5.0cm\epsfbox{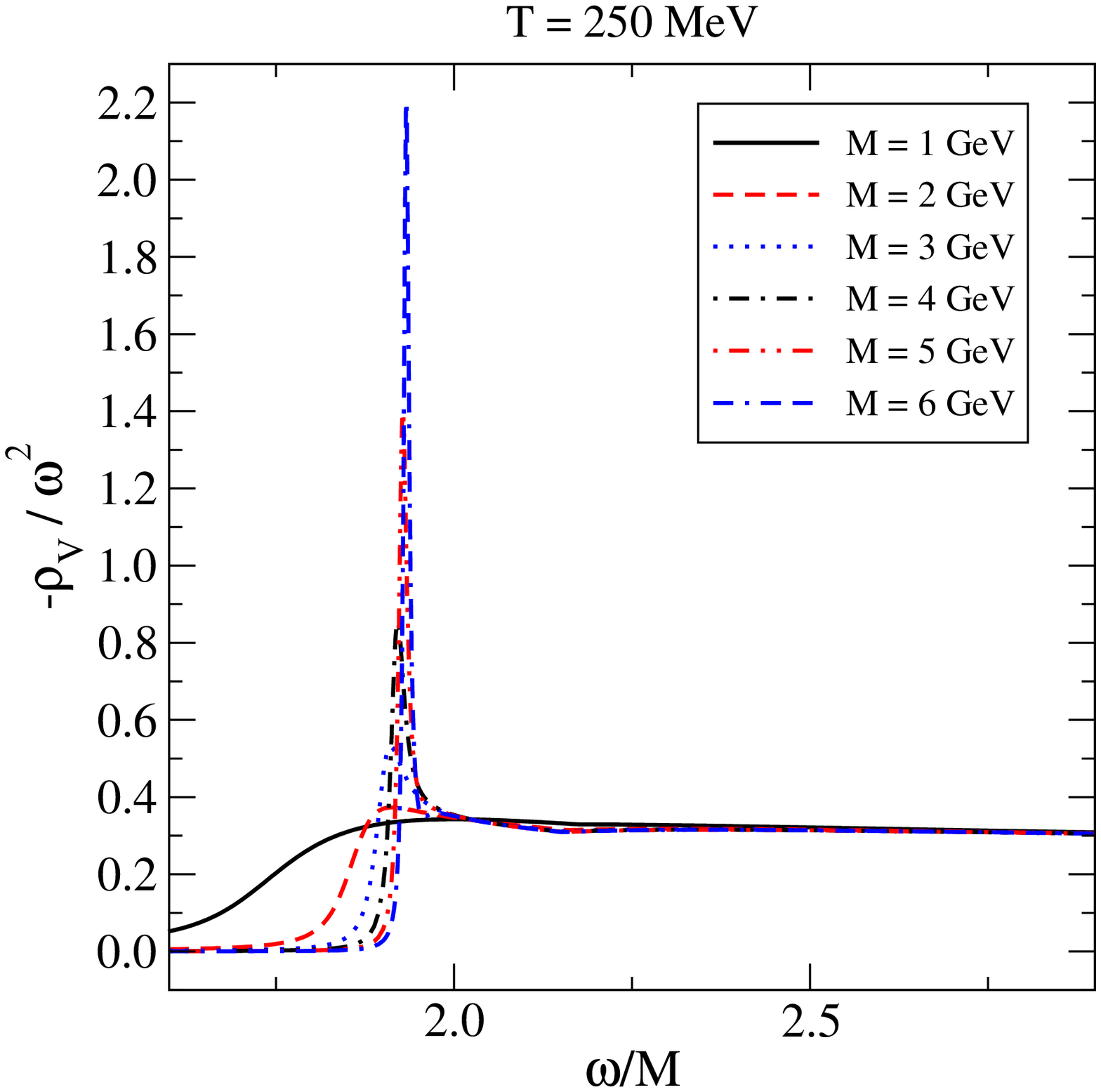}%
 ~~\epsfysize=5.0cm\epsfbox{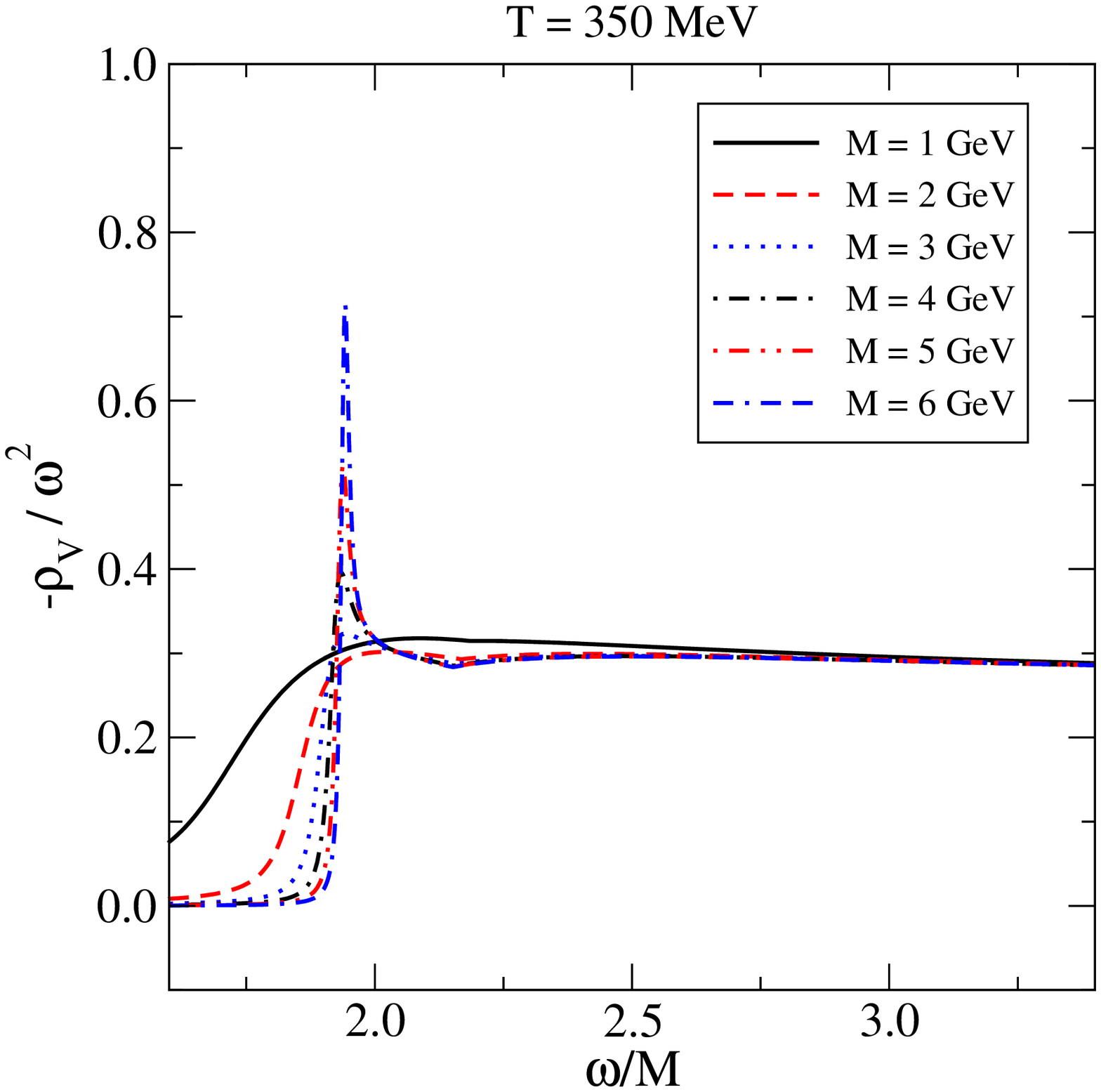}%
 ~~\epsfysize=5.0cm\epsfbox{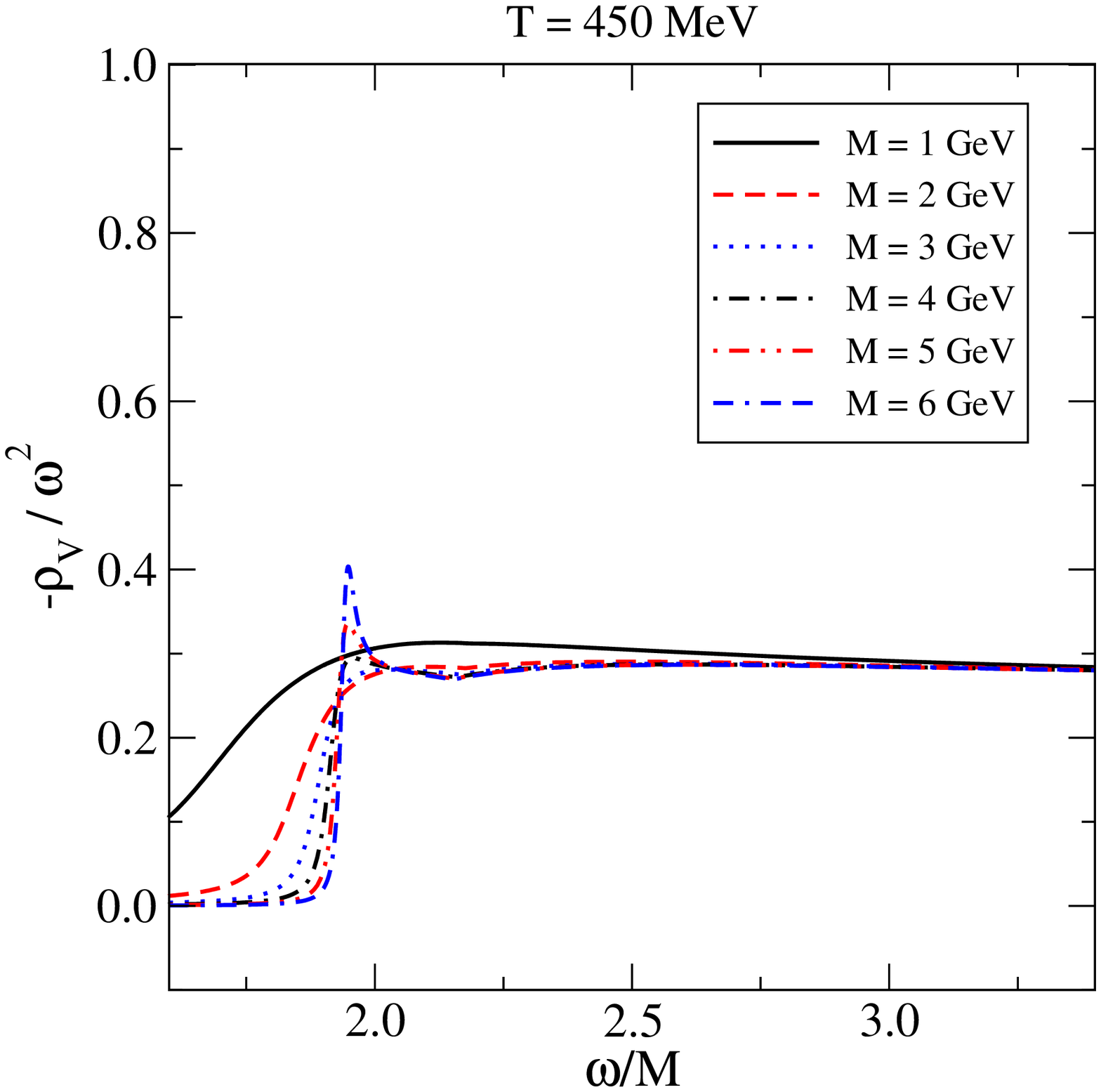}%
}

\vspace*{0.5cm}


\caption[a]{\small
The phenomenologically assembled vector channel
spectral function $\rho_V(\omega)$, in units of $-\omega^2$, 
for $T = 250, 350, 450$~MeV (from left to right).
To the order considered, $M$ is the heavy quark pole mass. Note that
for better visibility, the axis ranges are different in the leftmost 
figure. 
}

\la{fig:rhoV_T}
\end{figure}

\begin{figure}[t]

\centerline{%
 \epsfysize=7.5cm\epsfbox{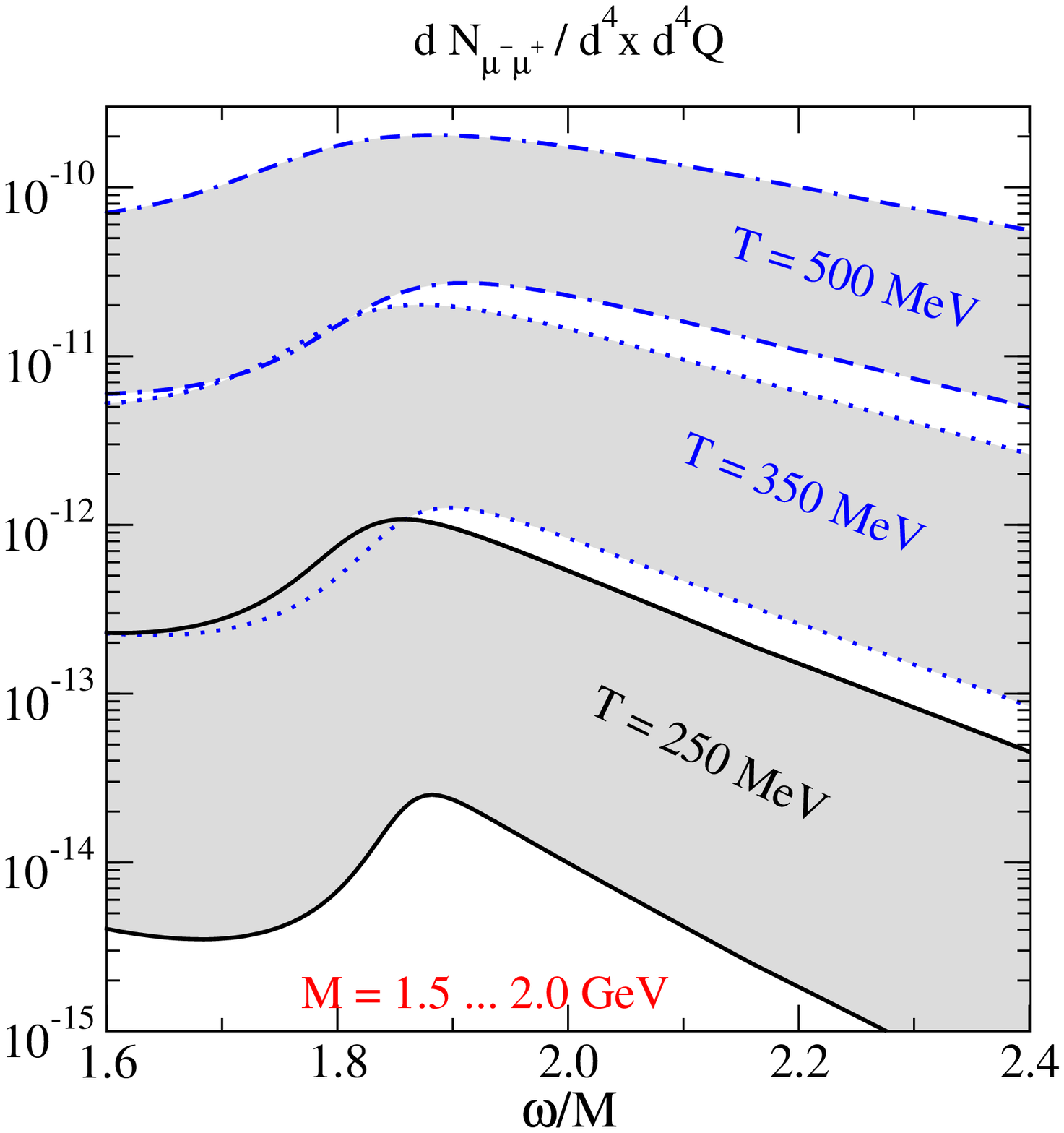}%
 ~~\epsfysize=7.5cm\epsfbox{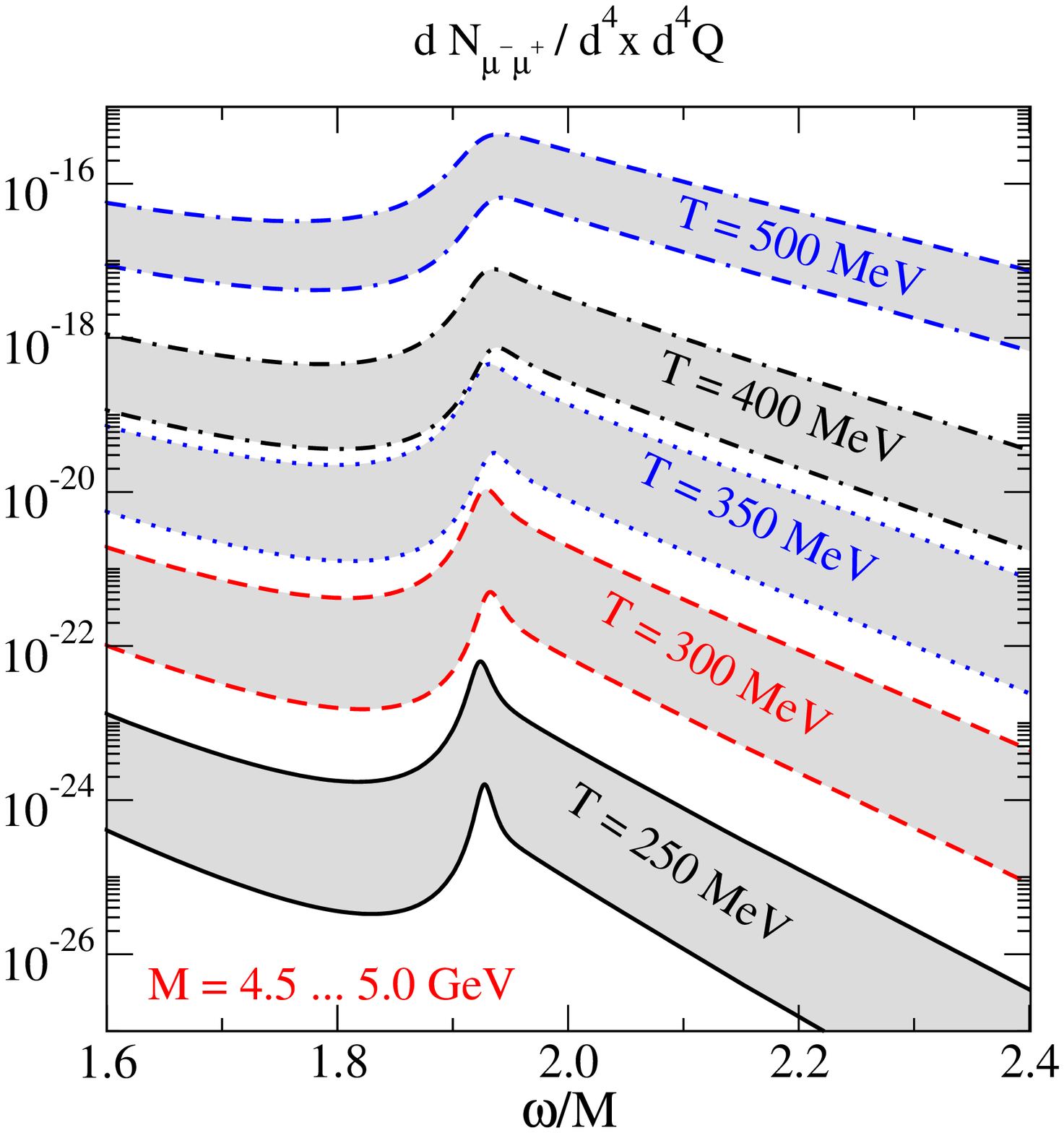}%
}

\vspace*{0.5cm}


\caption[a]{\small
The physical dilepton production rate, \eq\nr{dilepton}, 
from charmonium (left) and bottomonium (right), 
as a function of the energy, for various temperatures. 
The mass $M$ corresponds 
to the pole mass, and is subject to uncertainties of several
hundred MeV; we use the intervals 1.5...2.0 GeV and 4.5...5.0 GeV
to illustrate the uncertainties. The low mass corresponds to the 
upper edge of each band. Compared with ref.~\cite{peskin}, 
the main change is a 10 -- 30\% reduction of the 
overall magnitude.}
\la{fig:rate}

\end{figure}

The resulting full spectral function is shown in 
\figs\ref{fig:rhoV_M}, \ref{fig:rhoV_T} for various
masses and temperatures, as a function of $\omega$. 
The corresponding dilepton production rate from 
\eq\nr{dilepton} is shown in \fig\ref{fig:rate}. Compared
with the results in ref.~\cite{peskin}, the absolute
magnitude of the rate has decreased by about 10 -- 30\%, 
due to the inclusion of the normalization factor. 
We should again stress that particularly the charmonium 
case contains large uncertainties, and our results are 
to be trusted on the qualitative level only.

%
\section{Conclusions}
\la{se:concl}

The purpose of this paper has been to compute the heavy quark
contribution to the spectral function of the electromagnetic 
current at next-to-leading order in thermal QCD. The result consists
of a well-known vacuum part, \eq\nr{full_vac}, and a new thermal 
part, \eq\nr{full_T}. The thermal part is illustrated numerically 
in \fig\ref{fig:nlo} in comparison with the vacuum part.  

The thermal corrections in our result arise exclusively from 
the gluons with which the heavy quarks interact. Although these
contributions are not exponentially suppressed, they turn out 
to be power-suppressed at large energies $\omega \gg 2 M$: their 
general magnitude is $\rmO(g^2 T^2)$, and given that $T < g M$
is the phenomenologically interesting temperature range 
(cf.\ \eq\nr{range}), they can in principle be omitted 
in comparison with the next-to-leading order zero-temperature 
corrections, of $\rmO(g^2 M^2)$. This also means that the asymptotic
behaviour of the spectral function, needed as input for lattice 
studies, could (in the continuum limit) be extracted from the well-studied
zero-temperature computations (see, e.g., ref~\cite{hs}). At zero temperature
the next-to-leading order correction could, perhaps, even be worked out
at a finite lattice spacing. 

On the other hand, decreasing the energy towards the threshold, 
the thermal corrections become increasingly 
important. In fact, at next-to-leading
order, the vacuum spectral function vanishes at $\omega < 2 M$, while
the thermal correction stays finite. The result emerges from phase
space integrals associated with the energy constraint
$\delta(\omega+k-E_1-E_2)$, where $\omega$ is the photon energy; 
$k$ is the gluon energy; and $E_1, E_2$ are the energies of a heavy
quark and anti-quark. Graphically, the process corresponds to 
the annihilation of quarkonium into a gluon and a photon, the former
of which remains within the thermal medium. Since large values
of $k$ are Boltzmann suppressed, the thermal corrections
are substantial only for $|\omega - 2 M| \lsim T$.

Combining our new results, valid far enough away from the threshold,
with previously determined resummed expressions, valid close to the 
threshold, we have subsequently assembled phenomenological estimates for 
the spectral function in a macroscopic energy range 
(\figs\ref{fig:rhoV_M}, \ref{fig:rhoV_T}). The corresponding 
dilepton production rate is shown in \fig\ref{fig:rate}. 
Analogous results and plots for the spectral function 
in the scalar channel have been given in appendix~C. The computations of
the present paper play an important role in these plots particularly in that 
they fix the overall normalization of the assembled curves. 
We hope that these results can eventually be incorporated in 
a simulation including an expanding and cooling thermal fireball, 
which would then allow for a direct comparison with the dilepton
production rate measured in heavy ion collision experiments. 

We note, finally, that we have restricted to $\omega > 0$ in this paper. 
There is a lot of interesting structure in the vector channel spectral 
function also around $\omega \approx 0$, related to the heavy quark
diffusion coefficient. However, that structure is suppressed by 
$\exp(-\beta M)$, and a non-trivial result also only arises 
at the order $\rmO(\alpha_s^2)$~\cite{diff}, 
so that our present computation at $\rmO(\alpha_s)$ 
cannot add anything to the known results~\cite{chm}.

%
\section*{Acknowledgements}

We thank S.~Caron-Huot, D.~B\"odeker and Y.~Schr\"oder 
for useful discussions, and are grateful to the BMBF 
for financial support under project
{\em Hot Nuclear Matter from Heavy Ion Collisions 
     and its Understanding from QCD}.



\appendix
\renewcommand{\thesection}{Appendix~\Alph{section}}
\renewcommand{\thesubsection}{\Alph{section}.\arabic{subsection}}
\renewcommand{\theequation}{\Alph{section}.\arabic{equation}}

%
\section{Intermediate steps for a master sum-integral}

We elaborate in this appendix on the steps outlined in 
\se\ref{se:steps}. The starting point is the expression
in \eq\nr{nlo}.

%
\subsection{Matsubara sums and the spectral function}

The first step is to carry out 
the Matsubara sums $T \sum_{k_n}$, $T \sum_{\{p_n\}}$. 
The sum $T \sum_{k_n}$ is complicated by the appearance 
of the functions $\Pi_E$, $\Pi_T$ in the gluon propagators. 
The reason for their introduction was that there could in principle 
be infrared divergences associated with the gluons; in the Euclidean
formalism, these would come from small spatial momenta 
$\vec{k}$ for the Matsubara zero mode $k_n = 0$, and could then
be regulated by the fact that $\Pi_E(0,\vec{k}) = \mD^2 > 0$.
Our strategy in the following will be to {\em assume} that 
there are no infrared divergences, whereby we can 
set $\Pi_T = \Pi_E = 0$; the absence of divergences 
will be verified {\em a posteriori}. 
Nevertheless, it has still been important to keep 
$\Pi_E \neq \Pi_T$ in \eq\nr{nlo}, because it could happen 
that the structure multiplied by 
$1/(K^2 + \Pi_T) - 1/(K^2 + \Pi_E)$, 
which vanishes in the free limit, contains 
infrared sensitive parts which do not completely cancel against each other 
in the presence of $\Pi_E \neq \Pi_T$. 

In order to allow
for an eventual introduction of $\Pi_E$ and $\Pi_T$, we write for the moment
the gluon propagators in the spectral representation, 
\be
 \frac{1}{K^2+\Pi(K)} = 
 \int_{-\infty}^\infty \!\frac{\dd k^0}{\pi} 
 \frac{\rho(k^0,\mathbf{k})}{k^0-ik_n}
 \;,  \la{spectral}
\ee
and carry out the Matsubara sum $T \sum_{k_n}$
with the kernel $1/(k^0 - i k_n)$. 
In the free case, when the spectral function reads 
\be
 \rho_\rmi{free}(k^0,\vec{k}) = \frac{\pi}{2 k} 
 \Bigl[ 
   \delta(k^0-k) - \delta(k^0 + k)
 \Bigr]
 \;, \la{free}
\ee
with $k\equiv |\vec{k}|$, the whole procedure is obviously
just a rewriting of the decomposition
\be
 \frac{1}{k_n^2 + k^2} = \frac{1}{2 k}
 \biggl[ 
   \frac{1}{k - i k_n} + \frac{1}{k + i k_n}
 \biggr]
 \;. 
\ee
Note that the procedure is rather versatile and could also be 
interpreted as 
\ba
 & & \hspace*{-5mm} \frac{1}{K^2+\Pi(K)}  =   
 \biggl[ 1 - \frac{\Pi(K)}{K^2+\Pi(K)} \biggr]\frac{1}{K^2} 
  \la{rep} \\ & = & \!\!\!\!
 \int_{-\infty}^\infty \!\frac{\dd k^0}{\pi} \biggl\{ 
 \frac{\rho_\rmi{free}(k^0,\mathbf{k})}{k^0 - i k_n} 
  + 
 \frac{\bar\rho(k^0,\mathbf{k})}{2k} 
 \biggl( 
   \frac{1}{k^0+ik_n} - \frac{1}{k + i k_n}
 + 
   \frac{1}{k^0-ik_n} - \frac{1}{k - i k_n}
 \biggr) \frac{1}{k- k^0} \biggr\}
 \;,  \nonumber
\ea
where $\bar\rho(k^0,\vec{k})$ is the spectral function corresponding
to $-\Pi / (K^2 + \Pi)$, and we made use of the spectral function's 
antisymmetry in $k^0\to -k^0$. All the sums 
over $k_n$ are now with the same kernel 
as the one following from \eq\nr{spectral}.
The representation in \eq\nr{rep} would be 
relevant for $\Pi_E$, in which case $\rho_E$ would have a pole 
at $k^0 = k$~\cite{leb}.

After this lengthy introduction,  we are ready to carry out the sums. 
We describe the procedure in some detail for one of the master
sum-integrals appearing in \eq\nr{nlo}; for the others, the results
are listed in appendix B. 

The case we choose to consider in detail is 
\be
 {S_4^0}(\omega) \equiv \disc \biggl[  
 \int_{-\infty}^\infty \!\frac{\dd k^0}{\pi} 
 \Tint{K\{P\}}
 \frac{{\rho(k^0,\mathbf{k})} }{k^0-ik_n}
 \frac{1}{\Delta(P)\Delta(P-Q)\Delta(P-K)}
 \biggr]_{Q=(\omega_n\to -i\omega,\vec{0})}
 \;. \la{S40}
\ee
Denoting 
\be
 E_p \equiv \sqrt{\vec{p}^2 + M^2}
 \;, \quad
 E_{p-k} \equiv \sqrt{(\vec{p-k})^2 + M^2}
 \;, 
\ee
we can rewrite the sums as 
\ba
 & & \hspace*{-7mm}
 T\sum_{k_n} T\sum_{\{p_n\}} \frac{1}{[k^0 - i k_n][p_n^2 + E_p^2]
 [(p_n-\omega_n)^2+E_p^2][(p_n-k_n)^2 + E_{p-k}^2]}
 \nn & = & 
 T^4 \sum_{k_n}\sum_{\{p_n\}}\sum_{\{r_n\}}\sum_{\{s_n\}}
 \frac{\beta \delta_{r_n-p_n+\omega_n,0}
 \beta \delta_{s_n - p_n + k_n,0}}
 {[k^0-ik_n][p_n^2 + E_p^2][r_n^2 + E_p^2][s_n^2 + E_{p-k}^2]}
 \nn & = & 
 \int_0^{\beta}\!\!{\rm d}\tau\!\! \int_0^{\beta}\!\!{\rm d}\sigma\,
 e^{i\omega_n\tau}
 T^4 \sum_{k_n}\sum_{\{p_n\}}\sum_{\{r_n\}}\sum_{\{s_n\}}
 \frac{e^{i k_n\sigma}}{k^0 - i k_n}
 \frac{e^{-ip_n(\tau+\sigma)}}{p_n^2 + E_p^2}
 \frac{e^{i r_n\tau}}{r_n^2 + E_p^2}
 \frac{e^{i s_n\sigma}}{s_n^2 + E_{p-k}^2}
 \;, \hspace*{7mm}
\ea
where in the last step we used 
a representation of the Kronecker delta-function, 
$
 \beta \delta_{t_n,0} = \int_0^{\beta}\!{\rm d}\tau \, e^{i t_n\tau}
$.
The sums have factorized and can now be carried out: 
\ba
  T\sum_{k_n} \frac{e^{i k_n\sigma} }{k^0 - i k_n} & = &
 \nB{}(k^0) e^{(\sigma\,\rmi{mod}\,\beta)k^0} \;,
 \quad 0 < \sigma\,\mbox{mod}\,\beta < \beta \la{bsum} \;, 
 \\ 
  T\sum_{\{r_n\} } \frac{e^{\pm i r_n\tau} }{r_n^2 + E_p^2} & = & 
 \frac{\nF{}(E_p)}{2 E_p}
 \Bigl[
   e^{(\beta - |\tau\,\rmi{mod}\,2 \beta |) E} 
 - e^{|\tau\,\rmi{mod}\,2 \beta| E} 
 \Bigr] \;, 
 \quad -\beta \le \tau\,\mbox{mod}\,2 \beta \le \beta 
 \;, \nn[-3mm] \la{fsum}
\ea
where $\nF{}(\omega)\equiv 1/[\exp(\beta\omega)+1]$ and 
$\nB{}(\omega)\equiv 1/[\exp(\beta\omega)-1]$.\footnote{
 The sum in \eq\nr{bsum} is discontinuous at 
 $\sigma = 0 \,\mbox{mod}\,\beta$, and defining
 its value at the discontinuity requires care; although 
 of no importance in the present context, we note that 
 the expression with the correct antisymmetry in $k^0$
 corresponds to the ``average'',  
 $n_\rmii{B}(k^0)(1 + e^{\beta k^0})/2 = n_\rmii{B}(k^0) + \fr12$.}
The subsequent integrals over $\tau$ and $\sigma$ are elementary; 
we simply need to split 
$
 \int_0^{\beta}\!\!{\rm d}\sigma = 
 \int_0^{\beta-\tau}\!\!{\rm d}\sigma + 
 \int_{\beta-\tau}^{\beta}\!\!{\rm d}\sigma 
$, 
and note that in the latter range, 
$
 |\tau + \sigma \,\mbox{mod}\,2 \beta| = 2\beta - \tau - \sigma
$.
Setting $e^{i\omega_n\beta} \equiv 1$ after the 
integrations, the $\omega_n$-dependence of 
the result appears only in structures like $1/(i\omega_n + \sum_i E_i)$, 
and we can read off the discontinuity: 
\be
 \disc \biggl[ \frac{1}{i\omega_n + \sum_i E_i} \biggr]_{\omega_n\to -i\omega}
 = -\pi \delta(\omega + \sum_i E_i)
 \;. 
\ee
Implementing these steps in practice, and restricting the 
$k^0$-integral to positive values by making use of the 
antisymmetry of $\rho(k^0,\vec{k})$, we arrive at 
\ba
 S_4^0(\omega) & = & \int_0^\infty \!\frac{{\rm d}k^0}{\pi}
 \int_{\vec{k,p}} \rho(k^0,\vec{k}) \frac{\pi}{4 E_p E_{p-k}} \biggl\{ 
 \la{S40_Matsu} \\ & & 
 \frac{1}{2 E_p} 
 \Bigl[\delta(\omega - 2 E_p) - \delta(\omega+2 E_p) \Bigr]
 (1-2 \nF{1}) \times 
 \nn & & \times \Bigl[(\Delta^{-1}_{++} + \Delta^{-1}_{-+})
 (1+\nB{0} - \nF{2}) - (\Delta^{-1}_{--}+\Delta^{-1}_{+-})
 (\nB{0} + \nF{2}) \Bigr]
 \nn & & -
 \Bigl[\delta(\omega - \Delta_{++}) - \delta(\omega+\Delta_{++}) \Bigr]
 \Delta^{-1}_{++} \Delta^{-1}_{-+}
 \Bigl[(1+\nB{0})(1-\nF{1}-\nF{2}) + \nF{1}\nF{2} \Bigr]  
 \nn & & - 
 \Bigl[\delta(\omega - \Delta_{--}) - \delta(\omega+\Delta_{--}) \Bigr]
 \Delta^{-1}_{--}\Delta^{-1}_{+-}
 \Bigl[-\nB{0}(1-\nF{1}-\nF{2}) + \nF{1}\nF{2} \Bigr]  
 \nn & & -
 \Bigl[\delta(\omega - \Delta_{+-}) - \delta(\omega+\Delta_{+-}) \Bigr]
 \Delta^{-1}_{--}\Delta^{-1}_{+-}
 \Bigl[\nB{0}\nF{1}-(1+\nB{0})\nF{2} + \nF{1}\nF{2} \Bigr]  
 \nn & & -
 \Bigl[\delta(\omega - \Delta_{-+}) - \delta(\omega+\Delta_{-+}) \Bigr]
 \Delta^{-1}_{++}\Delta^{-1}_{-+}
 \Bigl[\nB{0}\nF{2}-(1+\nB{0})\nF{1} + \nF{1}\nF{2} \Bigr]  
 \biggr\} \;. \nonumber
\ea
To simplify the expression somewhat, we have introduced 
the shorthands
\ba
 & &  \Delta_{\sigma\tau} \equiv k^0 + \sigma E_{p} + \tau E_{p-k}
 \;, \quad \sigma, \tau = \pm
 \;, \la{sh1} \\[1mm]
 & & n_\rmii{B0} \equiv \nB{}(k^0)
 \;, \quad
 n_\rmii{F1} \equiv \nF{}(E_p)
 \;, \quad
 n_\rmii{F2} \equiv \nF{}(E_{p-k}) 
 \;. \la{sh2}
\ea
Note that the result in \eq\nr{S40_Matsu}
is antisymmetric in $\omega\to -\omega$, 
as must be the case. 

Inspecting \eq\nr{S40_Matsu}, we note the appearance of structures
in the denominator, $\Delta_{+-}$ etc, which look like they might 
vanish for some $\vec{k,p}$. In fact, in one of
the other master sum-integrals, even the structure $1/(E_{p-k}-E_p)$
appears, which certainly 
vanishes, for $2\vec{p}\cdot\vec{k} = \vec{k}^2$. 
It can be verified, however, that such poles always cancel between 
the various types of terms in the expression, and do not hinder 
the actual integration. (If integration variables are changed 
in a subset of the expression, $\vec{p}\to \vec{-p+k}$, to remove
an apparent symmetry in $E_p\leftrightarrow E_{p-k}$, 
then such terms do not in general cancel any 
more; nevertheless their contribution remains finite and correct
if the poles are interpreted as principal values.)

%
\subsection{Spatial momentum integrals}
\la{ss:spatial}

The result so far, \eq\nr{S40_Matsu}, 
contains integrals with two types of delta-functions: 
ones with $\delta(\omega\pm 2E_p)$, which we call ``factorized'' (fz) 
integrals, because the gluon momentum $\vec{k}$ does not 
appear inside the $\delta$-functions; 
and ones with more complicated $\delta$-functions, which we call 
``phase space'' (ps) integrals. In both cases, our
strategy is to first carry out the integral over
the quark momentum $p\equiv |\vec{p}|$ 
and over the 
angle between $\vec{p}$ and $\vec{k}$; the integral 
over the gluon momentum $k\equiv |\vec{k}|$ is left
for later (it is this integral which could potentially suffer from 
infrared divergences). 

We start by considering the phase space integrals, 
which are ultraviolet finite, so that we can set $d=3$. In order to simplify
the task, we ignore from now on terms suppressed by $\exp(-\beta M) \ll 1$.
This means that all appearances of 
$\nF{}(E_p)$ and $\nF{}(E_{p-k})$ can be omitted. Furthermore, 
restricting to $\omega > 0$, we note that the delta-function 
$
 \delta(\omega - \Delta_{--}) = \delta(\omega + E_p + E_{p-k} - k^0)
$
can only be realized for $k^0 > 2 M$, and will then lead to an exponentially
small contribution due to the appearance of 
the Bose distribution $\nB{}(k^0)$. The delta-function 
$
 \delta(\omega + \Delta_{++}) = \delta(\omega + k^0 + E_p + E_{p-k})
$
does not get realized at all. Thereby only two of the eight 
delta-functions in \eq\nr{S40_Matsu} remain non-zero, 
and the integral simplifies to 
\ba
  \left. S_4^0(\omega) \right|_\rmii{ps} 
 \!\! & = & \!\! \int_0^\infty \!\frac{{\rm d}k^0}{\pi}
 \int \! \frac{{\rm d}^{3}\vec{k}}{(2\pi)^{3}} 
 \,\rho(k^0,\vec{k})
 \int \! \frac{{\rm d}^{3}\vec{p}}{(2\pi)^{3}}
 \frac{\pi}{E_p E_{p-k}} \biggl\{  
 \nn &  & \hspace*{-2cm}  
  \delta(\omega - k^0 - E_p - E_{p-k})[1+\nB{}(k^0)] \phi(k^0) 
 + \delta(\omega + k^0 - E_p - E_{p-k})\nB{}(k^0) \phi(-k^0) 
 \biggr\} \;, \hspace*{1.0cm}
 \la{S40_Matsu_2}
\ea
where 
\be
 \phi(k^0) \equiv 
 \frac{-1}{4 (k^0 + E_p + E_{p-k})(k^0 - E_p + E_{p-k})} 
 \;. 
\ee
Fixing $\vec{k}$ and denoting $z\equiv - \vec{p}\cdot\vec{k}/pk$, so 
that $E_{p-k} = \sqrt{p^2 + k^2 + 2 p k z + M^2}$, 
we can change integration variables from $p,z$ to $E_p,E_{p-k}$: 
\be
 \int\! {\rm d}^3\vec{p} 
 = 2\pi \int_0^\infty \!\! {\rm d}p \; p^2 \int_{-1}^{+1} \! {\rm d}z 
 = 2 \pi \int_{M}^{\infty} \! {\rm d}E_p 
 \int_{E_{p-k}^-}^{E_{p-k}^+} \! {\rm d}E_{p-k} 
 \frac{E_p E_{p-k}}{k}
 \;, 
\ee
where $E_{p-k}^{\pm} \equiv \sqrt{p^2 \pm 2 p k + k^2 + M^2}$. 
The hard task is to figure out when the $\delta$-functions get 
realized. For $\delta(\omega - k^0 - E_p - E_{p-k})$
this happens provided that
$
 E_{p-k}^{-} < \omega - k^0 - E_p < E_{p-k}^{+}
$,  
which leads to  
\ba
 & & \omega >  2 M \;, \quad 
 k^0 <  \omega - 2 M \;, \quad
 k  <  \sqrt{(\omega - k^0)^2 - 4 M^2} \;, \la{ps1} \\  
 & &  \frac{\omega-k^0}{2}-\frac{k}{2}
  \sqrt{1- \frac{4M^2}{(\omega-k^0)^2-k^2}}
    <  E_p  < 
  \frac{\omega-k^0}{2}+\frac{k}{2}
 \sqrt{1 - \frac{4M^2}{(\omega-k^0)^2-k^2}}
 \;. \hspace*{1cm} \la{ps2}
\ea
In the case of the free gluon spectral function, with $k^0 = k$, 
these simplify to 
\ba
 & & \omega > 2 M \;, \quad
 k  <  \frac{\omega^2 - 4 M^2}{2\omega} \;, \la{simp1} \\  
 & &  \frac{\omega-k}{2}-\frac{k}{2}
  \sqrt{1- \frac{4M^2}{\omega(\omega - 2k)}}
    < E_p  < 
  \frac{\omega-k}{2}+\frac{k}{2}
 \sqrt{1 - \frac{4M^2}{\omega(\omega-2 k)}}
 \;. \hspace*{1cm} \la{simp2}
\ea
For $\delta(\omega + k^0 - E_p - E_{p-k})$, we simply need to 
set $k^0\to -k^0$ in \eqs\nr{ps1}, \nr{ps2};  
putting subsequently $k^0 = k$, the explicit expressions read
\ba
 & & \omega > 0 \;, \quad
 k  >  \mbox{max}\biggl(0, \frac{4 M^2 - \omega^2 }{2\omega} \biggr) 
 \;, \la{simp3} \\  
 & &  \frac{\omega+k}{2}-\frac{k}{2}
  \sqrt{1- \frac{4M^2}{\omega(\omega + 2k)}}
    < E_p  < 
  \frac{\omega+k}{2}+\frac{k}{2}
 \sqrt{1 - \frac{4M^2}{\omega(\omega+2 k)}}
 \;. \hspace*{1cm} \la{simp4}
\ea
Note also that the function $\phi$ evaluates 
to $-1/[4 \omega(\omega - 2 E_p)]$ after integration 
over $E_{p-k}$, for both delta functions
in \eq\nr{S40_Matsu_2}.

Inserting the free 
gluon spectral function from \eq\nr{free} and using the simplified
formulae from \eqs\nr{simp1}--\nr{simp4}, 
the integrals over $E_{p-k}$ and $E_p$ can now be carried out. For the thermal
part, i.e.\ the one proportional to $\nB{0}$, this yields 
\ba
  \left. S_4^0(\omega) \right|_\rmii{ps}^\rmii{$T$} 
  \!\! & = & \!\! \frac{1}{(4\pi)^3 \omega}
  \biggl\{  
 \int_0^\infty \! \dd k \, \nB{}(k) 
 \biggl[ 
   \theta(\omega) \, 
   \theta\Bigl(k- 
    \frac{4M^2-\omega^2}{2\omega}
    \Bigr)
   \,\mathrm{acosh}
   \sqrt{ \frac{\omega(\omega+2 k)}{4M^2} }
     \la{S40_ps_T} \\ & & \phantom{ \int_0^\infty \! \dd k \, } 
   -
   \theta(\omega - 2M) \, 
   \theta\Bigl(\frac{\omega^2-4M^2}{2\omega}-k\Bigr)
   \,\mathrm{acosh}
   \sqrt{ \frac{\omega(\omega-2 k)}{4M^2} }
 \; \biggr] \biggr\} + \rmO(e^{-\beta M})
 \;. \nonumber
\ea
The vacuum part, on the other hand, is given by the latter row
of \eq\nr{S40_ps_T}, but just without the function $\nB{}(k)$; then 
the final $k$-integral is doable as well, and we end up with  
\ba
  \left. S_4^0(\omega) \right|_\rmii{ps}^\rmii{vac} 
  \!\! & = & \!\! \frac{1}{(4\pi)^3 }
  \theta(\omega - 2 M )
  \biggl[ \frac{({\omega^2 - 4 M^2})^{\fr12}}{4\omega}
  + \frac{2 M^2 - \omega^2 }{2 \omega^2}\,\mathrm{acosh}
  \biggl(\frac{\omega}{2 M}\biggr)
  \biggr] 
  \;. \la{S40_ps_vac}
\ea

Consider next the factorized integrals, 
i.e.\ the first term inside the curly brackets 
in \eq\nr{S40_Matsu}. 
Again we start by integrating
over $p,z$, and leave the integration over $k$ for later. This time it
is useful to view $z$ as part of the $\vec{k}$-integral, i.e.\ 
\be
 \mu^{2\epsilon}\int\!\frac{\dd^d \vec{k}}{(2\pi)^d} = 
 \frac{4\mu^{2\epsilon}}{(4\pi)^\frac{d+1}{2}\Gamma(\frac{d-1}{2})} 
 \int_0^\infty \! \dd k \, k^{d-1} \int_{-1}^1 \! \dd z \, (1-z^2)^{(d-3)/2}
 \;, \la{z_measure}
\ee
where $d\equiv 3 - 2\epsilon$. The factorized integrals are, in general, 
ultraviolet divergent, and necessitate keeping track of $\epsilon\neq 0$.
As always, a helpful strategy is to add and subtract a simple infrared 
finite regulator, such as $1/(k^2 + M^2)^\alpha$, where $\alpha$ is 
so chosen that the complicated expression becomes ultraviolet finite after the 
subtraction, and can be worked out at $\epsilon = 0$, while the ultraviolet
divergent integral with the measure of \eq\nr{z_measure} 
is taken over the simple
regulator. In the complicated but ultraviolet finite integral, it is 
useful to change integration variables from $z$ to $E_{p-k}$, using 
\be
 \int_{-1}^{+1} \frac{{\rm d}z }{ E_{p-k} } = \int_{E_{p-k}^-}^{E_{p-k}^+}
 \frac{ {\rm d}E_{p-k} }{ p k }
 \;. \la{subst_z}
\ee
We should remark that in our particular example, $S_4^0$, the trick 
of adding and subtracting a regulator is superfluous, given that the 
divergent integral can be directly identified as a known case, but
in the general case we have found it to be very helpful. 

Now,
because of the constraint $\delta(\omega - 2 E_p)$ (for $\omega > 0$)
in the factorized integrals, 
the integral over $p$ can be carried out trivially. 
In fact,  
comparing \eq\nr{S40_Matsu} with \nr{S1_Matsu}, which 
defines a corresponding
1-loop integral (denoted by $S_1(\omega)$ and given explicitly 
in \eq\nr{S1_final}), we arrive at
\ba
 \left. S_4^0(\omega) \right|_\rmii{fz}
 & = & S_1(\omega) \int_0^\infty \!\frac{{\rm d}k^0}{\pi}
 \; \mu^{2\epsilon}\!\! \int\! \frac{{\rm d}^d\vec{k}}{(2\pi)^d 2 E_{p-k}} 
 \rho(k^0,\vec{k}) \biggl\{ 
 \nn & & \hspace*{-1cm} \Bigl[(\Delta^{-1}_{++} + \Delta^{-1}_{-+})
 (1+\nB{0} - \nF{2}) - (\Delta^{-1}_{--}+\Delta^{-1}_{+-})
 (\nB{0} + \nF{2}) \Bigr] \biggr\}_{p = \sqrt{\omega^2 - 4 M^2}/2}
 \;. \hspace*{1cm} \la{S40_fz_all}
\ea
Let us first inspect the vacuum ($T=0$) part hereof, 
i.e.\ the term without $\nB{0}$ or $\nF{2}$. Inserting the free 
gluon spectral function from \eq\nr{free},  
the multiplier of $S_1(\omega)$ becomes 
\be
 \mathcal{B}_0 \equiv 
  \mu^{2\epsilon} \!\!
 \int\! \frac{{\rm d}^d\vec{k}}{(2\pi)^d} \frac{1}{4 k E_{p-k}} 
  \biggl[ \frac{1}{k + E_p + E_{p-k}} +
  \frac{1}{k - E_p + E_{p-k}}
  \biggr]_{p = \sqrt{\omega^2 - 4 M^2}/2}
  \;. 
\ee
This can be compared with the integral 
\ba
 B_0(P^2;0,M^2) & \equiv & 
 \mu^{2\epsilon}\!\! \int\!\frac{{\rm d}^D\! K}{(2\pi)^D}
 \frac{1}{K^2[(P-K)^2 + M^2]} \la{B0} \\ 
 & = &  
 \mu^{2\epsilon}\!\!
 \int\! \frac{{\rm d}^d\vec{k}}{(2\pi)^d} \frac{1}{4 k E_{p-k}}
 \biggl[\frac{1}{ip_0 + k + E_{p-k}} + \frac{1}{-ip_0 + k + E_{p-k}}\biggr]
 \;,  
\ea
where we denoted $K=(k_0,\vec{k})$ and carried out the integral over $k_0$.
In other words, $\mathcal{B}_0 = B_0(P_E^2;0,M^2)$, 
where 
\be
 P_E \equiv  \left. (-iE_p,p\, \hat{\vec{e}})
 \right|_{p=\sqrt{\omega^2 - 4 M^2}/2}
 \;, \quad
 P_E^2 = - M^2 
 \;, 
\ee
and $\hat{\vec{e}}$ is a unit vector; the value of this standard
vacuum integral reads
\be
 B_0(-M^2;0,M^2) 
 = \frac{1}{(4\pi)^2} \biggl[ \frac{1}{\epsilon} + \ln\frac{\bmu^2}{M^2} + 2
 + \rmO(\epsilon) 
 \biggr]
 \;.
\ee 
Combining this with \eq\nr{S1_final}, the factorized vacuum part becomes
\ba
 \left. S_4^0(\omega) \right|_\rmii{fz}^\rmii{vac} 
 \!\! & = & \!\! 
 S_1(\omega) B_0(-M^2;0,M^2) 
 \nn  
 \!\! & = & \!\!
 \theta(\omega - 2 M)
 \frac{(\omega^2 - 4 M^2)^{\fr12}}{4 \omega (4\pi)^3}
 \tanh\Bigl(\frac{\beta\omega}{4}\Bigr)
 \biggl[
  \frac{1}{\epsilon} + 
  \ln\frac{\bmu^4}{M^2(\omega^2 - 4 M^2)}
  + 4 + \rmO(\epsilon) 
 \biggr]
 \;. \nn \la{S40_fz_vac}
\ea
For completeness, we have even kept exponentially small thermal terms
in the coefficient of $1/\epsilon$, given that it is useful to crosscheck
the exact cancellation of ultraviolet poles; after this check, we set 
$
 \tanh({\beta\omega}/{4}) =  1 + \rmO(\exp(-\beta M))
$, given that $\omega\ge 2 M$.

Consider then the thermal part of \eq\nr{S40_fz_all}. 
Again, we omit exponentially small 
terms $\sim \exp(-\beta M)$, and use the free gluon spectral function. 
Because of the remaining factor $\nB{0}$, the $k$-integral 
is exponentially convergent, and we can set $\epsilon = 0$. 
Employing \eq\nr{subst_z} the thermal part becomes 
\ba
 \left. S_4^0(\omega) \right|_\rmii{fz}^\rmii{$T$} & = & 
 \frac{S_1(\omega)}{(4\pi)^2} \int_0^\infty\!\!\! \dd k \, k \, \nB{}(k) 
 \int_{-1}^{+1} \! \frac{{\rm d}z}{E_{p-k}} \times
 \nn & & \hspace*{-1.5cm} \times
 \biggl[ 
   \frac{1}{k+E_p+E_{p-k}}
   +\frac{1}{k-E_p+E_{p-k}}
   -\frac{1}{k-E_p-E_{p-k}}
   -\frac{1}{k+E_p-E_{p-k}}
 \biggr]_{p = \sqrt{\omega^2 - 4 M^2}/2}  
 \nn & = & 
 \frac{S_1(\omega)}{(4\pi)^2 p} \int_0^\infty\!\!\! \dd k \, \nB{}(k) 
 \times
 \nn & & \hspace*{-1.5cm} \times
 \ln\left| 
 \frac{(k + E_p + E_{p-k}^+)(k - E_p + E_{p-k}^+)
       (k - E_p - E_{p-k}^+)(k + E_p - E_{p-k}^+)}
      {(k + E_p + E_{p-k}^-)(k - E_p + E_{p-k}^-)
       (k - E_p - E_{p-k}^-)(k + E_p - E_{p-k}^-)}
 \right|_{p = \sqrt{\omega^2 - 4 M^2}/2}  
 \;. \nn 
\ea
Making use of
\be
 (k + \sigma E_p)^2 - (E_{p-k}^\tau)^2 = 
 2 k [\sigma E_p - \tau\, p]
 \;, \quad \sigma, \tau = \pm \;, 
\ee
it can be seen that the argument of the logarithm evaluates to unity.
Hence, $\left. S_4^0(\omega) \right|_\rmii{fz}^\rmii{$T$} = 0$.

To summarize, combining \eqs\nr{S40_ps_T}, 
\nr{S40_ps_vac}, \nr{S40_fz_vac}, we get 
\be
 S_4^0(\omega) = 
 \left. S_4^0(\omega) \right|_\rmii{ps}^\rmii{$T$}
 + \left. S_4^0(\omega) \right|_\rmii{ps}^\rmii{vac}
 + \left. S_4^0(\omega) \right|_\rmii{fz}^\rmii{vac} 
 \;. \la{S40_final}
\ee 
The other master sum-integrals can be worked out in the same way, 
and the final results are listed in appendix B. 

\newpage

%
\section{General results for all master sum-integrals}

We collect in this appendix the results for all the master sum-integrals
entering the computation, obtained with the methods explained
in appendix~A. In each case, we list the definition
of the sum-integral; an intermediate result obtained after carrying
our the Matsubara sums and taking the discontinuity; and the final 
result, obtained after restricting to the free gluon spectral function,
omitting terms suppressed by $\exp(-\beta M)$ (except from the 
ultraviolet divergences), and carrying out the final spatial integrations.  
As before, the integration measure for the spatial integrations is defined as 
\be
 \int_{\vec{p}} \equiv
 \mu^{2\epsilon}  \int \! 
 \frac{{\rm d}^{3-2\epsilon}\vec{p}}
 {(2\pi)^{3-2\epsilon}}
 \;,  
\ee
and 
$\bmu^2 = 4\pi\mu^2 e^{-\gamma_\rmii{E}}$ 
denotes the $\msbar$ scale parameter. 
To simplify the expressions somewhat, 
we also make use of the shorthands
listed in \eqs\nr{sh1}, \nr{sh2}.
The subscripts ``ps'' and ``fz'' denote ``phase space'' and
``factorized'' integrations, respectively, in the sense
of \se\ref{ss:spatial}.

%
\subsection{$S_1$}

The sum-integral $S_1$ is defined as 
\be
 {S_1}(\omega) \equiv \disc \biggl[  
 \Tint{\{P\}}
 \frac{1}{\Delta(P)\Delta(P-Q)}
 \biggr]_{Q=(-i\omega,\vec{0})}
 \;. \la{S1}
\ee
Carrying out the Matsubara sum and taking the discontinuity leads to 
\be
 S_1(\omega) = \int_{\vec{p}}\frac{\pi}{4 E_p^2}
 \Bigl[1 - 2 \nF{}(E_p) \Bigr]
 \Bigl[
  \delta(\omega - 2 E_p) - \delta(\omega + 2 E_p) 
 \Bigr]
 \;.  \la{S1_Matsu}
\ee
The remaining integral is trivial due to the $\delta$-function and, 
restricting to $\omega > 0$, we arrive at 
\be
 S_1(\omega) = 
 \theta(\omega - 2 M)
 \frac{(\omega^2 - 4 M^2)^{\fr12}}{16\pi\omega}
 \tanh\Bigl(\frac{\beta\omega}{4}\Bigr)
 \biggl[
  1 + \epsilon \biggl(
  \ln\frac{\bmu^2}{\omega^2 - 4 M^2} + 2 \biggr) + \rmO(\epsilon^2) 
 \biggr]
 \;. \la{S1_final}
\ee

%
\subsection{$S_2$}

The sum-integral $S_2$ is defined as
\be
 {S_2}(\omega) \equiv \disc \biggl[  
 \Tint{\{P\}}
 \frac{1}{\Delta^2(P)\Delta(P-Q)}
 \biggr]_{Q=(-i\omega,\vec{0})}
 \;. \la{S2}
\ee
It is easy to see that $S_2 = -\fr12 {\rm d} S_1/{\rm d}M^2$.
Therefore, from \eq\nr{S1_final}, we obtain
\be
 S_2(\omega) = 
 \theta(\omega - 2 M)
 \frac{(\omega^2 - 4 M^2)^{-\fr12}}{16\pi\omega}
 \tanh\Bigl(\frac{\beta\omega}{4}\Bigr)
 \biggl[
  1 + \epsilon 
  \ln\frac{\bmu^2}{\omega^2 - 4 M^2}  + \rmO(\epsilon^2) 
 \biggr]
 \;. \la{S2_final}
\ee

%
\subsection{$S_3$}

The sum-integral $S_3$ is defined as 
\be
 {S_3}(\omega) \equiv \disc \biggl[  
 \int_{-\infty}^\infty \!\frac{\dd k^0}{\pi} 
 \Tint{K\{P\}}
 \frac{{\rho(k^0,\mathbf{k})} }{k^0-ik_n}
 \frac{1}{\Delta(P)\Delta(P-Q-K)}
 \biggr]_{Q=(-i\omega,\vec{0})}
 \;. \la{S3}
\ee
Performing the Matsubara sums, taking the discontinuity, and
making use of the antisymmetry of $\rho(k^0,\vec{k})$ yields
\ba
 S_3(\omega) & = & \int_0^\infty \!\frac{{\rm d}k^0}{\pi}
 \int_{\vec{k,p}} \rho(k^0,\vec{k}) \frac{\pi}{4 E_p E_{p-k}} \biggl\{ 
 \\ & & \phantom{+}
 \Bigl[\delta(\omega - \Delta_{++}) - \delta(\omega+\Delta_{++}) \Bigr]
 \Bigl[(1+\nB{0})(1-\nF{1}-\nF{2}) + \nF{1}\nF{2} \Bigr]  
 \nn & & + 
 \Bigl[\delta(\omega - \Delta_{--}) - \delta(\omega+\Delta_{--}) \Bigr]
 \Bigl[-\nB{0}(1-\nF{1}-\nF{2}) + \nF{1}\nF{2} \Bigr]  
 \nn & & +
 \Bigl[\delta(\omega - \Delta_{+-}) - \delta(\omega+\Delta_{+-}) \Bigr]
 \Bigl[\nB{0}\nF{1}-(1+\nB{0})\nF{2} + \nF{1}\nF{2} \Bigr]  
 \nn & & +
 \Bigl[\delta(\omega - \Delta_{-+}) - \delta(\omega+\Delta_{-+}) \Bigr]
 \Bigl[\nB{0}\nF{2}-(1+\nB{0})\nF{1} + \nF{1}\nF{2} \Bigr]  
 \biggr\} \;. \nonumber
\ea
Inserting the free gluon spectral function, and omitting exponentially
small terms, yields 
\ba
  S_3(\omega)  
  \!\! & = & \!\! \frac{1}{(4\pi)^3}
  \biggl\{  
  \theta(\omega - 2 M )
  \biggl[ \frac{({\omega^2 - 4 M^2})^{\fr12}(\omega^2 + 2 M^2)}{8\omega}
  + \frac{M^2 (M^2 - \omega^2)}{\omega^2}\,\mathrm{acosh}
  \biggl(\frac{\omega}{2 M}\biggr)
  \biggr]  
 \nn & + & \!\!\!
 \int_0^\infty \!\! \dd k \, k\, \nB{}(k) 
 \biggl[ 
   \theta(\omega) \, 
   \theta\Bigl(k- 
    \frac{4M^2-\omega^2}{2\omega}
    \Bigr)
   \sqrt{1 - \frac{4M^2}{\omega(\omega+2 k)}}
   \nn & & \hspace*{2.1cm} + 
   \theta(\omega - 2M) \, 
   \theta\Bigl(\frac{\omega^2-4M^2}{2\omega}-k\Bigr)
   \sqrt{1 - \frac{4M^2}{\omega(\omega-2 k)}}
 \; \biggr] \biggr\} + \rmO(e^{-\beta M})
 \;.   \nn  
\ea

%
\subsection{$S_4^0$}

The sum-integral $S_4^0$ is defined in \eq\nr{S40}; 
its value after the Matsubara sums is given in \eq\nr{S40_Matsu}; 
the result after the phase space integrals is the
sum of \eqs\nr{S40_ps_T}, \nr{S40_ps_vac}, \nr{S40_fz_vac}. 

%
\subsection{$S_4^1$}

The sum-integral $S_4^1$ is defined as
\be
 {S_4^1}(\omega) \equiv \disc \biggl[  
 \int_{-\infty}^\infty \!\frac{\dd k^0}{\pi} 
 \Tint{K\{P\}}
 \frac{{\rho(k^0,\mathbf{k})} }{k^0-ik_n}
 \frac{Q\cdot K}{\Delta(P)\Delta(P-Q)\Delta(P-K)}
 \biggr]_{Q=(-i\omega,\vec{0})}
 \;. \la{S41}
\ee
Performing the Matsubara sums, taking the discontinuity, and
making use of the antisymmetry of $\rho(k^0,\vec{k})$ yields
\ba
 S_4^1(\omega) & = & \int_0^\infty \!\frac{{\rm d}k^0}{\pi}
 \int_{\vec{k,p}} \rho(k^0,\vec{k}) 
 \frac{\pi k^0 \omega}{4 E_p E_{p-k}} \biggl\{ 
 \la{S41_Matsu}
 \\ & & 
 \frac{1}{2 E_p} 
 \Bigl[\delta(\omega - 2 E_p) + \delta(\omega+2 E_p) \Bigr]
 (1-2 \nF{1}) \times 
 \nn & & \times \Bigl[(\Delta^{-1}_{++} - \Delta^{-1}_{-+})
 (1+\nB{0} - \nF{2}) + (\Delta^{-1}_{--}-\Delta^{-1}_{+-})
 (\nB{0} + \nF{2}) \Bigr]
 \nn & & +
 \Bigl[\delta(\omega - \Delta_{++}) + \delta(\omega+\Delta_{++}) \Bigr]
 \Delta^{-1}_{++} \Delta^{-1}_{-+}
 \Bigl[(1+\nB{0})(1-\nF{1}-\nF{2}) + \nF{1}\nF{2} \Bigr]  
 \nn & & + 
 \Bigl[\delta(\omega - \Delta_{--}) + \delta(\omega+\Delta_{--}) \Bigr]
 \Delta^{-1}_{--}\Delta^{-1}_{+-}
 \Bigl[-\nB{0}(1-\nF{1}-\nF{2}) + \nF{1}\nF{2} \Bigr]  
 \nn & & +
 \Bigl[\delta(\omega - \Delta_{+-}) + \delta(\omega+\Delta_{+-}) \Bigr]
 \Delta^{-1}_{--}\Delta^{-1}_{+-}
 \Bigl[\nB{0}\nF{1}-(1+\nB{0})\nF{2} + \nF{1}\nF{2} \Bigr]  
 \nn & & +
 \Bigl[\delta(\omega - \Delta_{-+}) + \delta(\omega+\Delta_{-+}) \Bigr]
 \Delta^{-1}_{++}\Delta^{-1}_{-+}
 \Bigl[\nB{0}\nF{2}-(1+\nB{0})\nF{1} + \nF{1}\nF{2} \Bigr]  
 \biggr\} \;. \nonumber 
\ea
Inserting the free gluon spectral function, the ultraviolet divergent
factorized vacuum part reads
\ba
 \left. S_4^1(\omega) \right|_\rmii{fz}^\rmii{vac} 
 \!\! & = & \!\!
 - \theta(\omega - 2 M)
 \frac{\omega (\omega^2 - 4 M^2)^{\fr12}}{16 (4\pi)^3}
 \tanh\Bigl(\frac{\beta\omega}{4}\Bigr)
 \biggl[
  \frac{1}{\epsilon} + 
  \ln\frac{\bmu^4}{M^2(\omega^2 - 4 M^2)}
  + 3 + \rmO(\epsilon) 
 \biggr]
 \;, \nn \la{S41_fz_vac}
\ea
where in the coefficient of the divergence we have 
accounted even for exponentially small terms. 
The vacuum part from the phase space integrals reads 
\ba
  \left. S_4^1(\omega) \right|_\rmii{ps}^\rmii{vac} 
  \!\! & = & \!\!  \frac{ \theta(\omega - 2 M )}{(4\pi)^3 }
  \biggl[ \frac{3({\omega^2 - 4 M^2})^{\fr12}(2 M^2 - \omega^2)}{32\omega}
  + \frac{\omega^4 - 4 \omega^2 M^2 + 6 M^4 }{8 \omega^2}\,\mathrm{acosh}
  \biggl(\frac{\omega}{2 M}\biggr)
  \biggr] 
  \;, \nn \la{S41_ps_vac}
\ea
while the thermal parts amount to 
\ba
  \left. S_4^1(\omega) \right|_\rmii{fz}^\rmii{$T$} 
  \!\! & = & \!\! \frac{1}{(4\pi)^3}
  \biggl\{  
 \int_0^\infty \! \dd k \, k\,\nB{}(k) 
   \theta(\omega - 2M) \, 
   \biggl[ - 2 \,
   \,\mathrm{acosh}\biggl( \frac{\omega}{2M} \biggr)
 \; \biggr] \biggr\} + \rmO(e^{-\beta M})
 \;, \la{S41_fz_T} \\
  \left. S_4^1(\omega) \right|_\rmii{ps}^\rmii{$T$} 
  \!\! & = & \!\! \frac{1}{(4\pi)^3}
  \biggl\{  
 \int_0^\infty \! \dd k \, k\,\nB{}(k) 
 \biggl[ 
   \theta(\omega) \, 
   \theta\Bigl(k- 
    \frac{4M^2-\omega^2}{2\omega}
    \Bigr)
   \,\mathrm{acosh}
   \sqrt{ \frac{\omega(\omega+2 k)}{4M^2} }
   \la{S41_ps_T} \\ & & \phantom{ \int_0^\infty \! \dd k \, } 
   +
   \theta(\omega - 2M) \, 
   \theta\Bigl(\frac{\omega^2-4M^2}{2\omega}-k\Bigr)
   \,\mathrm{acosh}
   \sqrt{ \frac{\omega(\omega-2 k)}{4M^2} }
 \; \biggr] \biggr\} + \rmO(e^{-\beta M})
 \;. \nonumber
\ea

\pagebreak

%
\subsection{$S_4^2$}

The sum-integral $S_4^2$ is defined as 
\be
 {S_4^2}(\omega) \equiv \disc \biggl[  
 \Tint{K\{P\}}
 \frac{1}{K^2} 
 \frac{K^2}{\Delta(P)\Delta(P-Q)\Delta(P-K)}
 \biggr]_{Q=(-i\omega,\vec{0})}
 \;. \la{S42}
\ee
Because of the ultraviolet divergent factor in the numerator, the use 
of the spectral representation requires care in this case, and
we rather proceed directly with the sum, having gone over into 
free gluons to start with. Carrying out the shift $K\to P-K$, 
the summation factorizes, 
\be
  {S_4^2}(\omega) = 
  \disc \biggl[  
 \Tint{\{P\}}
 \frac{1}{\Delta(P)\Delta(P-Q)}
 \biggr]_{Q=(-i\omega,\vec{0})}
 \times
 \Tint{\{K\}}\frac{1}{\Delta(K)}
 = S_1(\omega) \, I_0(M^2)
 \;,  
\ee
where $S_1(\omega)$ is given in \eq\nr{S1_final}, while 
$I_0$ is a basic tadpole integral generalized to finite 
temperature. In fact, 
the finite temperature effects in $I_0$ are exponentially small and
can be omitted:
\ba
 I_0(M^2) & = & 
 \int_{\vec{k}} \frac{1}{2k}
 \Bigl[ 1 - 2 \nF{}(E_k) \Bigr]
 = 
 -\frac{M^2}{(4\pi)^2}
 \biggl[ \frac{1}{\epsilon} + \ln\frac{\bmu^2}{M^2} + 1  
 \biggr]  + \rmO(\epsilon, e^{-\beta M})
 \;. \la{I0}
\ea
Keeping exponentially small terms in the coefficient of 
the divergence, though, we arrive at 
\be
 S_4^2(\omega) = 
 - \theta(\omega - 2 M)
 \frac{(\omega^2 - 4 M^2)^{\fr12} M^2}{4\omega (4\pi)^3}
 \tanh\Bigl(\frac{\beta\omega}{4}\Bigr)
 \biggl[
  \frac{1}{\epsilon} + 
  \ln\frac{\bmu^4}{M^2(\omega^2 - 4 M^2)}
  + 3 + \rmO(\epsilon, e^{-\beta M})
 \biggr]
 \;. \la{S42_final}
\ee

%
\subsection{$S_5^0$}

The sum-integral $S_5^0$ is defined as
\be
 {S_5^0}(\omega) \equiv \disc \biggl[  
 \int_{-\infty}^\infty \!\frac{\dd k^0}{\pi} 
 \Tint{K\{P\}}
 \frac{{\rho(k^0,\mathbf{k})} }{k^0-ik_n}
 \frac{1}{\Delta^2(P)\Delta(P-Q)\Delta(P-K)}
 \biggr]_{Q=(-i\omega,\vec{0})}
 \;. \la{S50}
\ee
Performing the Matsubara sums, taking the discontinuity, and
making use of the antisymmetry of $\rho(k^0,\vec{k})$ yields
\ba
 S_5^0(\omega) & = & \int_0^\infty \!\frac{{\rm d}k^0}{\pi}
 \int_{\vec{k,p}} \rho(k^0,\vec{k}) 
 \frac{\pi}{8 E_p^2 E_{p-k}} \biggl\{ 
 \la{S50_Matsu}
 \\ & & 
 \frac{1}{2 E_p} 
 \Bigl[\delta(\omega - 2 E_p) - \delta(\omega+2 E_p) \Bigr]
 (1-2 \nF{1}) \times 
 \nn & & \times \Bigl[(\Delta^{-1}_{++} + \Delta^{-1}_{-+})
 (\Delta^{-1}_{++} - \Delta^{-1}_{-+} + E_p^{-1})
 (1+\nB{0} - \nF{2})
 \nn & & \;\; + 
 (\Delta^{-1}_{--}+\Delta^{-1}_{+-})
 (\Delta^{-1}_{--}-\Delta^{-1}_{+-} - E_p^{-1})
 (\nB{0} + \nF{2}) \Bigr]
 \nn & - & 
 \frac{\beta}{2 E_p} 
 \Bigl[\delta(\omega - 2 E_p) - \delta(\omega+2 E_p) \Bigr]
 (1-\nF{1})\nF{1} \times 
 \nn & & \times \Bigl[(\Delta^{-1}_{++} + \Delta^{-1}_{-+})
 (1+\nB{0} - \nF{2})
  -
 (\Delta^{-1}_{--}+\Delta^{-1}_{+-})
 (\nB{0} + \nF{2}) \Bigr]
 \nn & + & 
 \frac{1}{2 E_p} 
 \Bigl[\delta'(\omega - 2 E_p) + \delta'(\omega+2 E_p) \Bigr]
 (1-2 \nF{1}) \times 
 \nn & & \times \Bigl[(\Delta^{-1}_{++} + \Delta^{-1}_{-+})
 (1+\nB{0} - \nF{2})
  -
 (\Delta^{-1}_{--}+\Delta^{-1}_{+-})
 (\nB{0} + \nF{2}) \Bigr]
 \nn & + & \;\;\,
 \Bigl[\delta(\omega - \Delta_{++}) - \delta(\omega+\Delta_{++}) \Bigr]
 \Delta^{-1}_{++} \Delta^{-1}_{-+} 
  ( \Delta^{-1}_{-+} - \Delta^{-1}_{++} ) 
 \times \nn & & \hspace*{2cm} \times
 \Bigl[(1+\nB{0})(1-\nF{1}-\nF{2}) + \nF{1}\nF{2} \Bigr]  
 \nn & & + 
 \Bigl[\delta(\omega - \Delta_{--}) - \delta(\omega+\Delta_{--}) \Bigr]
 \Delta^{-1}_{--}\Delta^{-1}_{+-}
  ( \Delta^{-1}_{--} - \Delta^{-1}_{+-} ) 
 \times \nn & & \hspace*{2cm} \times
 \Bigl[-\nB{0}(1-\nF{1}-\nF{2}) + \nF{1}\nF{2} \Bigr]  
 \nn & & +
 \Bigl[\delta(\omega - \Delta_{+-}) - \delta(\omega+\Delta_{+-}) \Bigr]
 \Delta^{-1}_{--}\Delta^{-1}_{+-} 
  ( \Delta^{-1}_{--} - \Delta^{-1}_{+-} ) 
 \times \nn & & \hspace*{2cm} \times
 \Bigl[\nB{0}\nF{1}-(1+\nB{0})\nF{2} + \nF{1}\nF{2} \Bigr]  
 \nn & & +
 \Bigl[\delta(\omega - \Delta_{-+}) - \delta(\omega+\Delta_{-+}) \Bigr]
 \Delta^{-1}_{++}\Delta^{-1}_{-+}
  ( \Delta^{-1}_{-+} - \Delta^{-1}_{++} ) 
 \times \nn & & \hspace*{2cm} \times
 \Bigl[\nB{0}\nF{2}-(1+\nB{0})\nF{1} + \nF{1}\nF{2} \Bigr]  
 \biggr\} \;. 
 \nonumber
\ea
In the factorized part, it is useful to carry out 
a partial integration in order to remove the structure 
$
\delta'(\omega - 2 E_p) + \delta'(\omega+2 E_p)
$: 
\ba
 & & \hspace*{-1cm} 
 \int\! \frac{{\rm d}^d\vec{p}}{(2\pi)^d}
 \delta'(\omega - 2 E_p)
 g(p,E_p,E_{p-k})
 \la{pi} \\ & = & 
 \int\! \frac{{\rm d}^d\vec{p}}{(2\pi)^d}
 \delta(\omega - 2 E_p)
 \biggl\{ 
   \frac{(d-2)E_p g}{2p^2}  + 
   \frac{E_p}{2 p} \frac{\partial g}{\partial p}  + 
   \frac{1}{2 E_p} \frac{\partial ( E_p g )}{\partial E_p} + 
   \frac{(p+kz)E_p}{2 p E_{p-k}}
  \frac{\partial g}{\partial E_{p-k}}
 \biggr\} 
 \;. \nonumber
\ea
The subsequent steps proceed as described in appendix~A. 

In contrast to $S_4^0$, $S_4^1$, however, it is not possible to give 
separate closed expressions for the factorized and phase space vacuum
parts of $S_5^0$, because the integrals are logarithmically divergent 
at the lower limit of the $k$-integration (in the thermal case, they 
are linearly divergent). Yet the sum is finite, and 
inserting the free gluon spectral function, we get 
\ba
 \left. S_5^0(\omega) \right|^\rmii{vac} 
 \!\! & = & \!\!
 \frac{\theta(\omega - 2 M)}
 {4\omega (4\pi)^3 (\omega^2 - 4 M^2)^{\fr12}}
 \biggl\{ \tanh\Bigl(\frac{\beta\omega}{4}\Bigr)
 \biggl[
  \frac{1}{\epsilon} + 
  \ln\frac{\bmu^4}{M^2(\omega^2 - 4 M^2)}
  + 2  
 \biggr]
  \la{S50_vac} \\ & & \hspace*{-1cm} + \;  
  \frac{\omega^2 - 4 M^2}{M^2} 
  \ln 
  \frac{\omega (\omega^2 - 4 M^2)}{ M^3 } 
  + \frac{(\omega^2 - 4M^2)^{\fr12}(4M^2-3\omega^2)}
  {\omega M^2}\,\mathrm{acosh}
  \biggl(\frac{\omega}{2 M}\biggr)
 \biggr\}
 + \rmO(\epsilon)  
 \;, \nonumber
\ea
where in  the coefficient of the divergence 
we have accounted even for exponentially small thermal corrections.
For the thermal part proper we obtain
\ba
  \left. S_5^0(\omega) \right|^\rmii{$T$} 
  \!\! & = & \!\! \frac{1}{4 \omega^2 M^2 (4\pi)^3}
 \int_0^\infty \! \dd k \, \frac{\nB{}(k)}{k} 
 \biggl[  
 \la{S50_T} \\ 
 & & \hspace*{1.5cm}
   \theta(\omega) \, 
   \theta\Bigl(k- 
    \frac{4M^2-\omega^2}{2\omega}
    \Bigr)
   \sqrt{\omega(\omega+2 k)}
   \sqrt{\omega(\omega+2 k)- 4M^2} 
   \nn & & \hspace*{1.2cm}
   +
   \theta(\omega - 2M) \, 
   \theta\Bigl(\frac{\omega^2-4M^2}{2\omega}-k\Bigr)
   \sqrt{\omega(\omega-2 k)}
   \sqrt{\omega(\omega-2 k) - 4M^2}
   \nn & & \hspace*{1.2cm} - \theta(\omega - 2M) \times 
   2 \omega \sqrt{\omega^2 - 4 M^2}
 \; \biggr]  + \rmO(e^{-\beta M})
 \;. \nonumber
\ea
The last line, which originates from the factorized 
integrals, subtracts the values of the first two lines at $k=0$ 
(for $\omega > 2 M$), rendering the integral infrared finite. 

%
\subsection{$\hat S_5^0$}

The sum-integral ${\hat S_5^0}$ is defined as
\be
 {\hat S_5^0}(\omega) \equiv \disc \biggl[  
 \int_{-\infty}^\infty \!\frac{\dd k^0}{\pi} 
 \Tint{K\{P\}}
 \frac{{\rho(k^0,\mathbf{k})} }{k^0-ik_n}
 \frac{\vec{p}^2 - (\vec{p}\cdot\hat{\vec{k}})^2}
 {\Delta^2(P)\Delta(P-Q)\Delta(P-K)}
 \biggr]_{Q=(-i\omega,\vec{0})}
 \;. \la{hS50}
\ee
Carrying out the Matsubara sums proceeds precisely like 
for $S_5^0$, and leads to an expression like \eq\nr{S50_Matsu}; 
it is also again useful to carry out the partial integration 
in \eq\nr{pi}. The subsequent steps lead to the vacuum part
\ba
 \left. {\hat S_5^0}(\omega) \right|^\rmii{vac} 
 \!\! & = & \!\!
 \frac{\theta(\omega - 2 M)(\omega^2 - 4 M^2)^{\fr12}}
 {8\omega (4\pi)^3 }
 \biggl\{ \tanh\Bigl(\frac{\beta\omega}{4}\Bigr)
 \biggl[
  \frac{1}{\epsilon} + 
  \ln\frac{\bmu^4}{M^2(\omega^2 - 4 M^2)}
  + 1  
 \biggr]
  \nonumber \\ & & \hspace*{-1cm} - \;  
  4 \ln 
  \frac{\omega (\omega^2 - 4 M^2)}{ M^3 } 
  +\frac{2(7\omega^2 -8 M^2) }
  {\omega (\omega^2 - 4M^2)^{\fr12} }\,\mathrm{acosh}
  \biggl(\frac{\omega}{2 M}\biggr)
 \nn & & \hspace*{-1cm} 
 + \; \frac{2 \omega} {(\omega^2 - 4M^2)^{\fr12} } 
 \alpha\biggl( \frac{\sqrt{\omega^2 - 4M^2}}{\omega} \biggr) 
 \biggr\}
 + \rmO(\epsilon)  
 \;. \la{hS50_vac}
\ea
Here the function 
\ba
 \alpha(v) & \equiv & 
 \int_0^\infty
 \! \frac{{\rm d}x}{x}
 \biggl[
 \theta(v^2-x)(1-x) 
 \ln\frac{\sqrt{1-x} + \sqrt{v^2-x}}
 {\sqrt{1-x} - \sqrt{v^2-x}} 
 \nn & & \hspace*{1cm}
 + \; \ln \frac{(1+x+\sqrt{1+2vx+x^2})(-1+x+\sqrt{1-2vx+x^2})}
 {(1+x+\sqrt{1-2vx+x^2})(-1+x+\sqrt{1+2vx+x^2})}
 \biggr] \;,
\ea
where the integration variable $x$ is related to $k$ through $k = x \omega/2$,
is finite, but we have not bothered to work out its analytic expression, 
given that it does not appear in our final result. The thermal part reads
\ba
  \left. {\hat S_5^0}(\omega) \right|^\rmii{$T$} 
  \!\! & = & \!\! \frac{1}{2 \omega^2 (4\pi)^3}
 \int_0^\infty \! \dd k \, \frac{\nB{}(k)}{k} 
 \biggl\{ 
   \la{hS50_T} \\ 
 & & \hspace*{0.5cm}
   \theta(\omega) \, 
   \theta\biggl(k- 
    \frac{4M^2-\omega^2}{2\omega}
    \biggr) \times \nn 
 & & \hspace*{0.5cm} \times
   \biggl[ 
   - \sqrt{\omega(\omega+2 k)}
   \sqrt{\omega(\omega+2 k)- 4M^2} 
   + \omega(\omega+2 k) 
   \;\mathrm{acosh}\sqrt{\frac{\omega(\omega+2 k)}{4M^2}} 
   \biggr]
   \nn & & \hspace*{0.2cm}
   + \;
   \theta(\omega - 2M) \, 
   \theta\biggl(\frac{\omega^2-4M^2}{2\omega}-k\biggr)
    \times \nn 
 & & \hspace*{0.5cm} \times
   \biggl[ 
   - \sqrt{\omega(\omega-2 k)}
   \sqrt{\omega(\omega-2 k)- 4M^2} 
   + \omega(\omega-2 k) 
   \;\mathrm{acosh}\sqrt{\frac{\omega(\omega-2 k)}{4M^2}} 
   \biggr] \hspace*{5mm}
   \nn & & \hspace*{0.2cm} 
  + \;
  \theta(\omega - 2M) \times 
  \biggl[  2 \omega \sqrt{\omega^2 - 4 M^2} - 2 \omega^2 
  \, \mathrm{acosh} \biggl( \frac{\omega}{2 M} \biggr) 
  \; \biggr] \biggr\} + \rmO(e^{-\beta M})
 \;. \nonumber
\ea
The last line, which originates from the factorized 
integrals, subtracts the values of the first two lines at $k=0$ 
(for $\omega > 2 M$), rendering the integral infrared finite.

%
\subsection{$S_5^2$}

The sum-integral $S_5^2$ is defined as 
\be
 {S_5^2}(\omega) \equiv \disc \biggl[  
 \Tint{K\{P\}}
 \frac{1}{K^2}
 \frac{K^2}{\Delta^2(P)\Delta(P-Q)\Delta(P-K)}
 \biggr]_{Q=(-i\omega,\vec{0})}
 \;. \la{S52}
\ee
Because of the ultraviolet divergent factor in the numerator, the use 
of the spectral representation requires care in this case, and
we rather proceed directly with the sum, as in the case of $S_4^2$.
Carrying out the shift $K\to P-K$, 
the summation factorizes, 
\be
  {S_5^2}(\omega) = 
  \disc \biggl[  
 \Tint{\{P\}}
 \frac{1}{\Delta^2(P)\Delta(P-Q)}
 \biggr]_{Q=(-i\omega,\vec{0})}
 \times \Tint{\{K\}}\frac{1}{\Delta(K)}
 = S_2(\omega) \, I_0(M^2)
 \;,  
\ee
where $S_2(\omega)$ is given in \eq\nr{S2_final}, while 
$I_0$ is given in \eq\nr{I0}.
Keeping exponentially small terms in the coefficient of 
the divergence, we arrive at 
\be
 S_5^2(\omega) = 
 - \frac{\theta(\omega - 2 M) M^2}
  {4\omega (\omega^2 - 4 M^2)^{\fr12} (4\pi)^3}
 \tanh\Bigl(\frac{\beta\omega}{4}\Bigr)
 \biggl[
  \frac{1}{\epsilon} + 
  \ln\frac{\bmu^4}{M^2(\omega^2 - 4 M^2)}
  + 1 + \rmO(\epsilon, e^{-\beta M})
 \biggr]
 \;. \la{S52_final}
\ee

%
\subsection{$S_6^0$}

The sum-integral $S_6^0$ is defined as
\be
 {S_6^0}(\omega) \equiv \disc \biggl[  
 \int_{-\infty}^\infty \!\frac{\dd k^0}{\pi} 
 \Tint{K\{P\}}
 \frac{{\rho(k^0,\mathbf{k})} }{k^0-ik_n}
 \frac{1}{\Delta(P)\Delta(P-Q)\Delta(P-K)\Delta(P-Q-K)}
 \biggr]_{Q=(-i\omega,\vec{0})}
 \;. \la{S60}
\ee
Performing the Matsubara sums, taking the discontinuity, and
making use of the antisymmetry of $\rho(k^0,\vec{k})$ yields
\ba
 S_6^0(\omega) \!\! & = & \!\! 
 \int_0^\infty \!\frac{{\rm d}k^0}{\pi}
 \int_{\vec{k,p}} \rho(k^0,\vec{k}) 
 \frac{\pi}{2 E_p E_{p-k}} \biggl\{ 
  \la{S60_Matsu}
 \\
 & & 
 \frac{1}{8 E_p^2} 
 \Bigl[\delta(\omega - 2 E_p) - \delta(\omega+2 E_p) \Bigr]
 (1-2 \nF{1}) \times 
 \nn & & \times \biggl[
 \biggl( 
 \frac{\Delta^{-1}_{--}}{E_p+E_{p-k}} + \frac{\Delta^{-1}_{+-}}{E_p-E_{p-k}}
 \biggr)
 (\nB{0} + \nF{2}) 
 \nn & & \;\; 
 -
 \biggl(
 \frac{\Delta^{-1}_{++}}{E_p+E_{p-k}} + 
 \frac{\Delta^{-1}_{-+}}{E_p-E_{p-k}}  
 \biggr)
 (1+\nB{0} - \nF{2})
 \biggr]
 \nn & + & 
 \frac{1}{8 E_{p-k}^2} 
 \Bigl[\delta(\omega - 2 E_{p-k}) - \delta(\omega+2 E_{p-k}) \Bigr]
 (1-2 \nF{2}) \times 
 \nn & & \times \biggl[
 \biggl( 
 \frac{\Delta^{-1}_{--}}{E_{p-k}+E_{p}} + \frac{\Delta^{-1}_{-+}}{E_{p-k}-E_{p}}
 \biggr)
 (\nB{0} + \nF{1}) 
 \nn & & \;\; 
 -
 \biggl(
 \frac{\Delta^{-1}_{++}}{E_{p-k}+E_{p}} + 
 \frac{\Delta^{-1}_{+-}}{E_{p-k}-E_{p}}  
 \biggr)
 (1+\nB{0} - \nF{1})
 \biggr]
 \nn & + & \;\;\,
 \Bigl[\delta(\omega - \Delta_{++}) - \delta(\omega+\Delta_{++}) \Bigr]
 \Delta^{-2}_{++} \Delta^{-1}_{+-} \Delta^{-1}_{-+}  
 \Bigl[(1+\nB{0})(1-\nF{1}-\nF{2}) + \nF{1}\nF{2} \Bigr]  
 \nn & & + 
 \Bigl[\delta(\omega - \Delta_{--}) - \delta(\omega+\Delta_{--}) \Bigr]
 \Delta^{-2}_{--}\Delta^{-1}_{+-} \Delta^{-1}_{-+}
 \Bigl[-\nB{0}(1-\nF{1}-\nF{2}) + \nF{1}\nF{2} \Bigr]  
 \nn & & +
 \Bigl[\delta(\omega - \Delta_{+-}) - \delta(\omega+\Delta_{+-}) \Bigr]
 \Delta^{-2}_{+-}\Delta^{-1}_{++} \Delta^{-1}_{--}
 \Bigl[\nB{0}\nF{1}-(1+\nB{0})\nF{2} + \nF{1}\nF{2} \Bigr]  
 \nn & & +
 \Bigl[\delta(\omega - \Delta_{-+}) - \delta(\omega+\Delta_{-+}) \Bigr]
 \Delta^{-2}_{-+}\Delta^{-1}_{++} \Delta^{-1}_{--}
 \Bigl[\nB{0}\nF{2}-(1+\nB{0})\nF{1} + \nF{1}\nF{2} \Bigr]  
 \biggr\} \;. \nonumber
\ea
In the factorized part, a change of integration variables 
$\vec{p}\to \vec{k-p}$ allows trivially to change the structure
with $\delta(\omega - 2 E_{p-k}) - \delta(\omega+2 E_{p-k})$
into the familiar one with 
$\delta(\omega - 2 E_p) - \delta(\omega+2 E_p)$. (The only 
complication is that then the difference $1/(E_p-E_{p-k})$ needs
to be interpreted as a principal value.)
The subsequent steps proceed as described in appendix~A. 

Like with $S_5^0$,  it is again not possible to give 
separate closed expressions for the factorized and phase space vacuum
parts of $S_6^0$, because the integrals are logarithmically divergent 
at the lower limit of the $k$-integration (in the thermal case, they 
are linearly divergent). The sum is infrared finite, however, and 
inserting the free gluon spectral function, yields~\cite{db}  
\ba
 \left. S_6^0(\omega) \right|^\rmii{vac} 
 \!\! & = & \!\!
 \frac{\theta(\omega - 2 M)}
 {\omega^2 (4\pi)^3}
 L_2 \Bigl( \frac{\omega - \sqrt{\omega^2 - 4 M^2}}
 {\omega + \sqrt{\omega^2 - 4 M^2}} \Bigr)
 + \rmO(\epsilon)  
 \;, \la{S60_vac}
\ea
where the function $L_2$ is defined in \eq\nr{L2}.
For the thermal parts we obtain, 
omitting exponentially small terms,
\ba
  \left. S_6^0(\omega) \right|^\rmii{$T$} 
  \!\! & = & \!\! \frac{2}{\omega^2 (4\pi)^3}
 \int_0^\infty \! \dd k \, \frac{\nB{}(k)}{k} 
 \biggl[ \nn 
 & & \hspace*{1.5cm}
   \theta(\omega) \, 
   \theta\Bigl(k- 
    \frac{4M^2-\omega^2}{2\omega}
    \Bigr)
   \,\mathrm{acosh}\sqrt{\frac{\omega(\omega+2 k)}{4M^2}} 
   \nn & & \hspace*{1.2cm}
   +
   \theta(\omega - 2M) \, 
   \theta\Bigl(\frac{\omega^2-4M^2}{2\omega}-k\Bigr)
   \,\mathrm{acosh}\sqrt{\frac{\omega(\omega-2 k)}{4M^2}}
   \nn & & \hspace*{1.2cm} - \theta(\omega - 2M) \times 
   2 \, \mathrm{acosh} \biggl( \frac{\omega}{2 M} \biggr)
 \; \biggr]  + \rmO(e^{-\beta M})
 \;. \la{S60_T}
\ea
The last line, which originates from the factorized 
integrals, subtracts the values of the first two lines at $k=0$ 
(for $\omega > 2 M$), rendering the integral infrared finite.

%
\subsection{$\hat S_6^0$}

The sum-integral ${\hat S_6^0}$ is defined as
\be
 {\hat S_6^0}(\omega) \equiv \disc \biggl[  
 \int_{-\infty}^\infty \!\frac{\dd k^0}{\pi} 
 \Tint{K\{P\}}
 \frac{{\rho(k^0,\mathbf{k})} }{k^0-ik_n}
 \frac{\vec{p}^2 - (\vec{p}\cdot\hat{\vec{k}})^2}
 {\Delta(P)\Delta(P-Q)\Delta(P-K)\Delta(P-Q-K)}
 \biggr]_{Q=(-i\omega,\vec{0})}
 \;. \la{hS60}
\ee
Carrying out the Matsubara sums proceeds precisely like 
for $S_6^0$, and leads to an expression like \eq\nr{S60_Matsu}. 
The subsequent steps lead to the vacuum part
\ba
 \left. {\hat S_6^0}(\omega) \right|^\rmii{vac} 
 \!\! & = & \!\!
 \frac{\theta(\omega - 2 M)}
 {\omega^2 (4\pi)^3 }
 \biggl[ - M^2  L_2 \Bigl( \frac{\omega - \sqrt{\omega^2 - 4 M^2}}
 {\omega + \sqrt{\omega^2 - 4 M^2}} \Bigr)
 + \frac{\omega^2}{2}
 \beta\biggl( \frac{\sqrt{\omega^2 - 4M^2}}{\omega} \biggr) 
 \biggr]
 + \rmO(\epsilon)  
 \;. \nn \la{hS60_vac}
\ea
Here the function 
\ba
 \beta(v) & \equiv & 
 \int_0^\infty
 \! \frac{{\rm d}x}{x}
 \biggl[
 \theta(v^2-x) \sqrt{1-x} \sqrt{v^2-x} - v 
 + \fr34 \Bigl( \sqrt{1+2vx+x^2} - \sqrt{1-2vx+x^2} \Bigr) 
 \nn & & \hspace*{1cm}
 + \; \frac{x^2-4}{8} 
 \ln \biggl| \frac{(1+\sqrt{1+2vx+x^2})(-1+\sqrt{1-2vx+x^2})}
 {(1+\sqrt{1-2vx+x^2})(-1+\sqrt{1+2vx+x^2})} \biggr|
 \; \biggr]
\ea
where the integration variable $x$ is related to $k$ through $k = x \omega/2$,
is finite, but we have not bothered to work out its analytic expression, 
given that it does not appear in our final result. The thermal part reads
\ba
  \left. {\hat S_6^0}(\omega) \right|^\rmii{$T$} 
  \!\! & = & \!\! \frac{1}{2 \omega^2 (4\pi)^3}
 \int_0^\infty \! \dd k \, \frac{\nB{}(k)}{k} 
 \biggl\{ 
 \la{hS60_T} \\
 & & \hspace*{0.5cm}
   \theta(\omega) \, 
   \theta\biggl(k- 
    \frac{4M^2-\omega^2}{2\omega}
    \biggr) \times \nn 
 & & \hspace*{0.5cm} \times
   \biggl[ 
    \sqrt{\omega(\omega+2 k)}
   \sqrt{\omega(\omega+2 k)- 4M^2} 
   - 4 M^2
   \;\mathrm{acosh}\sqrt{\frac{\omega(\omega+2 k)}{4M^2}} 
   \biggr]
   \nn & & \hspace*{0.2cm}
   + \;
   \theta(\omega - 2M) \, 
   \theta\biggl(\frac{\omega^2-4M^2}{2\omega}-k\biggr)
    \times \nn 
 & & \hspace*{0.5cm} \times
   \biggl[ 
    \sqrt{\omega(\omega-2 k)}
   \sqrt{\omega(\omega-2 k)- 4M^2} 
   -4 M^2
   \;\mathrm{acosh}\sqrt{\frac{\omega(\omega-2 k)}{4M^2}} 
   \biggr] \hspace*{20mm}
   \nn & & \hspace*{0.2cm} 
  + \;
  \theta(\omega - 2M) \times 
  \biggl[ -  2 \omega \sqrt{\omega^2 - 4 M^2} + 8 M^2
  \, \mathrm{acosh} \biggl( \frac{\omega}{2 M} \biggr) 
  \; \biggr] \biggr\} + \rmO(e^{-\beta M})
 \;. \nonumber
\ea
The last line, which originates from the factorized 
integrals, subtracts the values of the first two lines at $k=0$ 
(for $\omega > 2 M$), rendering the integral infrared finite. 

%
\subsection{$S_6^2$}

The sum-integral $S_6^2$ is defined as 
\be
 {S_6^2}(\omega) \equiv \disc \biggl[  
 \Tint{K\{P\}}
 \frac{1}{K^2}
 \frac{K^2}{\Delta(P)\Delta(P-Q)\Delta(P-K)\Delta(P-Q-K)}
 \biggr]_{Q=(-i\omega,\vec{0})}
 \;. \la{S62}
\ee
Like with $S_4^2$ and $S_5^2$, we proceed directly with free gluons
rather than using the spectral representation. 
Carrying out the shift $K\to P-K$, the summation factorizes, 
\ba
  {S_6^2}(\omega) \!\!\! & = & \!\!\!  
  \disc \biggl\{ \biggl[  
 \Tint{\{P\}}
 \frac{1}{\Delta(P)\Delta(P-Q)}
 \biggr]_{Q=(-i\omega,\vec{0})}
 \times \biggl[
 \Tint{\{K\}}
 \frac{1}{\Delta(K)\Delta(K-Q)}
 \biggr]_{Q=(-i\omega,\vec{0})}
 \biggr\}
 \nn & = & \!\!\! 
 2 
  \re \biggl[  
 \Tint{\{P\}}
 \frac{1}{\Delta(P)\Delta(P-Q)}
 \biggr]_{Q=(-i\omega,\vec{0})}
 \times \disc \biggl[
 \Tint{\{K\}}
 \frac{1}{\Delta(K)\Delta(K-Q)}
 \biggr]_{Q=(-i\omega,\vec{0})}
 \;,  \nn 
\ea
where $\re[...]$ denotes the regular (non-discontinuous) part, while
the discontinuous part can be identified with the function $S_1(\omega)$, 
given in \eq\nr{S1_final}. The Matsubara sum in the regular part can 
be carried out as before; the only difference with respect to the 
procedure in appendix~A is that taking the regular part
after the substitution $\omega_n\to -i \omega$ yields a principle
value rather than a delta-function: 
\ba
 & & \hspace*{-1cm} \re \biggl[  
 \Tint{\{P\}}
 \frac{1}{\Delta(P)\Delta(P-Q)}
 \biggr]_{Q=(-i\omega,\vec{0})} 
 \nn & = & 
 \int_{\vec{p}}
 \frac{1}{4 E_p^2}
 \biggl[
  P\biggl( \frac{1}{\omega + 2 E_p} \biggr)- 
  P\biggl( \frac{1}{\omega - 2 E_p} \biggr) 
 \biggr]
 \Bigl[ 1 - 2 \nF{}(E_p)
 \Bigr]
 \;. 
\ea
It is seen that the finite-temperature effects continue to be exponentially 
suppressed. The zero-temperature part, on the other hand, equals 
the real part of the general function $B_0$, another special
case of which was met in \eq\nr{B0}: 
\ba
 & & \hspace*{-1cm} 
 \re\biggl[  \mu^{2\epsilon}\!\!\int\!\frac{{\rm d}^D\! P}{(2\pi)^D}
 \frac{1}{(P^2 + M^2)[(P-Q)^2 + M^2]}
 \biggr]_{Q=(-i\omega,\vec{0})}  
 =  \re\Bigl[ B_0(-\omega^2;M^2,M^2) \Bigr] \nn 
 & = &   
 \frac{1}{(4\pi)^2}
 \biggl[
   \frac{1}{\epsilon} + \ln\frac{\bmu^2}{M^2} + 2 
  - \frac{2(\omega^2 - 4 M^2)^{\fr12}}{\omega}
  \,\mathrm{acosh}\biggl( \frac{\omega}{2 M}\biggr) 
 + \rmO(\epsilon) \biggr] 
 \;, \quad \omega > 2 M
 \;. \hspace*{1cm}
\ea
Combining this with $S_1(\omega)$, and keeping 
the exponentially small terms in the coefficient of 
the divergence, we arrive at 
\ba
 S_6^2(\omega) & = &  
  \theta(\omega - 2 M) 
  \frac{(\omega^2 - 4 M^2)^{\fr12}}
  {2\omega  (4\pi)^3}
 \tanh\Bigl(\frac{\beta\omega}{4}\Bigr) \times
 \la{S62_final} \\ & \times & 
 \biggl[
  \frac{1}{\epsilon} + 
  \ln\frac{\bmu^4}{M^2(\omega^2 - 4 M^2)}
  + 4  - \frac{2(\omega^2 - 4 M^2)^{\fr12}}{\omega}
  \,\mathrm{acosh}\biggl( \frac{\omega}{2 M}\biggr) 
  + \rmO(\epsilon, e^{-\beta M})
 \biggr]
 \;. \nonumber 
\ea

%
\subsection{$\hat S_6^2$}

The sum-integral ${\hat S_6^2}$ is defined as 
\be
 {\hat S_6^2}(\omega) \equiv \disc \biggl[  
 \Tint{K\{P\}}
 \frac{1}{K^2}
 \frac{K^2\, [\vec{p}^2 - (\vec{p}\cdot\hat{\vec{k}})^2]}
 {\Delta(P)\Delta(P-Q)\Delta(P-K)\Delta(P-Q-K)}
 \biggr]_{Q=(-i\omega,\vec{0})}
 \;. \la{hS62}
\ee
The summation factorizes into two independent parts like 
for $S_6^2$; however, the spatial integrations do not factorize
due to the additional structure in the numerator. 
Therefore the evaluation is somewhat more involved, yet the general 
techniques introduced in appendix A yield a solution: 
\ba
 {\hat S_6^2}(\omega) & = &  
  \theta(\omega - 2 M) 
  \frac{(\omega^2 - 4 M^2)^{\fr32}}
  {12\omega  (4\pi)^3}
 \tanh\Bigl(\frac{\beta\omega}{4}\Bigr) \times
  \la{hS62_final}  \\ & \times & 
 \biggl[
  \frac{1}{\epsilon} + 
  \ln\frac{\bmu^4}{M^2(\omega^2 - 4 M^2)}
  + 4 + \frac{\omega^2+2 M^2}{3(\omega^2-4M^2)}
 \nn & &   \hspace*{1.5cm}
 -\; \frac{2 (\omega^4-6 \omega^2 M^2 + 12 M^4)}
 {\omega (\omega^2 - 4 M^2)^{\fr32}}
  \,\mathrm{acosh}\biggl( \frac{\omega}{2 M}\biggr) 
  + \rmO(\epsilon, e^{-\beta M})
 \biggr]
 \;. \nonumber
\ea

%
\subsection{$S_6^4$}

The sum-integral $S_6^4$ is defined as 
\be
 {S_6^4}(\omega) \equiv \disc \biggl[  
 \Tint{K\{P\}}
 \frac{1}{K^2}
 \frac{(K^2)^2}{\Delta(P)\Delta(P-Q)\Delta(P-K)\Delta(P-Q-K)}
 \biggr]_{Q=(-i\omega,\vec{0})}
 \;. \la{S64}
\ee
Like with $S_4^2$, $S_5^2$ and $S_6^2$, we proceed directly with 
free gluons rather than using the spectral representation. Cancelling
one $K^2$ and carrying out the shift $K\to P-K$, we get 
\be
 {S_6^4}(\omega) = \disc \biggl[  
 \Tint{\{K,P\}}
 \frac{\Delta(P)+\Delta(K)-2 (M^2 + P\cdot K)}
 {\Delta(P)\Delta(P-Q)\Delta(K)\Delta(K-Q)}
 \biggr]_{Q=(-i\omega,\vec{0})}
 \;. \la{S64_2}
\ee
Another change of integration variables, $P\to Q-P$, shows that
\be
  \Tint{\{P\}}
 \frac{P}
 {\Delta(P)\Delta(P-Q)} = 
 \frac{Q}{2} 
 \Tint{\{P\}}
 \frac{1}
 {\Delta(P)\Delta(P-Q)}
 \;, 
\ee
and similarly for the term with $\Tinti{\{K\}}$. Thereby we arrive at
\ba
 S_6^4(\omega) & = &
 2 S_4^2(\omega) + \fr12 (\omega^2 - 4 M^2) S_6^2(\omega)
 \\ & = &
  \theta(\omega - 2 M) 
  \frac{(\omega^2 - 4 M^2)^{\fr12}}
  {4\omega  (4\pi)^3}
 \tanh\Bigl(\frac{\beta\omega}{4}\Bigr) \times
 \nn & \times & 
 \biggl\{
  (\omega^2 - 6 M^2) 
  \biggl[
  \frac{1}{\epsilon} + 
  \ln\frac{\bmu^4}{M^2(\omega^2 - 4 M^2)}
  + 3 \biggr]  
  \nn & & 
  + (\omega^2 - 4 M^2) 
  \biggl[ 
  1  - \frac{2(\omega^2 - 4 M^2)^{\fr12}}{\omega}
  \,\mathrm{acosh}\biggl( \frac{\omega}{2 M}\biggr) 
  \biggr]
  + \rmO(\epsilon, e^{-\beta M})
 \biggr\}
 \;. \la{S64_final} 
\ea

\newpage

%
\section{Spectral function in the scalar channel}

For completeness, we have worked out the spectral function
corresponding to the scalar channel with the same methods as 
described above for the vector channel. It seems, though, that 
the physical significance is not clear in the scalar case:
no direct relation to an observable, in the sense of \eq\nr{dilepton}, 
has been worked out, as far as we know, and 
the computation as such
appears to possess a number of ambiguities. In particular, 
the scalar density requires renormalization, and the renormalization 
factor cannot be uniquely specified; moreover the resummation of the
spectral function within a potential model near the threshold 
appears to lead to ambiguities~\cite{peskin}. Nevertheless,
on the lattice the correlator of (bare) scalar densities can be 
treated on the same footing as that of the vector 
currents~\cite{latt,latt_rev}.

Concerning the first of the issues, namely renormalization, 
the method we choose is to consider the object 
\be
 \hat {\mathcal{S}}
 \equiv
 M_B^{(\delta)} \, \hat{\bar\psi}\, \hat\psi  
 \;, \la{Sd_def}
\ee
where $M_B^{(\delta)}$ is essentially 
the bare quark mass defined in \eq\nr{MB}, 
only with a possible additional constant as a ``probe'', 
\be
 \Bigl(M_B^{(\delta)}\Bigr)^2 \equiv M^2 - \frac{6 g^2 C_F M^2}{(4\pi)^2} 
 \biggl( \frac{1}{\epsilon} + \ln\frac{\bmu^2}{M^2} + \fr43
 + \delta \biggr) 
 + \rmO(g^4)
 \;.  \la{MBd}
\ee
We then define 
\be
 \rho_S (\omega) 
 \equiv 
 \int_{-\infty}^\infty 
 \!\! {\rm d}t \,  e^{i \omega t}
 \!
 \int \! {\rm d}^{3-2\epsilon} \vec{x}\,
 \left\langle
  \fr12 {[ 
  \hat {\cal{S}}(t,\vec{x}), 
  \hat {\cal{S}}(0,\vec{0})
   ]}
 \right\rangle
 \;, \la{rhoS} 
\ee
which turns out to be finite. 
Starting again at 1-loop level, 
and omitting $Q$-independent terms which 
are killed by the discontinuity in \eq\nr{disc_def}, we get
\ba
 \nn[-10mm]
 \ConnectedC(\TAsc,\TAsc) 
 & = & 
 [Q-\mbox{indep.}] - 
 2 C_A M^2 \Tint{\{P\}} \frac{Q^2 + 4 M^2}
 {\Delta(P)\Delta(P-Q)}
 \;. \la{loS} \\[-10mm] \nonumber
\ea
The counterterm graph yields
\ba
 \nn[-10mm] 
 \ConnectedD(\TAsc,\TAsc) 
 & = & 
 [Q-\mbox{indep.}] + 
 \frac{12 g^2 C_A C_F M^2}{(4\pi)^2} \biggl\{ 
 \biggl( \frac{1}{\epsilon} + \ln\frac{\bmu^2}{M^2} + \fr43 \biggr) 
 \la{ctS}
 \\[-5mm] & \times & 
 \Tint{\{P\}} \biggl[ 
  \frac{Q^2 + 8 M^2} {\Delta(P)\Delta(P-Q)}
 - \frac{2 M^2 (Q^2 + 4 M^2)}
 {\Delta^2(P)\Delta(P-Q)}
 \biggr] + \delta\,  
  \Tint{\{P\}} 
  \frac{Q^2 + 4 M^2} {\Delta(P)\Delta(P-Q)}
 \biggr\}
 \;. \nonumber
\ea
Finally, the genuine 2-loop graphs add up to 
\ba
 \nn[-10mm]
 && \hspace*{-2.5cm}
 \ConnectedA(\TAsc,\TAsc,\TAgl) \; + 
 \ConnectedB(\TAsc,\TAsc,\TLgl) 
 =   [Q-\mbox{indep.}] + 4 g^2 C_A C_F M^2 \Tint{K\{P\}} 
 \biggl\{ \nn[-5mm] 
 &&  \hspace*{-2cm}
 \biggl( \frac{1}{K^2+\Pi_T}-\frac{1}{K^2+\Pi_E} \biggr)
	[\mathbf{p}^2-(\mathbf{p}\cdot \hat{\mathbf{k}})^2 ] \times  
 \nn
 && \hspace*{-0.8cm}
 \times \biggl[ 
     -\frac{4 (Q^2+4M^2)}{\Delta^2(P)\Delta(P-Q)\Delta(P-K)}
	-\frac{2 ( Q^2+4M^2) }
      {\Delta(P)\Delta(P-Q)\Delta(P-K)\Delta(P-Q-K)} \biggr] \nn 
 &\displaystyle +\;\frac{D-2}{K^2+\Pi_T}& \hspace*{-3mm} 
 \biggl[ \frac{Q^2 + 4 M^2}{\Delta^2(P)\Delta(P-Q)}
	+\frac{2 Q\cdot K }{\Delta(P)\Delta(P-Q)\Delta(P-K)}
	 \nn 
 &&  \hspace*{-1cm}  {}
 - \frac{(Q^2 + 4 M^2 ) K^2}{\Delta^2(P)\Delta(P-Q)\Delta(P-K)}
 - \frac{\fr12 (Q^2 + 4 M^2 ) K^2}
  {\Delta(P)\Delta(P-Q)\Delta(P-K)\Delta(P-Q-K)}
	 \biggr] \nn
 &\displaystyle +\;\frac{1}{K^2+\Pi_E}& \hspace*{-3mm}
 \biggl[ 
 -\frac{4(Q^2 +4M^2 )}{\Delta(P)\Delta(P-Q)\Delta(P-K)} 
 + \frac{4(Q^2 +4M^2 ) M^2 }{\Delta^2(P)\Delta(P-Q)\Delta(P-K)}
  \nn
 && \hspace*{-1cm} {}+\frac{(Q^2 + 2 M^2)
 (Q^2 + 4 M^2) + Q^2  K^2}
 {\Delta(P)\Delta(P-Q)\Delta(P-K)\Delta(P-Q-K)}
 \biggr] \; \biggr\}
 \;.  \la{nloS}
\ea
Again any dependence on the gauge parameter 
$\xi$ has disappeared, and $P^E_{\mu\nu}$ of \eq\nr{PE} could 
have been replaced with $\delta_{\mu\nu} - P^T_{\mu\nu}$. 

We note that the master sum-integrals appearing in \eq\nr{nloS}
are a subset of those in \eq\nr{nlo}. Therefore 
the discussion in \se\ref{se:IR} continues to hold, 
and there are no infrared divergences in the result, 
so that we can set $\Pi_T = \Pi_E = 0$ in \eq\nr{nloS}.
The full result can now be written as 
\ba
 & & \hspace*{-1.2cm}
 \left. \rho_S(\omega) \right|_\rmii{raw} \; = \; 
 2 C_A M^2 (\omega^2 - 4 M^2) S_1(\omega) + 4 g^2 C_A C_F M^2
 \biggl\{ \nn & & 
 \biggl[  
   - \frac{3}{(4\pi)^2}
   \biggl(
     \frac{1}{\epsilon} + \ln\frac{\bmu^2}{M^2} + \fr43 
   \biggr) 
 \biggr](\omega^2 - 8 M^2) S_1(\omega)
  - \frac{3\delta }{(4\pi)^2}
 (\omega^2 - 4 M^2) S_1(\omega)
 \nn[1mm] 
 & & -\;  
 \biggl[ \frac{T^2}{6} 
   - \frac{6 M^2}{(4\pi)^2}
   \biggl(
     \frac{1}{\epsilon} + \ln\frac{\bmu^2}{M^2} + \fr43 
   \biggr) 
 \biggr] (\omega^2 - 4 M^2) S_2(\omega)
 \nn[1mm] 
 & & +\; 4 (\omega^2 -4 M^2) S_4^0(\omega) 
 + 4 (1-\epsilon) S_4^1(\omega)
 - 2 (\omega^2 -4 M^2)
 \Bigl[ 2 M^2 S_5^0(\omega) - (1-\epsilon) S_5^2(\omega) \Bigr]
 \nn  & &
 +\; (\omega^2 - 2 M^2)(\omega^2 - 4 M^2) S_6^0(\omega) 
 - \Bigl[ \epsilon \omega^2 + 4(1- \epsilon) M^2 \Bigr] S_6^2(\omega)
 \biggr\} + \rmO(\epsilon) \;.
   \la{full_raw_S}
\ea
We have set here $\epsilon\to 0$ whenever the master sum-integral
that it multiplies is finite. 

Now, the structure of \eq\nr{full_raw_S} reveals
an ambiguity with regard to the treatment of the ``resummation'' 
of thermal mass corrections, which in the vector case lead 
to \eq\nr{M_thermal}. In the vector case, the need to resum 
is unambiguous, because 
anything else than \eq\nr{M_thermal} would lead to a thermal correction
diverging at the threshold. In the scalar case, we do not 
have this guidance: terms multiplied by $T^2$ vanish as 
$\theta(\omega-2 M)(\omega - 2M)^{\fr12}$ at the threshold, 
being thus subdominant with respect to the leading thermal 
corrections which remain non-zero. Nevertheless, we would like 
to apply a ``universal'' thermal resummation, 
i.e.\ precisely \eq\nr{M_thermal}; however, it may be questioned
whether it is valid to do this also in the term $M^2$, coming from 
the (``ultraviolet-completed'') definition of the scalar current, 
or only in more infrared sensitive parts.
It seems to us that this question can be fully
settled only through a next-to-next-to-leading order
computation; in the following, we assume that the resummation 
of \eq\nr{M_thermal} is only carried out in the Lagrangian, 
not in the definition of the scalar density. If so, a redefinition
of the mass according to  \eq\nr{M_thermal}
leads to the modified result
\ba
 \rho_S(\omega) \!\!\! & = & \!\!\!
 2 C_A M^2 (\omega^2 - 4 M^2) S_1(\omega) + 4 g^2 C_A C_F M^2
 \biggl\{ \nn & & \hspace*{-8mm} 
 \frac{T^2}{3} S_1(\omega)
   - \frac{3\delta }{(4\pi)^2}
 (\omega^2 - 4 M^2) S_1(\omega)
 \nn[1mm] 
 & &  \hspace*{-8mm}
   - \; \frac{3}{(4\pi)^2}
   \biggl(
     \frac{1}{\epsilon} + \ln\frac{\bmu^2}{M^2} + \fr43 
   \biggr) 
 \Bigl[ (\omega^2 - 8 M^2) S_1(\omega)
 - 2 M^2 (\omega^2 - 4 M^2) S_2(\omega) \Bigr] 
 \nn[1mm] 
 & &  \hspace*{-8mm} + \; 4 (\omega^2 -4 M^2) S_4^0(\omega) 
 + 4 (1-\epsilon) S_4^1(\omega)
 - 2 (\omega^2 -4 M^2)
 \Bigl[ 2 M^2 S_5^0(\omega) - (1-\epsilon) S_5^2(\omega) \Bigr]
 \nn  & &  \hspace*{-8mm}
 + \; (\omega^2 - 2 M^2)(\omega^2 - 4 M^2) S_6^0(\omega) 
 - \Bigl[ \epsilon \omega^2 + 4(1- \epsilon) M^2 \Bigr] S_6^2(\omega)
 \biggr\} + \rmO(\epsilon) \;.
   \la{full_raw_S2}
\ea
Unfortunately, the issue of what is resummed is not insignificant 
in the sense that the difference between \eqs\nr{full_raw_S} and 
\nr{full_raw_S2} is numerically of $\rmO(1)$ for $\omega > 2 M$.

Inserting the explicit expressions for the functions 
$S_i^j(\omega)$ from
appendix B into \eq\nr{full_raw_S2}, 
the final result for the vacuum part reads
\ba
 & & \left.
 \hspace*{-0.5cm}
  \rho_S(\omega) \right|^\rmii{vac} 
  = \theta(\omega - 2 M)   
  \frac{C_A M^2 (\omega^2 - 4 M^2)^{\fr32}}{8\pi\omega}
 + \theta(\omega - 2 M) 
 \frac{4 g^2 C_A C_F M^2}{(4\pi)^3 \omega^2} 
 \biggl\{  
 \la{full_vacS}
 \\ & & \!\!\!
 (\omega^2 - 2 M^2) (\omega^2 - 4 M^2)  
 L_2 \biggl( \frac{\omega - \sqrt{\omega^2 - 4 M^2}}
 {\omega + \sqrt{\omega^2 - 4 M^2}} \biggr)
 + \biggl(\fr32 \omega^4  - 2 \omega^2 M^2  - 13 M^4 \biggr)
  \,\mathrm{acosh} \biggl( \frac{\omega}{2 M} \biggr)
 \nn &  & \!\!\!
 -\; \omega (\omega^2 - 4 M^2)^{\fr12}
 \biggl[
   (\omega^2 - 4 M^2) \biggl( 
  \ln \frac{\omega (\omega^2 - 4 M^2)}{M^3}
  + \fr34 \delta \biggr)
  -\fr38 (3 \omega^2 - 14 M^2) 
 \biggr]
 \biggr\} + \rmO(\epsilon,g^4) \;, 
  \nonumber
\ea
where the function $L_2$ is defined in \eq\nr{L2}.
Let us note that although similar to the vector channel spectral
function in \eq\nr{full_vac} at first sight, \eq\nr{full_vacS} has
also some significant differences; in particular, logarithms of $\omega/M$
do {\em not} cancel at $\omega \gg M$ any more, but the asymptotic behaviour 
becomes
\be
 \left.
   \rho_S(\omega) \right|^\rmii{vac} 
 \;\; \stackrel{\omega \gg M}{\approx} \;\;
 - \frac{3 g^2 C_A C_F \omega^2 M^2 }{(4\pi)^3}
 \biggl(
  \ln\frac{\omega^2}{M^2} + \delta - \fr32 
 \biggr) 
 \;.
 \la{full_vacS_asympt}
\ee
As the dependence on $\delta$ and $\delta$'s definition through
\eq\nr{MBd} show, the logarithm is in some sense
related to the need to renormalize the scalar density
and its correlators.

\begin{figure}[t]

\centerline{%
\epsfysize=8.0cm\epsfbox{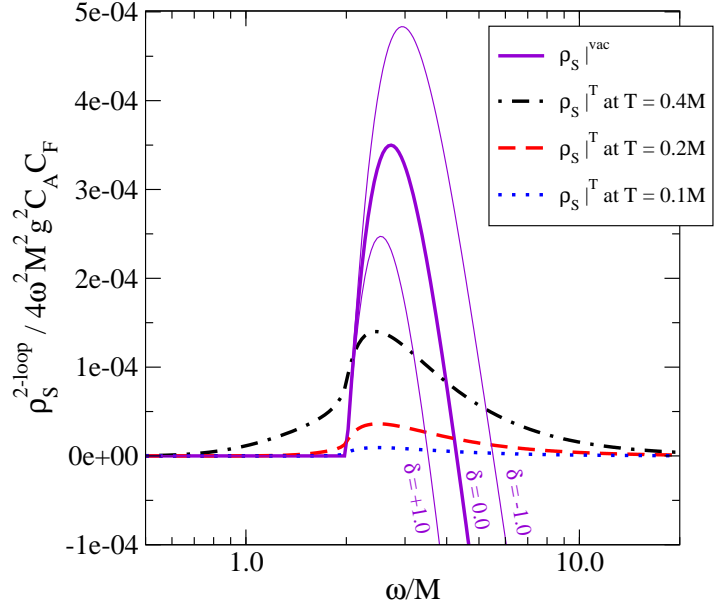}%
}

\vspace*{0.5cm}


\caption[a]{\small
The vacuum and thermal parts of the next-to-leading order 
correction in the scalar channel, normalized by dividing with
$4 \omega^2 M^2 g^2 C_A C_F$. The vacuum part can become
negative because the bare scalar correlator is multiplied with 
a bare mass parameter, cf.\ \eqs\nr{Sd_def}, \nr{MBd};
the constant $\delta$ illustrates how strong the dependence 
on the renormalization convention is.}

\la{fig:nloS}
\end{figure}

The thermal correction, in turn, reads, 
\ba
 \left. \rho_S(\omega) \right|^\rmii{$T$} \!\!\! & = & \!\!\!
 \frac{4 g^2 C_A C_F M^2}{(4\pi)^3 \omega^2}
 \int_0^\infty \! \dd k \, \frac{\nB{}(k)}{k} 
 \biggl\{
 \la{full_TS} \\
 & & \hspace*{-0.1cm}
   \theta(\omega) \, 
   \theta\Bigl(k- 
    \frac{4M^2-\omega^2}{2\omega}
    \Bigr)
   \biggl[ 
  -\;    (\omega^2 - 4 M^2) \sqrt{\omega(\omega+2 k)}
    \sqrt{\omega(\omega+2 k)- 4M^2}
  \nn & & \hspace*{0.1cm}
  +\; 2 \Bigl((\omega^2 - 2 M^2)(\omega^2 - 4 M^2) 
  + 2 \omega k (\omega^2 - 4 M^2) 
  + 2 \omega^2 k^2\Bigr)
  \, \mathrm{acosh} \sqrt{\frac{\omega(\omega+2k)}{4M^2}} 
   \biggr]
   \nn & + & \hspace*{-0.1cm}
   \theta(\omega - 2M) \, 
   \theta\Bigl(\frac{\omega^2-4M^2}{2\omega}-k\Bigr)
   \biggl[ 
   - \;   (\omega^2 - 4 M^2) \sqrt{\omega(\omega-2 k)}
    \sqrt{\omega(\omega-2 k)- 4M^2}
  \nn & & \hspace*{0.1cm}
   +\;  2 \Bigl((\omega^2 - 2 M^2)(\omega^2 - 4 M^2)
  - 2 \omega k (\omega^2 - 4 M^2) 
  + 2 \omega^2 k^2\Bigr)
  \, \mathrm{acosh} \sqrt{\frac{\omega(\omega-2k)}{4M^2}} 
   \biggr]
   \nn & + & \hspace*{-0.1cm}
  \theta(\omega - 2M) \biggl[ 
   2 (\omega^2 - 4 M^2 + 4 k^2)\, \omega \sqrt{\omega^2 - 4 M^2}
  \nn & & \hspace*{0.1cm}
   -\; 4 \Bigl((\omega^2 - 2 M^2)(\omega^2 - 4 M^2) + 2 \omega^2 k^2\Bigr)
   \, \mathrm{acosh} \biggl( \frac{\omega}{2M} \biggr) 
  \biggr] \biggr\}
 + \rmO(e^{-\beta M},g^4)
 \;, \nonumber
\ea
where we represented $T^2$ as
$
 \pi^2 T^2 = 6 \int_0^\infty {\rm d}k \, k \, \nB{}(k)
$.

A numerical evaluation of this result, compared 
with the vacuum part of \eq\nr{full_vacS}, is shown 
in \fig\ref{fig:nloS}. For small $\omega$ the thermal part 
appears to be somewhat more significant than in the case of the 
vector channel; this is because there is a cancellation
of positive and negative contributions in the vacuum part, before
the negative terms take over 
at large $\omega$ (cf.\ \eq\nr{full_vacS_asympt}).
The thermal part, in contrast,
stays positive and vanishes rapidly at large $\omega$.

We wish to draw attention to the amusing feature, 
already mentioned at the end of \se\ref{se:final}, that 
while the next-to-leading order vacuum part is continuous, 
the next-to-leading order thermal part appears even to have 
a continuous first derivative. In the vector channel, 
in contrast, the next-to-leading order 
vacuum part is discontinuous at the threshold, 
while the next-to-leading order thermal part 
appears to be continuous (cf.\ \fig\ref{fig:nlo}). 
In other words, the thermal part seems always
to be one degree smoother than the vacuum part.

\begin{figure}[t]

\centerline{%
 \epsfysize=5.0cm\epsfbox{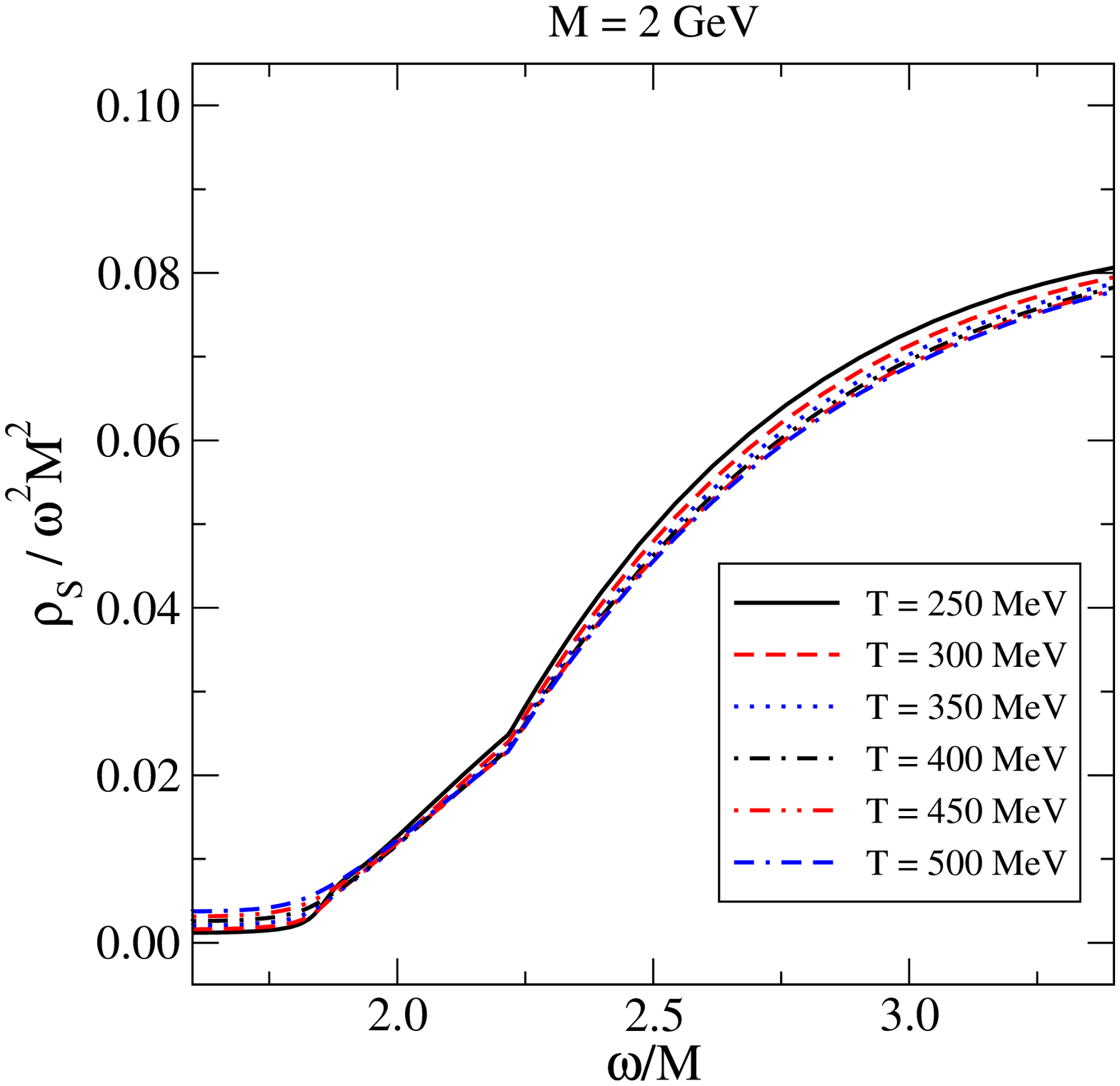}%
 ~~\epsfysize=5.0cm\epsfbox{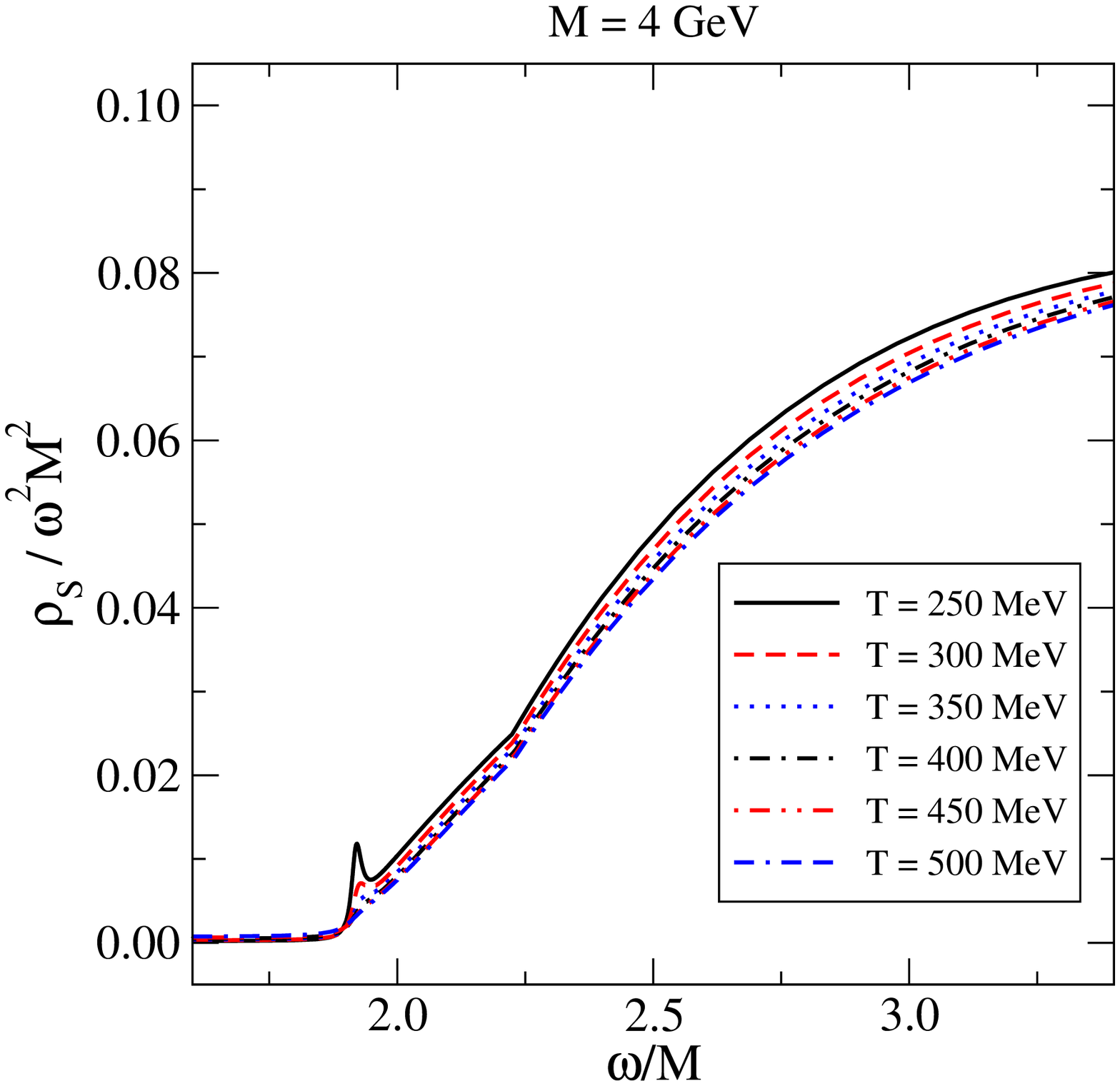}%
 ~~\epsfysize=5.0cm\epsfbox{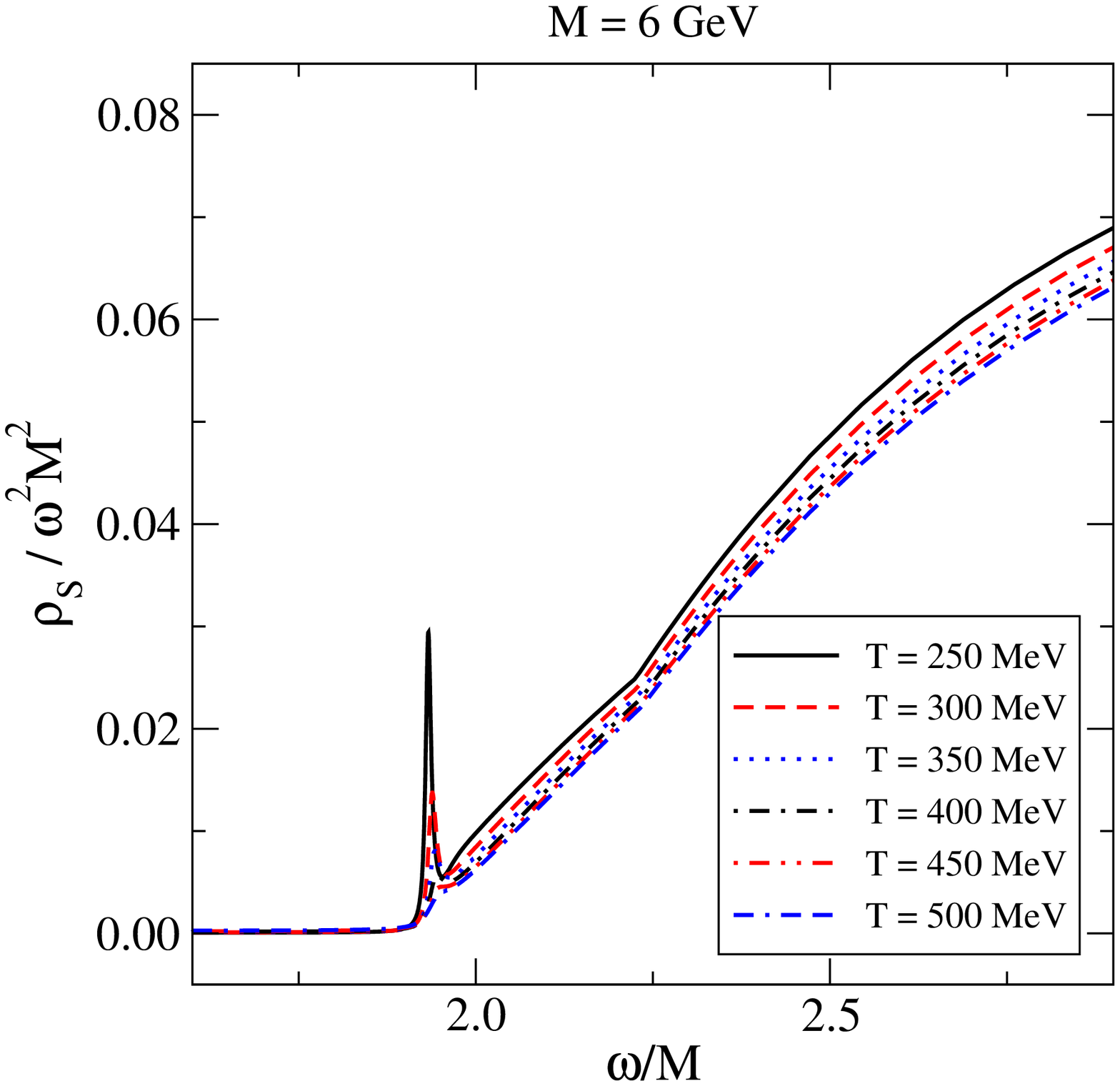}%
}

\vspace*{0.5cm}


\caption[a]{\small
The phenomenologically assembled scalar channel 
spectral function $\rho_S(\omega)$, in units of $\omega^2 M^2$, 
for $M = 2, 4, 6$~GeV (from left to right).
To the order considered, $M$ is the heavy quark pole mass. Note that
for better visibility, the axis ranges are different in the rightmost 
figure. As discussed after \eq\nr{norm_S}, we are not confident 
that these plots have a definite physical significance; 
the figures are meant for illustration only. 
}

\la{fig:rhoS_M}
\end{figure}

\begin{figure}[t]

\centerline{%
 \epsfysize=5.0cm\epsfbox{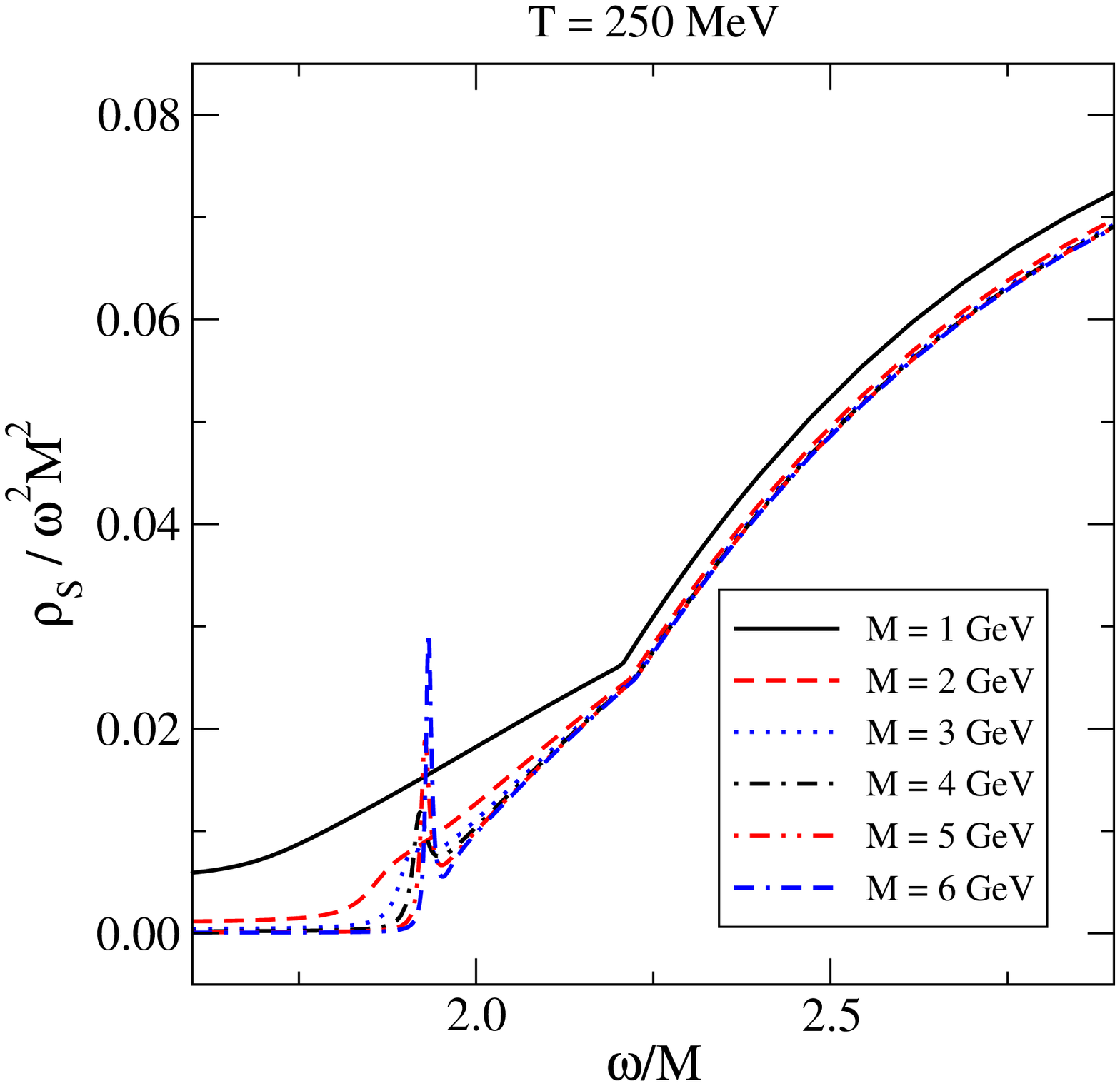}%
 ~~\epsfysize=5.0cm\epsfbox{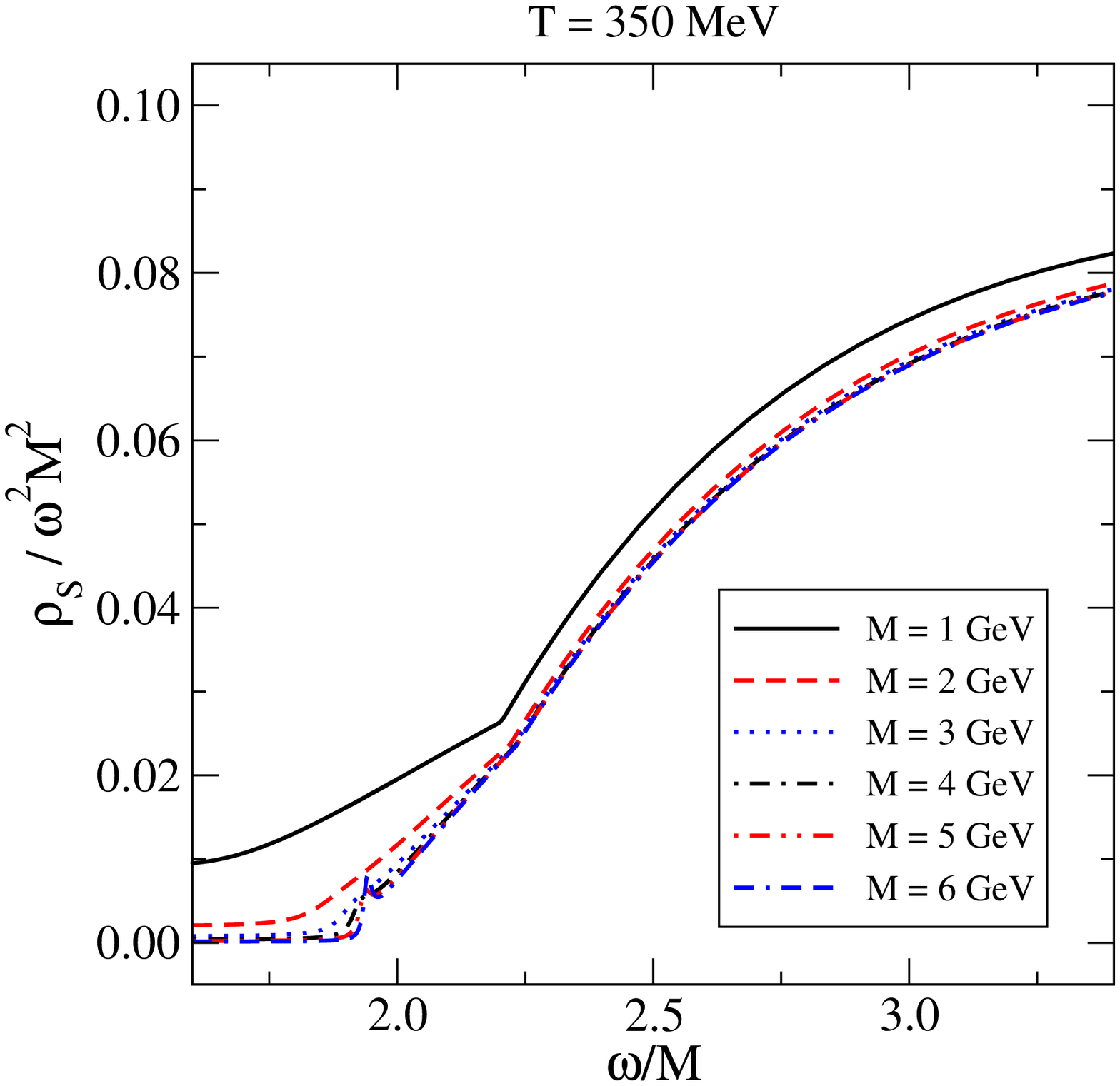}%
 ~~\epsfysize=5.0cm\epsfbox{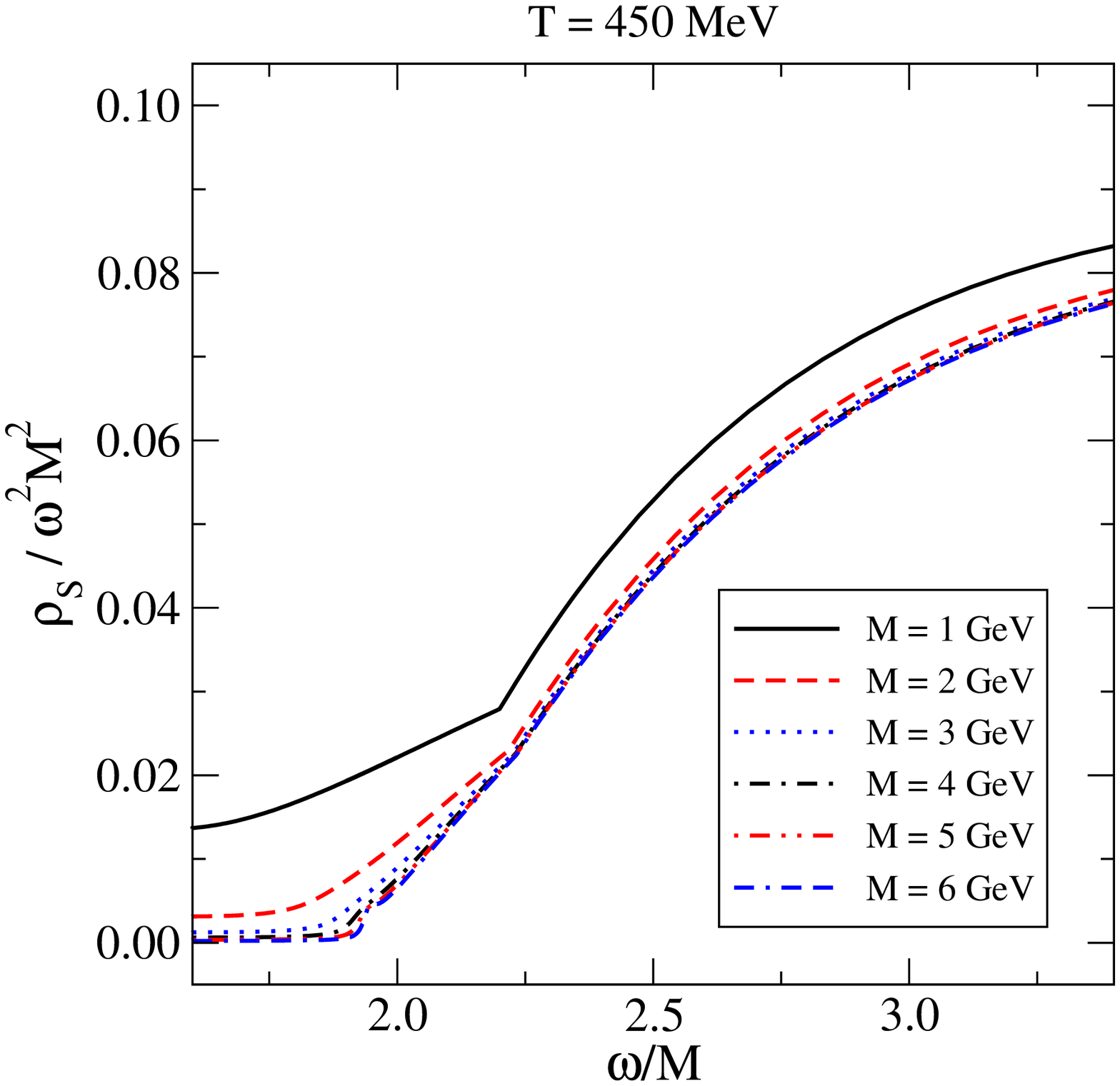}%
}

\vspace*{0.5cm}


\caption[a]{\small
The phenomenologically assembled scalar channel 
spectral function $\rho_S(\omega)$, in units of $\omega^2 M^2$, 
for $T = 250, 350, 450$~MeV (from left to right).
To the order considered, $M$ is the heavy quark pole mass. Note that
for better visibility, the axis ranges are different in the leftmost 
figure.  As discussed after \eq\nr{norm_S}, we are not confident 
that these plots have a definite physical significance; 
the figures are meant for illustration only. 
}

\la{fig:rhoS_T}
\end{figure}

In order to now combine our result with that obtained 
within a resummed framework in ref.~\cite{peskin}, 
we need to match the normalizations, in analogy 
with \eq\nr{V_norm}. Indeed, 
employing the notation of \eq\nr{v_def}, 
the leading order vacuum result in \eq\nr{full_vacS} becomes
\be
 \left. \frac{\rho_S(\omega)}{\omega^2 M^2} \right|_\rmii{LO} 
 = \theta(\omega - 2 M)
 \frac{C_A v^{3}}{8\pi}
 \;, \la{LO_exp_S} 
\ee
while the next-to-leading order result 
can be expanded as
\be
 \left. \frac{\rho_S(\omega)}{\omega^2 M^2} \right|_\rmii{NLO} 
 = 4 g^2 C_A C_F \theta(\omega - 2 M)
 \biggl[ \frac{v^2}{256\pi} - \frac{v^3}{128\pi^3}
 \biggl(1 + \fr32 \delta \biggr) + \rmO(v^4)
 \biggr]
 \;. \la{NLO_exp_S} 
\ee
Since radiative corrections within a non-relativistic potential
model always involve
a power of $v$, it is possible to account for the second term
in \eq\nr{NLO_exp_S}, equalling $-g^2 C_F(1 + 3\delta/2)/4 \pi^2$ 
times the leading
term in \eq\nr{LO_exp_V}, only by a multiplicative correction of 
the scalar density,  
\be
 {\cal S}_\rmii{QCD} = {\cal S}_\rmii{NRQCD}
 \biggl[1 - \frac{g^2 C_F}{8\pi^2}
 \biggl( 1 + \fr32 \delta \biggr) + ... \biggr]
 \;. \la{norm_S}
\ee
Even though closer to unity than in \eq\nr{V_norm}, the normalization factor 
could be numerically significant.  
In fact, if we leave the normalization factor open, 
and search for a value minimizing the squared difference of the 
resummed and QCD results (with $\delta = 0$)
in the range $(\omega-2M)/M = 0.0 - 0.4$ (thereby also accounting 
for thermal corrections), 
we find a best fit with an overall normalization 
factor $0.4 - 0.6$, i.e.\ with a {\em larger reduction}
than in the vector case, in contrast 
to what \eq\nr{norm_S} would suggest.\footnote{%
  We note, however, that if we introduce another fit parameter, 
  a horizontal energy shift, then the two results can be 
  matched smoothly, with a multiplicative factor close to unity. 
  We have not used 
  this method in the plots because we prefer  
  a universal procedure
  for the scalar and vector cases. 
  } 
This is perhaps another indication 
that the treatment of $\rho_S(\omega)$ within a 
Schr\"odinger-equation based resummed framework as 
in ref.~\cite{peskin} may not capture the correct physics. 

Nevertheless, putting this worry aside for a moment, we
again construct an ``assembled'' result as
$  
 \rho_\rmii{$S$}^\rmii{(assembled)} \equiv 
 \mbox{max}(\rho_\rmii{$S$}^\rmii{(QCD)},
 \rho_\rmii{$S$}^\rmii{(resummed)})
$.
The numerical value of 
the gauge coupling is taken from \eq\nr{numg2}.
The outcome is shown in 
\figs\ref{fig:rhoS_M}, \ref{fig:rhoS_T} for $\delta = 0$
and for various
masses and temperatures, as a function of $\omega$. 
Compared with the results in ref.~\cite{peskin}, the overall 
magnitude is smaller by about $40 - 60$\%. At the same time, as is 
obvious from the plots, the two results do not interpolate to 
each other well; we have no explanation for this at the moment, 
but wish to repeat our concerns on the validity 
of the resummed near-threshold
function $\rho_\rmii{$S$}^\rmii{(resummed)}$.

\newpage



\begin{thebibliography}{99}

\bibitem{old}
  G.~K\"all\'en and A.~Sabry,
  Kong.\ Dan.\ Vid.\ Sel.\ Mat.\ Fys.\ Med.\  {29N17} (1955) 1;
%
  J.S.~Schwinger,
  {\it Particles, Sources and Fields. Volume II}, p.~407
 (Addison-Wesley, 1973).

\bibitem{br} 
  R.~Barbieri and E.~Remiddi,
  Nuovo Cim.\  A {13} (1973) 99.

\bibitem{db}
  D.J.~Broadhurst, J.~Fleischer and O.V.~Tarasov,
  Z.\ Phys.\  C {60} (1993) 287
  [hep-ph/9304303].

\bibitem{ah} 
  A.H.~Hoang, V.~Mateu and S.~Mohammad Zebarjad,
  0807.4173.

\bibitem{dilepton}
  L.D.~McLerran and T.~Toimela,
  Phys.\ Rev.\ D {31} (1985) 545;
%
  H.A.~Weldon,
  Phys.\ Rev.\ D {42} (1990) 2384;
%
  C.~Gale and J.I.~Kapusta,
  Nucl.\ Phys.\ B {357} (1991) 65.

\bibitem{ms}
  T.~Matsui and H.~Satz,
  Phys.\ Lett.\ B {178} (1986) 416.

\bibitem{static}
  M.~Laine, O.~Philipsen, P.~Romatschke and M.~Tassler,
  JHEP {03} (2007) 054
  [hep-ph/0611300];
%
  M.~Laine,
  JHEP {05} (2007) 028
  [0704.1720];
%
  M.~Laine, O.~Philipsen and M.~Tassler,
  JHEP {09} (2007) 066
  [0707.2458].

\bibitem{peskin}
  Y.~Burnier, M.~Laine and M.~Veps\"al\"ainen,
  JHEP { 01} (2008) 043
  [0711.1743].

\bibitem{ab}
  A.~Beraudo, J.P.~Blaizot and C.~Ratti,
  Nucl.\ Phys.\  A { 806} (2008) 312
  [0712.4394].

\bibitem{es}
  M.A.~Escobedo and J.~Soto,
  0804.0691. 

\bibitem{nb3}
  N.~Brambilla, J.~Ghiglieri, A.~Vairo and P.~Petreczky,
  Phys.\ Rev.\  D {78} (2008) 014017
  [0804.0993].

\bibitem{sewm}
  M.~Laine,
  0810.1112. 

\bibitem{col}
  F.~Dominguez and B.~Wu,
  0811.1058.

\bibitem{phi_revs}
  R.~Rapp, D.~Blaschke and P.~Crochet,
  0807.2470. 


\bibitem{phi_revs2}
  A.~M\'ocsy,
  0811.0337. 

\bibitem{op}
  O.~Philipsen,
  0810.4685. 

\bibitem{ads}
  R.C.~Myers, A.O.~Starinets and R.M.~Thomson,
  JHEP {11} (2007) 091
  [0706.0162].


\bibitem{ir}
  A.D.~Linde,
  Phys.\ Lett.\ {B 96} (1980) 289;
%
  D.J.~Gross, R.D.~Pisarski and L.G.~Yaffe,
  Rev.\ Mod.\ Phys.\ {53} (1981) 43.

\bibitem{nspt_mass}
  F.~Di Renzo, M.~Laine, V.~Miccio, Y.~Schr\"oder and C.~Torrero,
  JHEP {07} (2006) 026
  [hep-ph/0605042].

\bibitem{fund}
  G.~Cuniberti, E.~De Micheli and G.A.~Viano,
  Commun.\ Math.\ Phys.\  {216} (2001) 59
  [cond-mat/0109175].


\bibitem{latt}
  A.~Jakov\'ac, P.~Petreczky, K.~Petrov and A.~Velytsky,
  Phys.\ Rev.\  {D 75} (2007) 014506
  [hep-lat/0611017];
%
  G.~Aarts, C.~Allton, M.B.~Oktay, M.~Peardon and J.I.~Skullerud,
  Phys.\ Rev.\  D {76} (2007) 094513
  [0705.2198].

\bibitem{latt_rev}
  P.~Petreczky,
  0810.0258. 

\bibitem{free_spectral}
  F.~Karsch, E.~Laermann, P.~Petreczky and S.~Stickan,
  Phys.\ Rev.\  D {68} (2003) 014504
  [hep-lat/0303017];
%
  G.~Aarts and J.M.~Mart\'{\i}nez Resco,
  Nucl.\ Phys.\  B {726} (2005) 93
  [hep-lat/0507004];
%
%
  A.~M\'ocsy and P.~Petreczky,
  Phys.\ Rev.\ D {73} (2006) 074007
  [hep-ph/0512156];
%
  G.~Aarts and J.~Foley, 
  JHEP {02} (2007) 062
  [hep-lat/0612007].

\bibitem{hs}
  R.V.~Harlander and M.~Steinhauser,
  Comput.\ Phys.\ Commun.\  {153} (2003) 244
  [hep-ph/0212294].


\bibitem{leb}
  M. Le Bellac, {\it Thermal Field Theory}
  (Cambridge University Press, Cambridge, 2000).

\bibitem{kg}
  J.I.~Kapusta and C.~Gale,
  {\it Finite-Temperature Field Theory: Principles and Applications} 
  (Cambridge University Press, Cambridge, 2006).

\bibitem{htlold}
  V.P.~Silin, 
  Sov.\ Phys.\ JETP {11} (1960) 1136
  [Zh.\ Eksp.\ Teor.\ Fiz.\ {38} (1960) 1577];
  %
  V.V.~Klimov,
  Sov.\ Phys.\ JETP {55} (1982) 199
  [Zh.\ Eksp.\ Teor.\ Fiz.\  {82} (1982) 336];
  %
  H.A.~Weldon,
  Phys.\ Rev.\ D {26} (1982) 1394.

\bibitem{htl}
  R.D.~Pisarski,
  Phys.\ Rev.\ Lett.\  {63} (1989) 1129;
  %
  J.~Frenkel and J.C.~Taylor,
  Nucl.\ Phys.\ B {334} (1990) 199;
  %
  E.~Braaten and R.D.~Pisarski,
  Nucl.\ Phys.\ B {337} (1990) 569;
  %
  J.C.~Taylor and S.M.H.~Wong,
  Nucl.\ Phys.\ B {346} (1990) 115.

\bibitem{dhr}
  J.F.~Donoghue, B.R.~Holstein and R.W.~Robinett,
  Annals Phys.\  { 164} (1985) 233
  [Erratum-ibid.\  { 172} (1986) 483].

\bibitem{cl}
  W.E.~Caswell and G.P.~Lepage,
  Phys.\ Lett.\ B {167} (1986) 437.


\bibitem{kt}
  J.G.~K\"orner and G.~Thompson,
  Phys.\ Lett.\  B {264} (1991) 185.

\bibitem{ps}
  A.~Pineda and J.~Soto,
  Nucl.\ Phys.\ B (Proc.\ Suppl.)\  {64} (1998) 428
  [hep-ph/9707481].

\bibitem{nb2}
  N.~Brambilla, A.~Pineda, J.~Soto and A.~Vairo,
  Nucl.\ Phys.\ B {566} (2000) 275
  [hep-ph/9907240].

\bibitem{bkp}
  M.~Beneke, Y.~Kiyo and A.A.~Penin,
  Phys.\ Lett.\  B { 653} (2007) 53
  [0706.2733].


\bibitem{current}
  A.~Czarnecki and K.~Melnikov,
  Phys.\ Rev.\ Lett.\  {80} (1998) 2531
  [hep-ph/9712222];
%
  M.~Beneke, A.~Signer and V.A.~Smirnov,
  Phys.\ Rev.\ Lett.\  {80} (1998) 2535
  [hep-ph/9712302].

\bibitem{gE2}
  M.~Laine and Y.~Schr\"oder,
  JHEP {03} (2005) 067
  [hep-ph/0503061].

\bibitem{adjoint}
  K.~Kajantie, M.~Laine, K.~Rummukainen and M.~Shaposhnikov,
  Nucl.\ Phys.\ B {503} (1997) 357
  [hep-ph/9704416].

\bibitem{diff}
  E.~Braaten and M.H.~Thoma,
  Phys.\ Rev.\  D {44} (1991) 2625;
%
  G.D.~Moore and D.~Teaney,
  Phys.\ Rev.\  C {71} (2005) 064904
  [hep-ph/0412346];
%
  P.~Petreczky and D.~Teaney,
  Phys.\ Rev.\  D {73} (2006) 014508
  [hep-ph/0507318].


\bibitem{chm}
  S.~Caron-Huot and G.D.~Moore,
  JHEP {02} (2008) 081
  [0801.2173].




\end{thebibliography}
\end{document}